\documentclass[manuscript,screen,authorversion, nonacm]{acmart}

\usepackage{tikz}
\usepackage[framemethod=tikz]{mdframed}
\usepackage{xcolor}
\usepackage{array}
\usepackage{multirow}
\usepackage{booktabs}
\usepackage{makecell}
\usepackage{url}
\usepackage{graphicx}
\usepackage{enumitem}
\usepackage{subcaption}

\AtBeginDocument{%
  }

\setcopyright{rightsretained}
\copyrightyear{2025}
\acmYear{2025}
\acmDOI{10.48550/arXiv.2501.08500}

\definecolor{summaryCol}{RGB}{0, 102, 204}

\tikzset{
  summarysymbol/.style={
    rectangle,
    rounded corners=2pt,
    inner sep=3pt,
    draw=summaryCol!50,
    fill=summaryCol,
    text=white,
    scale=1,
    overlay,
    font=\small\bfseries
  }
}
\usetikzlibrary{calc}
\newmdenv[
  linecolor=summaryCol,
  linewidth=2pt,
  leftline=true,
  topline=false,
  bottomline=false,
  rightline=false,
  backgroundcolor=summaryCol!5,
  innerleftmargin=8pt,
  innertopmargin=3pt,
  innerbottommargin=2pt,
  skipabove=8pt,
  skipbelow=0pt,
  firstextra={
    \path let \p1=(P), \p2=(O) in 
    node[summarysymbol] at ($(\x2,\y1)$)  {S};
  },
  singleextra={
    \path let \p1=(P), \p2=(O) in 
    node[summarysymbol] at ($(\x2,\y1)$)  {S};
  }
]{summarybox}

\begin{document}

\newcommand{\quot}[1]{``#1''}
\newcommand{\todo}[1]{\textcolor{red}{TODO: #1}}

\definecolor{urlcolor}{RGB}{0, 118, 196}
\newcommand{\websiteurl}{\textcolor{urlcolor}{{\href{https://networks-ie-survey.dbvis.de}{networks-ie-survey.dbvis.de}}}}

\newcommand{\apps}{\texttt{A}}
\newcommand{\studies}{\texttt{S}}
\newcommand{\appstudycount}[2]{\apps{}:~#1, \studies{}:~#2}

\definecolor{datacolordef}{RGB}{4, 143, 11}
\definecolor{techcolordef}{RGB}{219, 97, 15}

\newcommand{\datacolor}[1]{\textcolor{datacolordef}{#1}}
\newcommand{\techcolor}[1]{\textcolor{techcolordef}{#1}}
\newcommand{\datakeyword}[1]{\datacolor{\quot{#1}}}
\newcommand{\techkeyword}[1]{\techcolor{\quot{#1}}}


\definecolor{pipelineBlue}{RGB}{29, 76, 125}
\definecolor{pipelineGreen}{RGB}{48, 111, 29}
\definecolor{pipelineGray}{RGB}{94, 93, 89}

\newcommand{\filledRec}{\tikz\fill[black] (0,0) rectangle (4pt,4pt);}
\newcommand{\unfilledRec}{\tikz\draw[black] (0,0) rectangle (4pt,4pt);}
\newcommand{\filledTri}{\tikz\fill[black] (0,0) -- (4pt,0) -- (2pt,4pt) -- cycle;}
\newcommand{\unfilledTri}{\tikz\draw[black] (0,0) -- (4pt,0) -- (2pt,4pt) -- cycle;}

\newcommand{\filledRecText}{\tikz\fill[black] (0,0) rectangle (6pt,6pt);}
\newcommand{\unfilledRecText}{\tikz\draw[black] (0,0) rectangle (6pt,6pt);}
\newcommand{\filledTriText}{\tikz\fill[black] (0,0) -- (6pt,0) -- (3pt,6pt) -- cycle;}
\newcommand{\unfilledTriText}{\tikz\draw[black] (0,0) -- (6pt,0) -- (3pt,6pt) -- cycle;}

\newcommand{\domainabstract}{\raisebox{-0.2\height}{\includegraphics[height=1em]{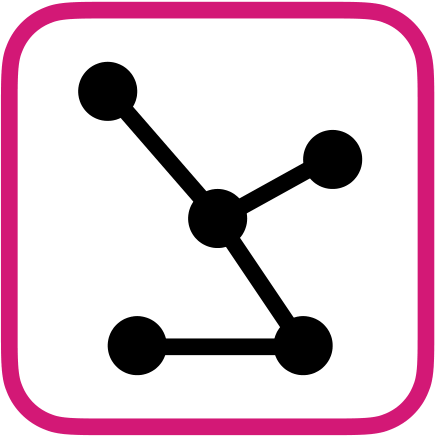}}}
\newcommand{\domainabstractheadline}{\includegraphics[bb=0pt 20pt 100pt 110pt, height=1.0em]{img/glyphs/domain_abstract.png}}
\newcommand{\domainaai}{\raisebox{-0.2\height}{\includegraphics[height=1em]{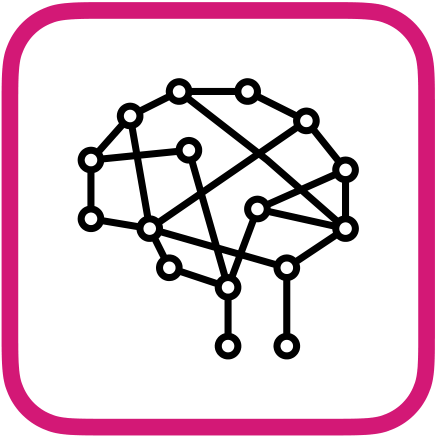}}}
\newcommand{\domainaaiheadline}{\includegraphics[bb=0pt 20pt 100pt 110pt, height=1.0em]{img/glyphs/domain_ai.png}}
\newcommand{\domainbio}{\raisebox{-0.2\height}{\includegraphics[height=1em]{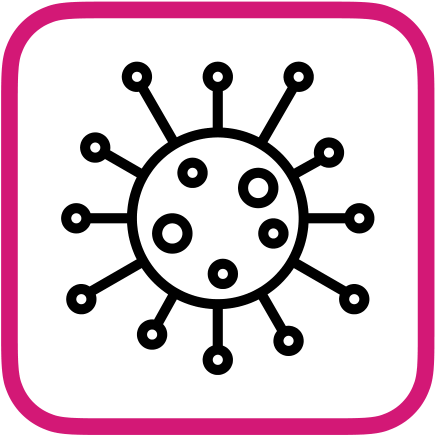}}}
\newcommand{\domainbioheadline}{\includegraphics[bb=0pt 20pt 100pt 110pt, height=1.0em]{img/glyphs/domain_bio.png}}
\newcommand{\domainknowledge}{\raisebox{-0.2\height}{\includegraphics[height=1em]{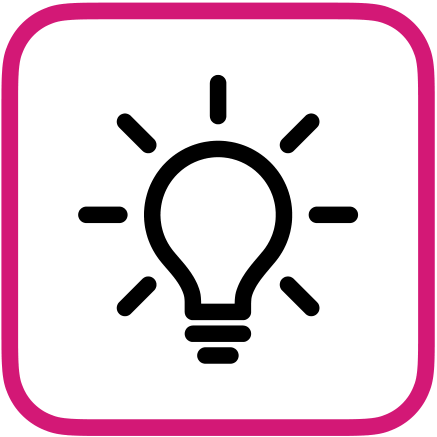}}}
\newcommand{\domainknowledgeheadline}{\includegraphics[bb=0pt 20pt 100pt 110pt, height=1.0em]{img/glyphs/domain_knowledge.png}}
\newcommand{\domainnetwork}{\raisebox{-0.2\height}{\includegraphics[height=1em]{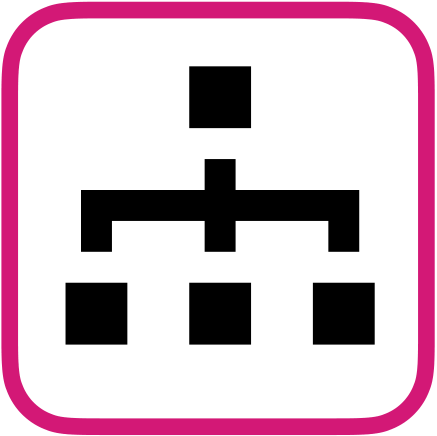}}}
\newcommand{\domainnetworkheadline}{\includegraphics[bb=0pt 20pt 100pt 110pt, height=1.0em]{img/glyphs/domain_network.png}}
\newcommand{\domainother}{\raisebox{-0.2\height}{\includegraphics[height=1em]{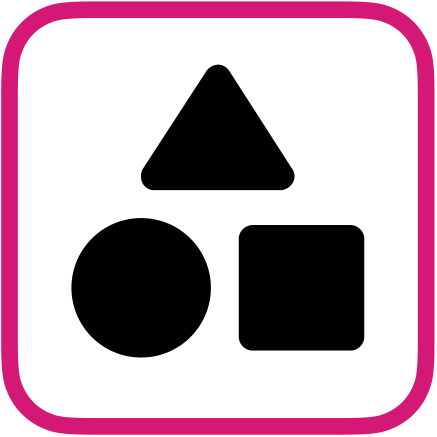}}}
\newcommand{\domainotherheadline}{\includegraphics[bb=0pt 20pt 100pt 110pt, height=1.0em]{img/glyphs/domain_other.png}}
\newcommand{\domainsocial}{\raisebox{-0.2\height}{\includegraphics[height=1em]{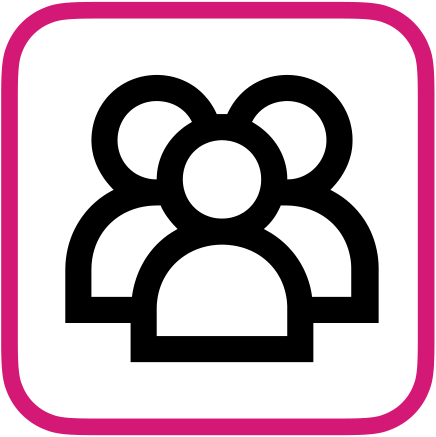}}}
\newcommand{\domainsocialheadline}{\includegraphics[bb=0pt 20pt 100pt 110pt, height=1.0em]{img/glyphs/domain_social.png}}
\newcommand{\domainsoftware}{\raisebox{-0.2\height}{\includegraphics[height=1em]{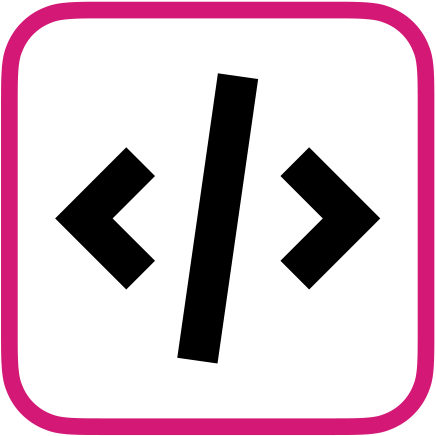}}}
\newcommand{\domainsoftwareheadline}{\includegraphics[bb=0pt 20pt 100pt 110pt, height=1.0em]{img/glyphs/domain_software.png}}

\newcommand{\displayhmd}{\raisebox{-0.2\height}{\includegraphics[height=1em]{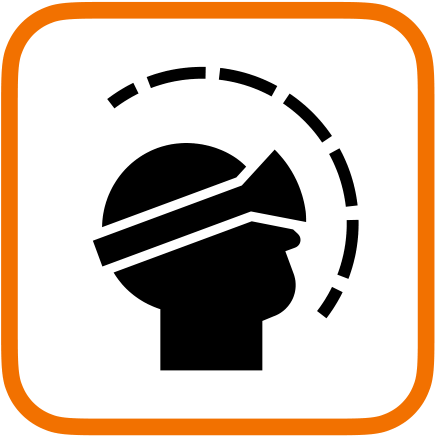}}}
\newcommand{\displayhmdheadline}{\includegraphics[bb=0pt 20pt 100pt 110pt, height=1.0em]{img/glyphs/display_hmd.png}}
\newcommand{\displaytwodsmall}{\raisebox{-0.2\height}{\includegraphics[height=1em]{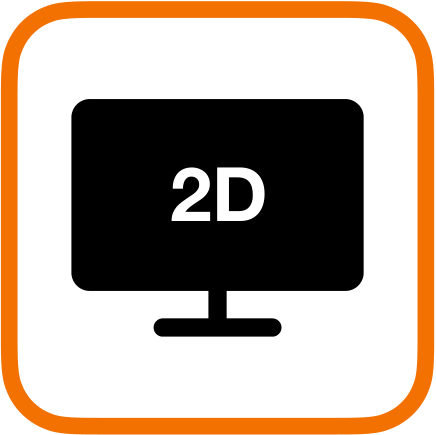}}}
\newcommand{\displaytwodsmallheadline}{\includegraphics[bb=0pt 20pt 100pt 110pt, height=1.0em]{img/glyphs/display_2d_small.png}}
\newcommand{\displaycave}{\raisebox{-0.2\height}{\includegraphics[height=1em]{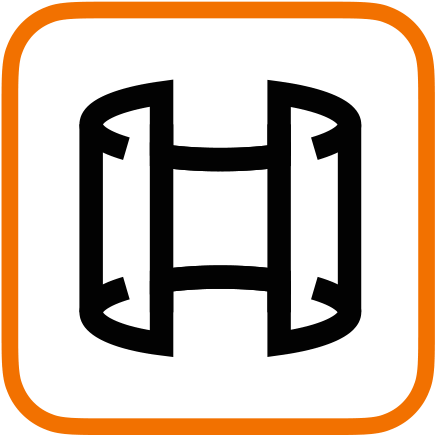}}}
\newcommand{\displaycaveheadline}{\includegraphics[bb=0pt 20pt 100pt 110pt, height=1.0em]{img/glyphs/display_cave.png}}
\newcommand{\displaymobile}{\raisebox{-0.2\height}{\includegraphics[height=1em]{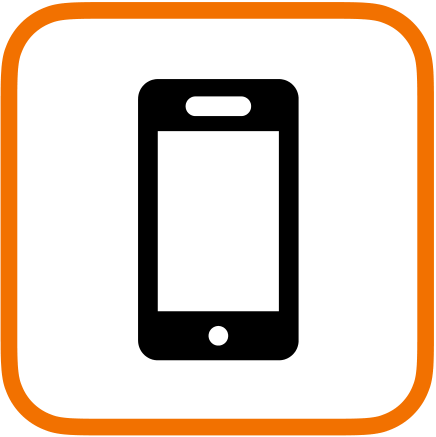}}}
\newcommand{\displaymobileheadline}{\includegraphics[bb=0pt 20pt 100pt 110pt, height=1.0em]{img/glyphs/display_mobile.png}}
\newcommand{\displayphysicalisation}{\raisebox{-0.2\height}{\includegraphics[height=1em]{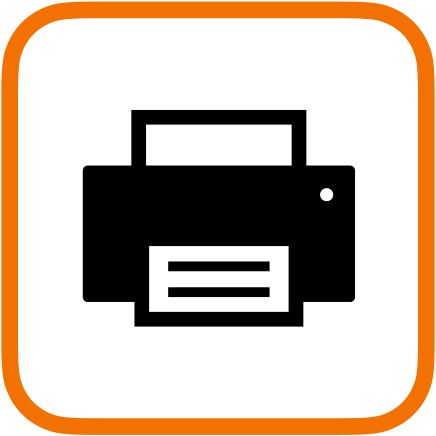}}}
\newcommand{\displayphysicalisationheadline}{\includegraphics[bb=0pt 20pt 100pt 110pt, height=1.0em]{img/glyphs/display_physicalisation.png}}
\newcommand{\displaylargetwod}{\raisebox{-0.2\height}{\includegraphics[height=1em]{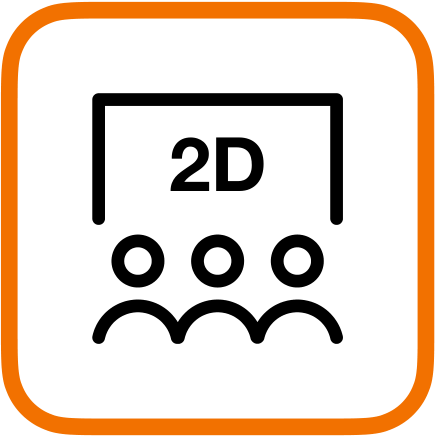}}}
\newcommand{\displaylargetwodheadline}{\includegraphics[bb=0pt 20pt 100pt 110pt, height=1.0em]{img/glyphs/display_screen_2D.png}}
\newcommand{\displaythreed}{\raisebox{-0.2\height}{\includegraphics[height=1em]{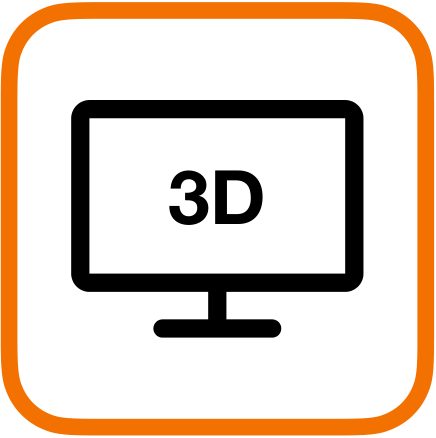}}}
\newcommand{\displaythreedheadline}{\includegraphics[bb=0pt 20pt 100pt 110pt, height=1.0em]{img/glyphs/display_screen_3D.png}}

\newcommand{\interactioncontroller}{\raisebox{-0.2\height}{\includegraphics[height=1em]{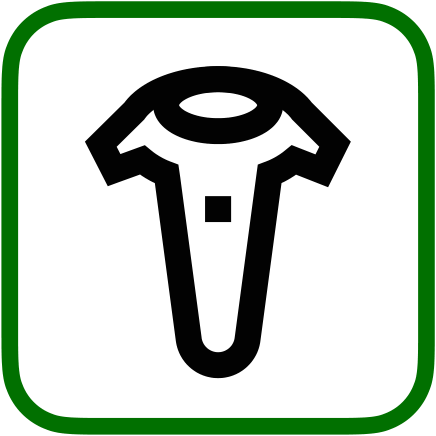}}}
\newcommand{\interactioncontrollerheadline}{\includegraphics[bb=0pt 20pt 100pt 110pt, height=1.0em]{img/glyphs/interaction_controller.png}}
\newcommand{\interactiongestures}{\raisebox{-0.2\height}{\includegraphics[height=1em]{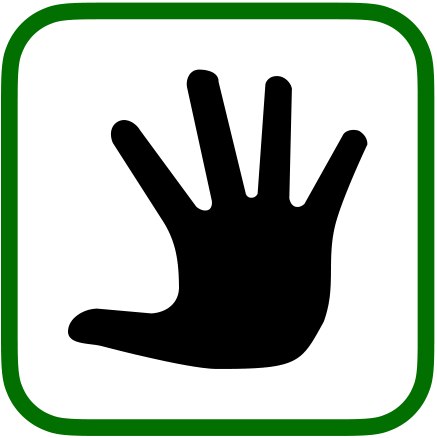}}}
\newcommand{\interactiongesturesheadline}{\includegraphics[bb=0pt 20pt 100pt 110pt, height=1.0em]{img/glyphs/interaction_gesture.png}}
\newcommand{\interactiontouch}{\raisebox{-0.2\height}{\includegraphics[height=1em]{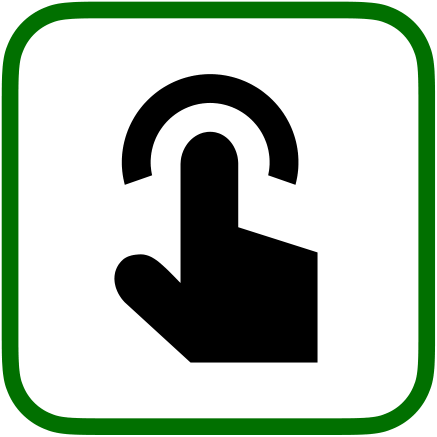}}}
\newcommand{\interactiontouchheadline}{\includegraphics[bb=0pt 20pt 100pt 110pt, height=1.0em]{img/glyphs/interaction_touch.png}}
\newcommand{\interactionmousekeyboard}{\raisebox{-0.2\height}{\includegraphics[height=1em]{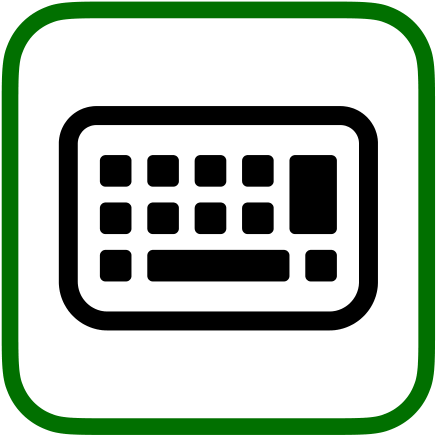}}}
\newcommand{\interactionmousekeyboardheadline}{\includegraphics[bb=0pt 20pt 100pt 110pt, height=1.0em]{img/glyphs/interaction_keyboard-mouse.png}}
\newcommand{\interactiongaze}{\raisebox{-0.2\height}{\includegraphics[height=1em]{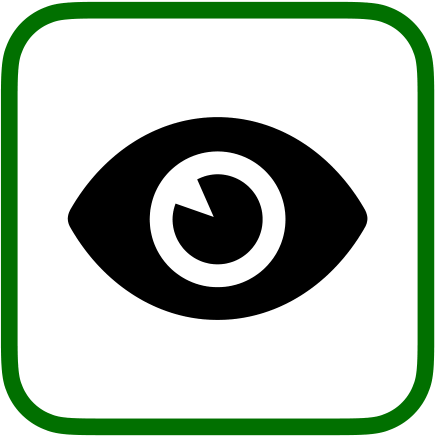}}}
\newcommand{\interactiongazeheadline}{\includegraphics[bb=0pt 20pt 100pt 110pt, height=1.0em]{img/glyphs/interaction_gaze.png}}
\newcommand{\interactionspeech}{\raisebox{-0.2\height}{\includegraphics[height=1em]{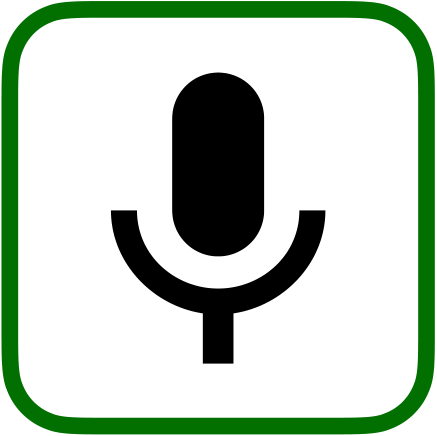}}}
\newcommand{\interactionspeechheadline}{\includegraphics[bb=0pt 20pt 100pt 110pt, height=1.0em]{img/glyphs/interaction_voice.png}}
\newcommand{\interactionmovement}{\raisebox{-0.2\height}{\includegraphics[height=1em]{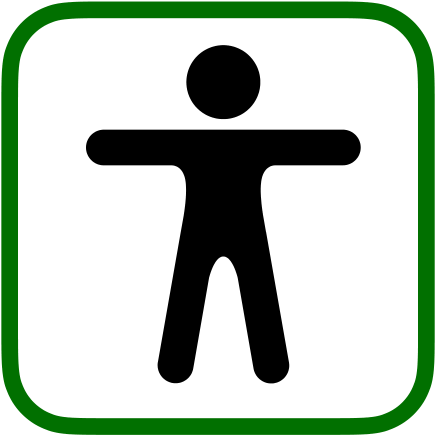}}}
\newcommand{\interactionmovementheadline}{\includegraphics[bb=0pt 20pt 100pt 110pt, height=1.0em]{img/glyphs/interaction_movement.png}}

\newcommand{\nodescircle}{\raisebox{-0.2\height}{\includegraphics[height=1em]{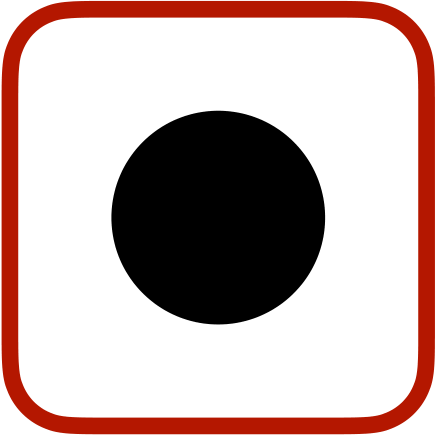}}}
\newcommand{\nodescircleheadline}{\includegraphics[bb=0pt 20pt 100pt 110pt, height=1.0em]{img/glyphs/nodes_circle.png}}
\newcommand{\nodeplane}{\raisebox{-0.2\height}{\includegraphics[height=1em]{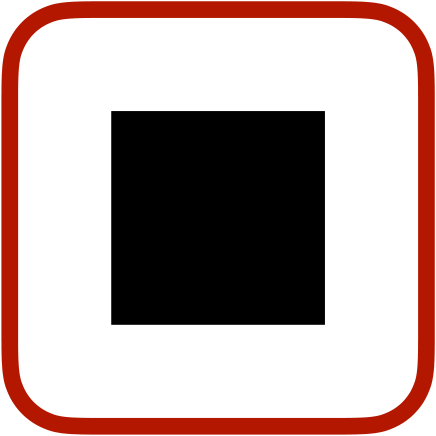}}}
\newcommand{\nodeplaneheadline}{\includegraphics[bb=0pt 20pt 100pt 110pt, height=1.0em]{img/glyphs/nodes_plane.png}}
\newcommand{\nodescube}{\raisebox{-0.2\height}{\includegraphics[height=1em]{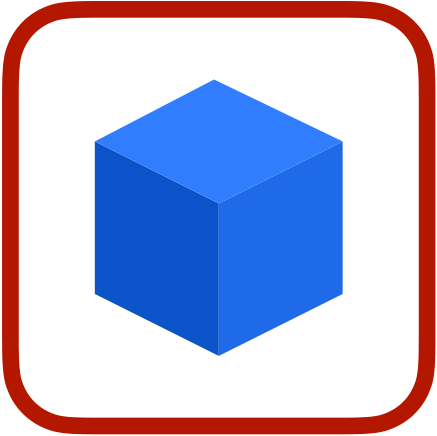}}}
\newcommand{\nodescubeheadline}{\includegraphics[bb=0pt 20pt 100pt 110pt, height=1.0em]{img/glyphs/nodes_cube.png}}
\newcommand{\nodessphere}{\raisebox{-0.2\height}{\includegraphics[height=1em]{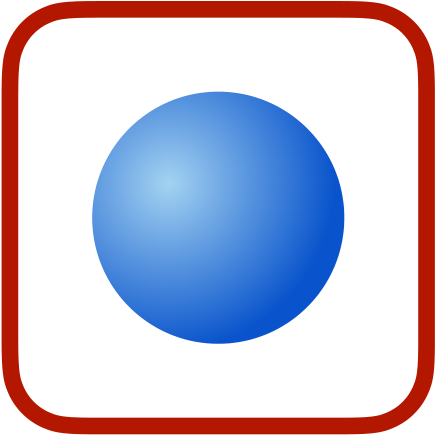}}}
\newcommand{\nodessphereheadline}{\includegraphics[bb=0pt 20pt 100pt 110pt, height=1.0em]{img/glyphs/nodes_sphere.png}}
\newcommand{\nodesother}{\raisebox{-0.2\height}{\includegraphics[height=1em]{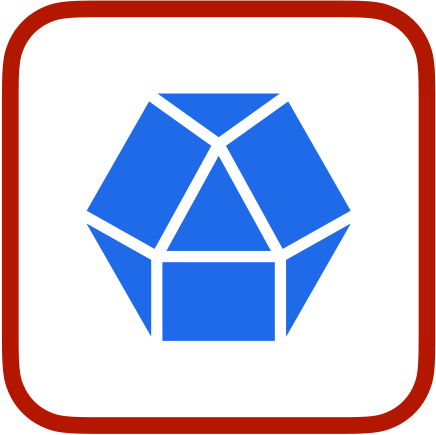}}}
\newcommand{\nodesotherheadline}{\includegraphics[bb=0pt 20pt 100pt 110pt, height=1.0em]{img/glyphs/nodes_other.png}}

\newcommand{\edgeslines}{\raisebox{-0.2\height}{\includegraphics[height=1em]{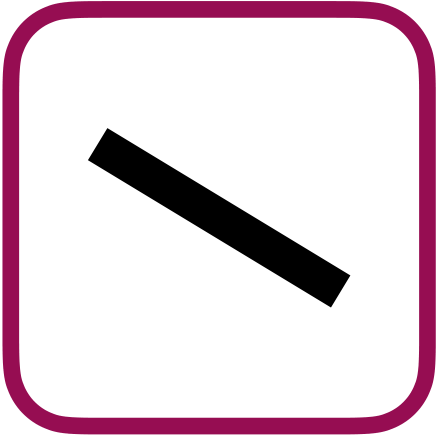}}}
\newcommand{\edgeslinesheadline}{\includegraphics[bb=0pt 20pt 100pt 110pt, height=1.0em]{img/glyphs/edges_line.png}}
\newcommand{\edgestubes}{\raisebox{-0.2\height}{\includegraphics[height=1em]{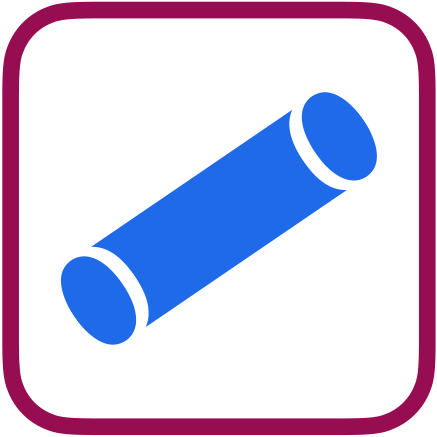}}}
\newcommand{\edgestubesheadline}{\includegraphics[bb=0pt 20pt 100pt 110pt, height=1.0em]{img/glyphs/edges_tube.png}}

\newcommand{\layoutdimtwo}{\raisebox{-0.2\height}{\includegraphics[height=1em]{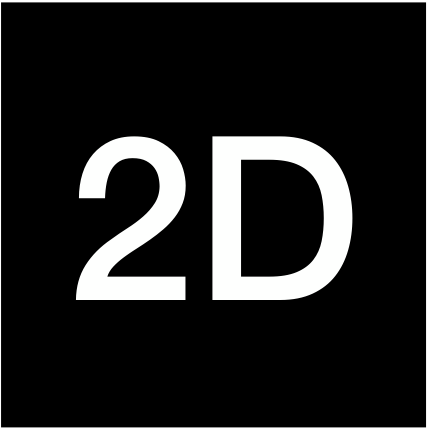}}}
\newcommand{\layoutdimtwoheadline}{\includegraphics[bb=0pt 20pt 100pt 110pt, height=1.0em]{img/glyphs/layout_2d.png}}
\newcommand{\layoutdimthree}{\raisebox{-0.2\height}{\includegraphics[height=1em]{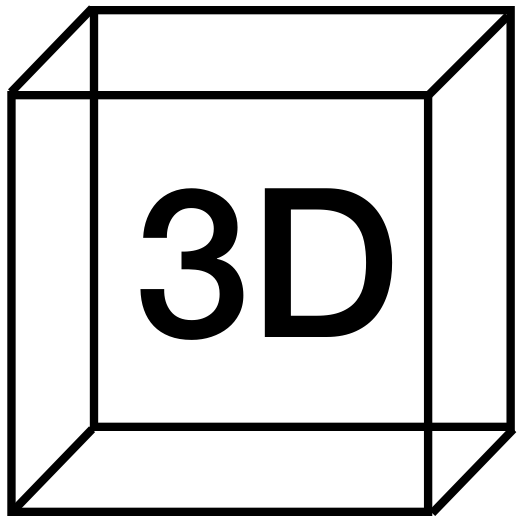}}}
\newcommand{\layoutdimthreeheadline}{\includegraphics[bb=0pt 20pt 100pt 110pt, height=1.0em]{img/glyphs/layout_3d.png}}

\newcommand{\layoutperspexo}{\raisebox{-0.2\height}{\includegraphics[height=1em]{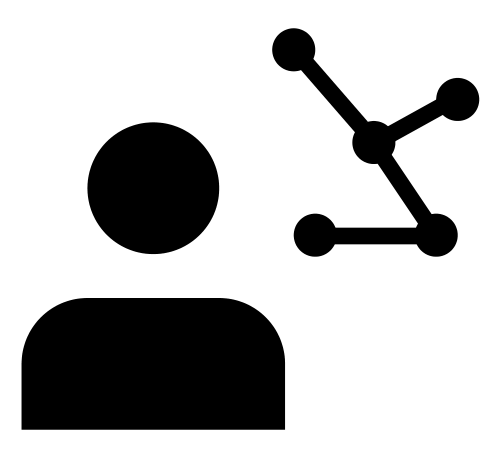}}}
\newcommand{\layoutperspexoheadline}{\includegraphics[bb=0pt 20pt 100pt 110pt, height=1.0em]{img/glyphs/layout_perspective_exocentric.png}}
\newcommand{\layoutperspego}{\raisebox{-0.2\height}{\includegraphics[height=1em]{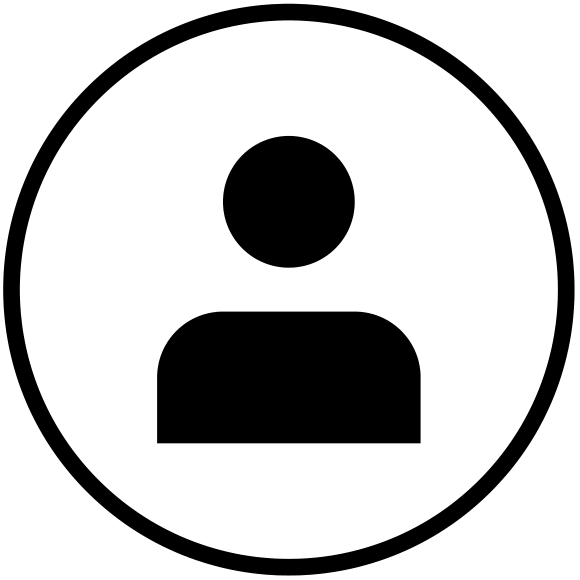}}}
\newcommand{\layoutperspegoheadline}{\includegraphics[bb=0pt 20pt 100pt 110pt, height=1.0em]{img/glyphs/layout_perspective_egocentric.png}}

\newcommand{\layoutmethforce}{\raisebox{-0.2\height}{\includegraphics[height=1em]{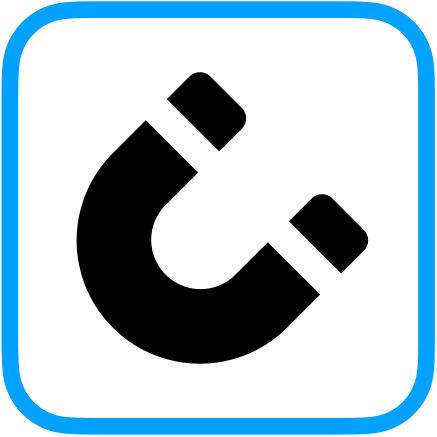}}}
\newcommand{\layoutmethforceheadline}{\includegraphics[bb=0pt 20pt 100pt 110pt, height=1.0em]{img/glyphs/layout_method_force.png}}
\newcommand{\layoutmethsemantic}{\raisebox{-0.2\height}{\includegraphics[height=1em]{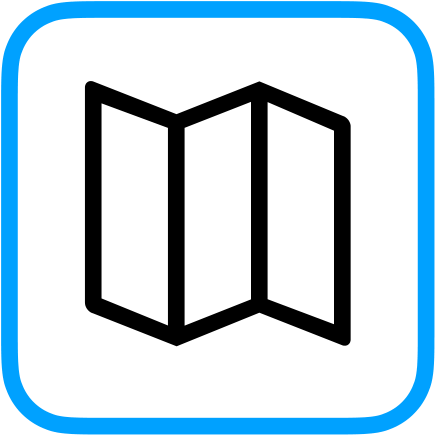}}}
\newcommand{\layoutmethsemanticheadline}{\includegraphics[bb=0pt 20pt 100pt 110pt, height=1.0em]{img/glyphs/layout_method_semantic.png}}
\newcommand{\layoutmethprocess}{\raisebox{-0.2\height}{\includegraphics[height=1em]{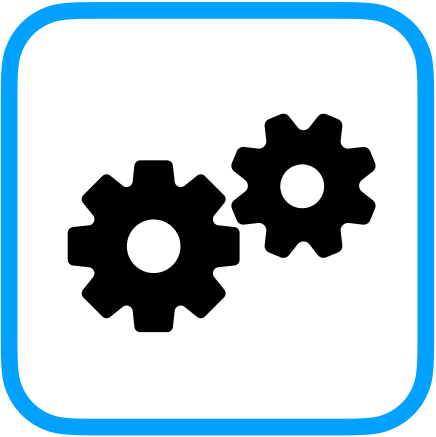}}}
\newcommand{\layoutmethprocessheadline}{\includegraphics[bb=0pt 20pt 100pt 110pt, height=1.0em]{img/glyphs/layout_method_process.png}}
\newcommand{\layoutmethlayered}{\raisebox{-0.2\height}{\includegraphics[height=1em]{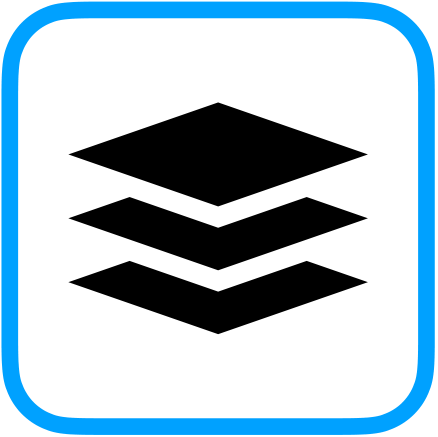}}}
\newcommand{\layoutmethlayeredheadline}{\includegraphics[bb=0pt 20pt 100pt 110pt, height=1.0em]{img/glyphs/layout_method_layer.png}}
\newcommand{\layoutmethcircular}{\raisebox{-0.2\height}{\includegraphics[height=1em]{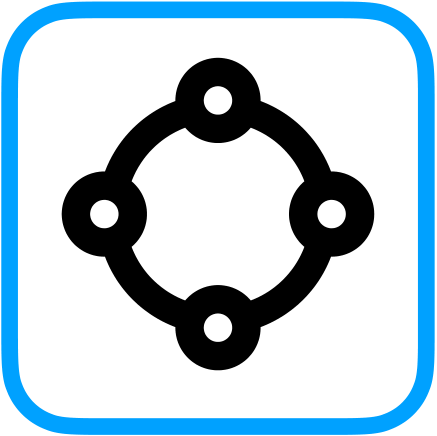}}}
\newcommand{\layoutmethcircularheadline}{\includegraphics[bb=0pt 20pt 100pt 110pt, height=1.0em]{img/glyphs/layout_method_circular.png}}
\newcommand{\layoutmethcustom}{\raisebox{-0.2\height}{\includegraphics[height=1em]{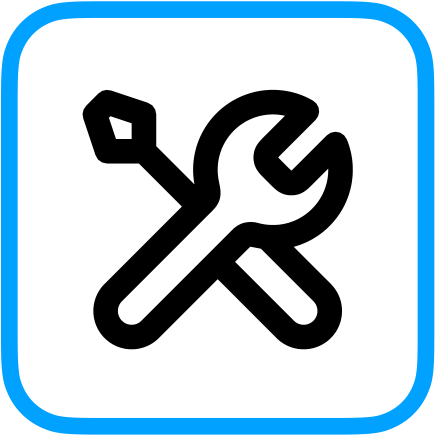}}}
\newcommand{\layoutmethcustomheadline}{\includegraphics[bb=0pt 20pt 100pt 110pt, height=1.0em]{img/glyphs/layout_method_custom.png}}
\newcommand{\layoutmethinput}{\raisebox{-0.2\height}{\includegraphics[height=1em]{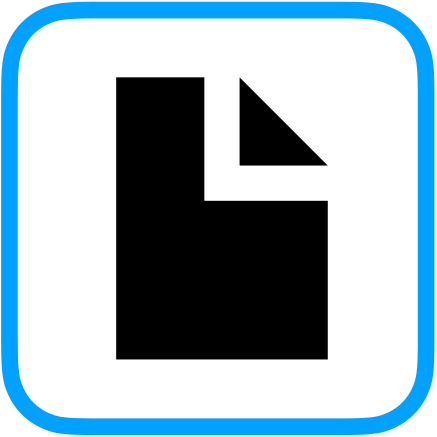}}}
\newcommand{\layoutmethinputheadline}{\includegraphics[bb=0pt 20pt 100pt 110pt, height=1.0em]{img/glyphs/layout_method_data.png}}

\newcommand{\navigationhead}{\raisebox{-0.2\height}{\includegraphics[height=1em]{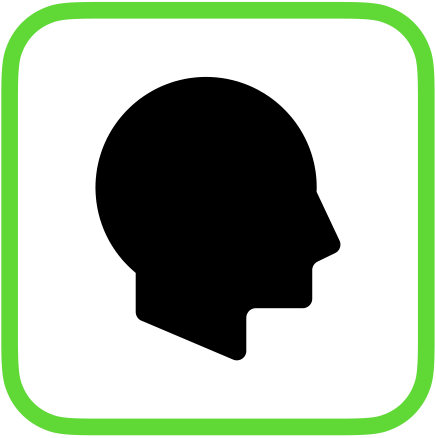}}}
\newcommand{\navigationheadheadline}{\includegraphics[bb=0pt 20pt 100pt 110pt, height=1.0em]{img/glyphs/navigation_head.png}}
\newcommand{\navigationbody}{\raisebox{-0.2\height}{\includegraphics[height=1em]{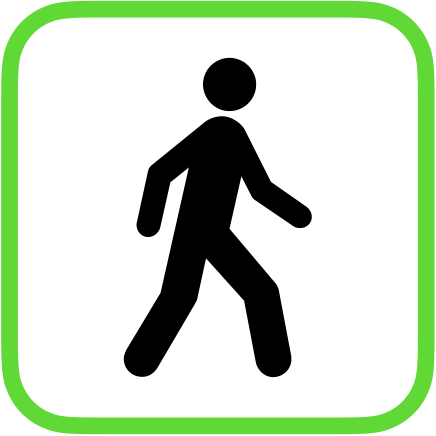}}}
\newcommand{\navigationbodyheadline}{\includegraphics[bb=0pt 20pt 100pt 110pt, height=1.0em]{img/glyphs/navigation_body_movement.png}}
\newcommand{\navigationvirtual}{\raisebox{-0.2\height}{\includegraphics[height=1em]{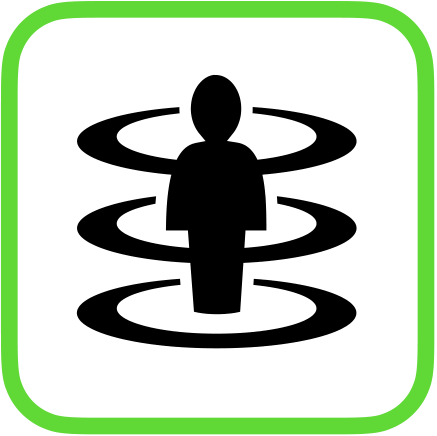}}}
\newcommand{\navigationvirtualheadline}{\includegraphics[bb=0pt 20pt 100pt 110pt, height=1.0em]{img/glyphs/navigation_virtual_movement.png}}
\newcommand{\navigationmanipulation}{\raisebox{-0.2\height}{\includegraphics[height=1em]{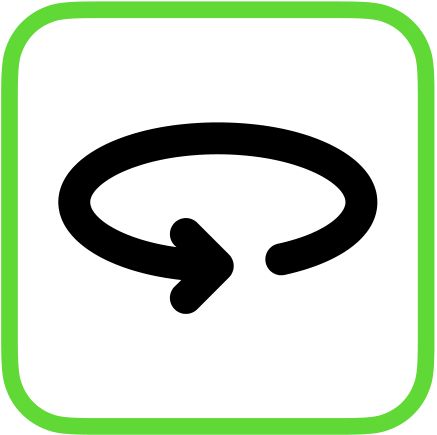}}}
\newcommand{\navigationmanipulationheadline}{\includegraphics[bb=0pt 20pt 100pt 110pt, height=1.0em]{img/glyphs/navigation_rotate.png}}
\newcommand{\navigationguided}{\raisebox{-0.2\height}{\includegraphics[height=1em]{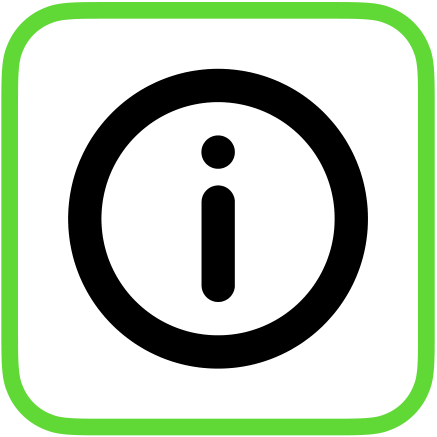}}}
\newcommand{\navigationguidedheadline}{\includegraphics[bb=0pt 20pt 100pt 110pt, height=1.0em]{img/glyphs/navigation_guided.png}}
\newcommand{\navigationdisplay}{\raisebox{-0.2\height}{\includegraphics[height=1em]{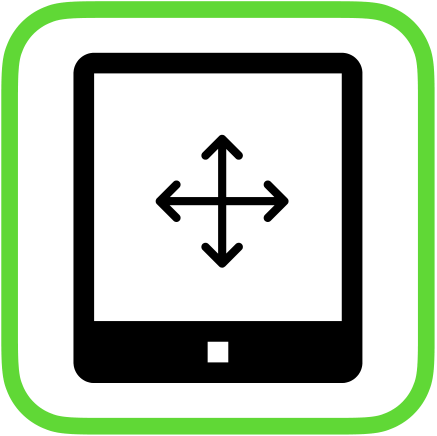}}}
\newcommand{\navigationdisplayheadline}{\includegraphics[bb=0pt 20pt 100pt 110pt, height=1.0em]{img/glyphs/navigation_display_movement.png}}

\newcommand{\taskselect}{\raisebox{-0.2\height}{\includegraphics[height=1em]{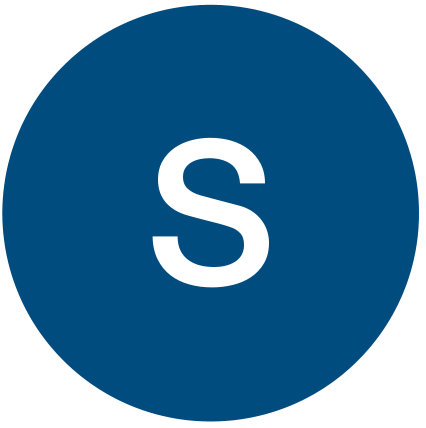}}}
\newcommand{\taskselectheadline}{\includegraphics[bb=0pt 20pt 100pt 110pt, height=1.0em]{img/glyphs/task_S.png}}
\newcommand{\taskchange}{\raisebox{-0.2\height}{\includegraphics[height=1em]{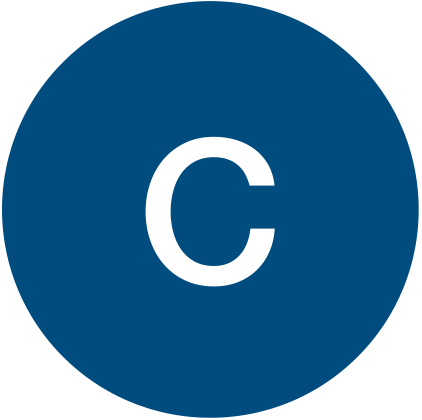}}}
\newcommand{\taskchangeheadline}{\includegraphics[bb=0pt 20pt 100pt 110pt, height=1.0em]{img/glyphs/task_C.png}}
\newcommand{\taskfilter}{\raisebox{-0.2\height}{\includegraphics[height=1em]{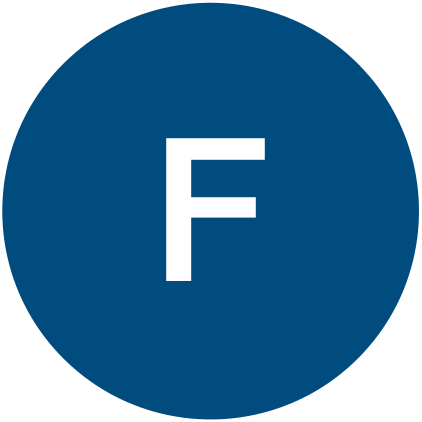}}}
\newcommand{\taskfilterheadline}{\includegraphics[bb=0pt 20pt 100pt 110pt, height=1.0em]{img/glyphs/task_F.png}}
\newcommand{\taskimport}{\raisebox{-0.2\height}{\includegraphics[height=1em]{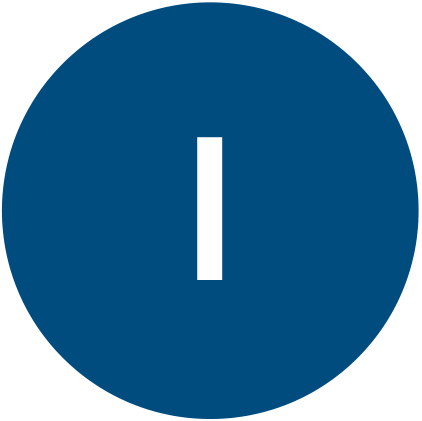}}}
\newcommand{\taskimportheadline}{\includegraphics[bb=0pt 20pt 100pt 110pt, height=1.0em]{img/glyphs/task_I.png}}
\newcommand{\taskarrange}{\raisebox{-0.2\height}{\includegraphics[height=1em]{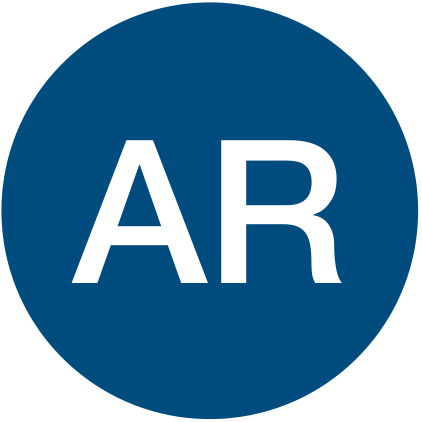}}}
\newcommand{\taskarrangeheadline}{\includegraphics[bb=0pt 20pt 100pt 110pt, height=1.0em]{img/glyphs/task_AR.png}}
\newcommand{\taskaggregate}{\raisebox{-0.2\height}{\includegraphics[height=1em]{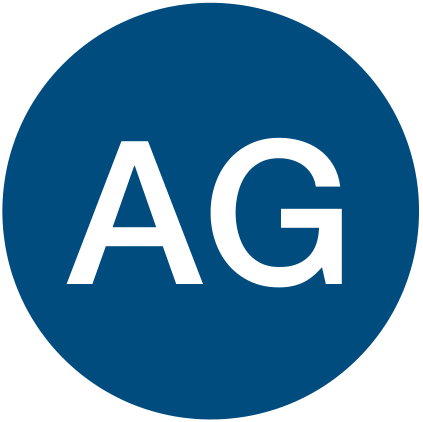}}}
\newcommand{\taskaggregateheadline}{\includegraphics[bb=0pt 20pt 100pt 110pt, height=1.0em]{img/glyphs/task_AG.png}}
\newcommand{\taskrecord}{\raisebox{-0.2\height}{\includegraphics[height=1em]{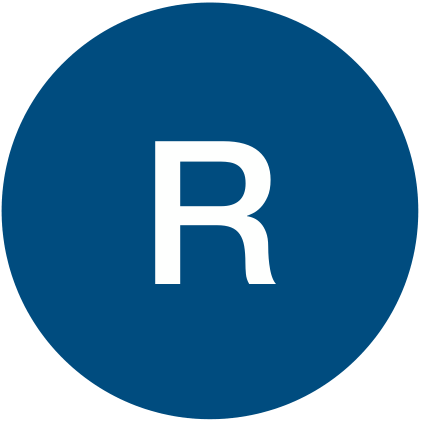}}}
\newcommand{\taskrecordheadline}{\includegraphics[bb=0pt 20pt 100pt 110pt, height=1.0em]{img/glyphs/task_R.png}}
\newcommand{\taskannotate}{\raisebox{-0.2\height}{\includegraphics[height=1em]{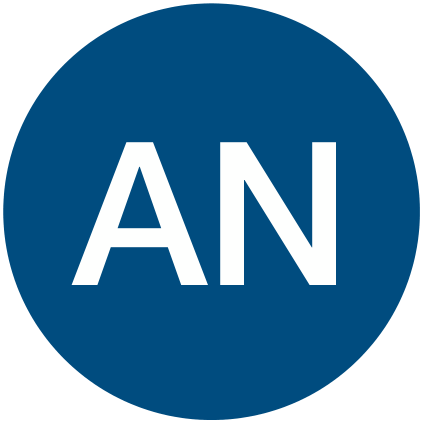}}}
\newcommand{\taskannotateheadline}{\includegraphics[bb=0pt 20pt 100pt 110pt, height=1.0em]{img/glyphs/task_AN.png}}
\newcommand{\taskderive}{\raisebox{-0.2\height}{\includegraphics[height=1em]{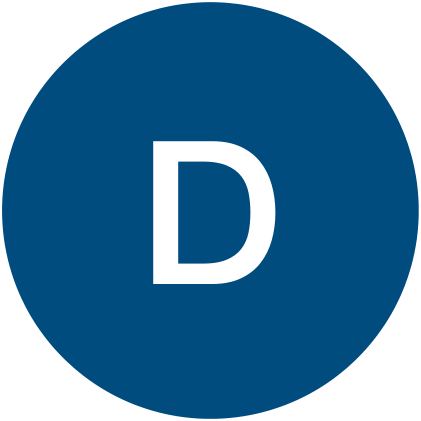}}}
\newcommand{\taskderiveheadline}{\includegraphics[bb=0pt 20pt 100pt 110pt, height=1.0em]{img/glyphs/task_D.png}}

\newcommand{\taskoverview}{\raisebox{-0.2\height}{\includegraphics[height=1em]{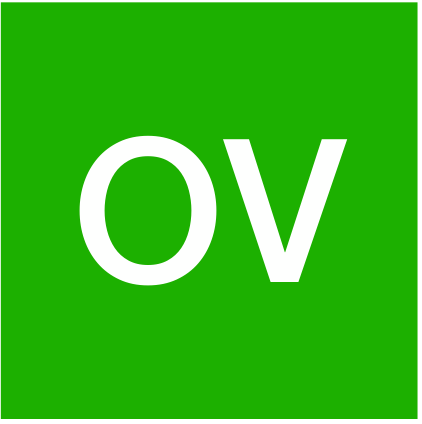}}}
\newcommand{\taskoverviewheadline}{\includegraphics[bb=0pt 20pt 100pt 110pt, height=1.0em]{img/glyphs/task_OV.png}}
\newcommand{\tasktopology}{\raisebox{-0.2\height}{\includegraphics[height=1em]{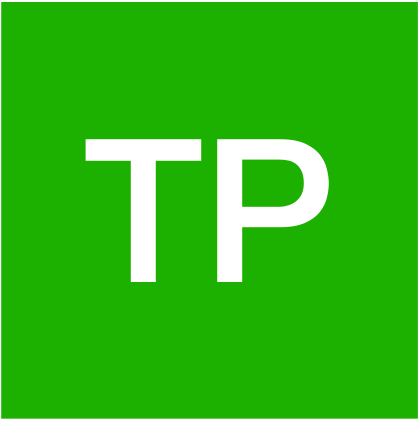}}}
\newcommand{\tasktopologyheadline}{\includegraphics[bb=0pt 20pt 100pt 110pt, height=1.0em]{img/glyphs/task_TP.png}}
\newcommand{\taskattributes}{\raisebox{-0.2\height}{\includegraphics[height=1em]{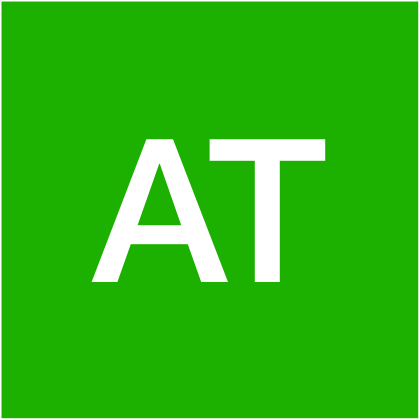}}}
\newcommand{\taskattributesheadline}{\includegraphics[bb=0pt 20pt 100pt 110pt, height=1.0em]{img/glyphs/task_AT.png}}
\newcommand{\taskbrowsing}{\raisebox{-0.2\height}{\includegraphics[height=1em]{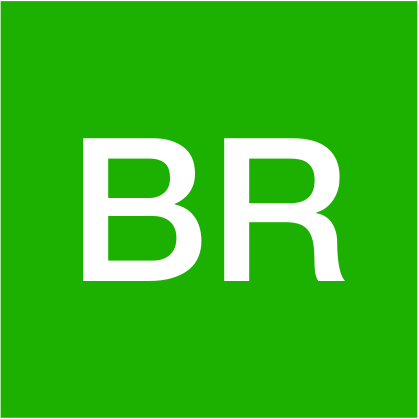}}}
\newcommand{\taskbrowsingheadline}{\includegraphics[bb=0pt 20pt 100pt 110pt, height=1.0em]{img/glyphs/task_BR.png}}
\newcommand{\taskhighlevel}{\raisebox{-0.2\height}{\includegraphics[height=1em]{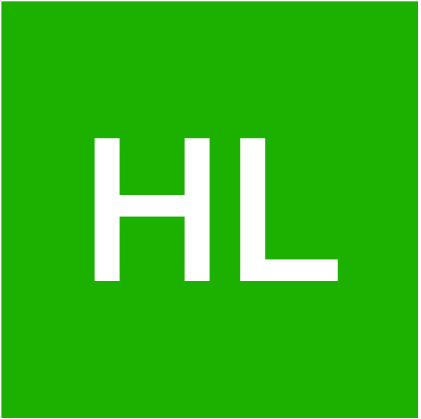}}}
\newcommand{\taskhighlevelheadline}{\includegraphics[bb=0pt 20pt 100pt 110pt, height=1.0em]{img/glyphs/task_HL.png}}
\newcommand{\tasknavigate}{\raisebox{-0.2\height}{\includegraphics[height=1em]{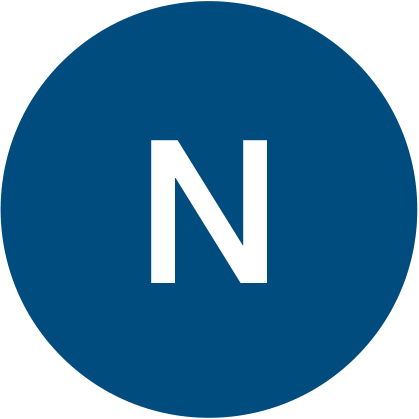}}}
\newcommand{\tasknavigateheadline}{\includegraphics[bb=0pt 20pt 100pt 110pt, height=1.0em]{img/glyphs/task_N.png}}
\newcommand{\taskmemory}{\raisebox{-0.2\height}{\includegraphics[height=1em]{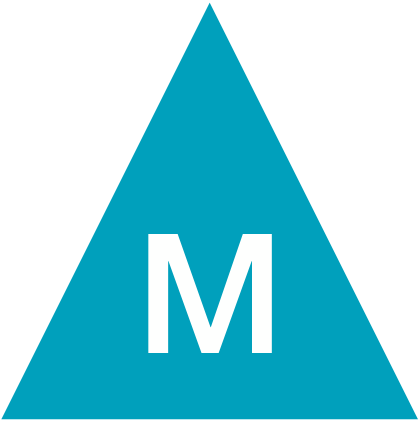}}}
\newcommand{\taskmemoryheadline}{\includegraphics[bb=0pt 20pt 100pt 110pt, height=1.0em]{img/glyphs/task_MEM.png}}
\newcommand{\taskdetectchange}{\raisebox{-0.2\height}{\includegraphics[height=1em]{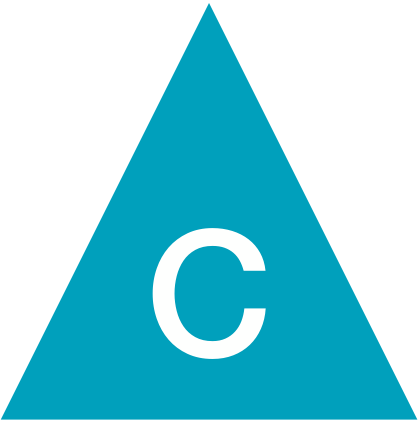}}}
\newcommand{\taskdetectchangeheadline}{\includegraphics[bb=0pt 20pt 100pt 110pt, height=1.0em]{img/glyphs/task_Detect_Change.png}}

\newcommand{\taskPlainAnalysis}{\raisebox{-0.15\height}{\includegraphics[height=0.8em]{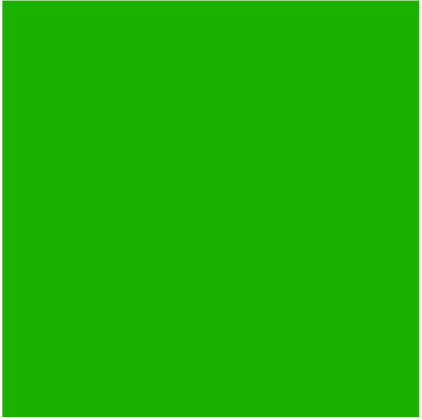}}}
\newcommand{\taskPlainInteraction}{\raisebox{-0.15\height}{\includegraphics[height=0.8em]{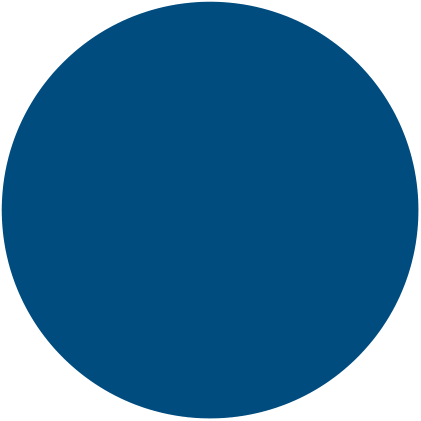}}}
\newcommand{\taskPlainMentalMap}{\raisebox{-0.15\height}{\includegraphics[height=0.8em]{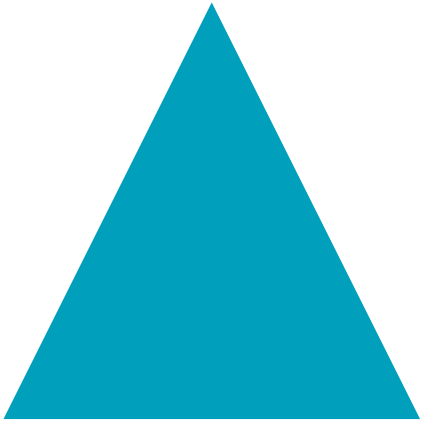}}}

\newcommand{\datasizes}{\raisebox{-0.2\height}{\includegraphics[height=1em]{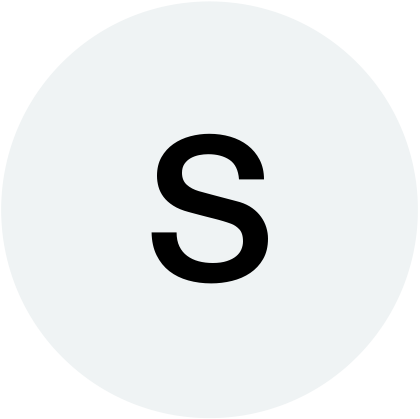}}}
\newcommand{\datasizesheadline}{\includegraphics[bb=0pt 20pt 100pt 110pt, height=1.0em]{img/glyphs/data_s.png}}
\newcommand{\datasizem}{\raisebox{-0.2\height}{\includegraphics[height=1em]{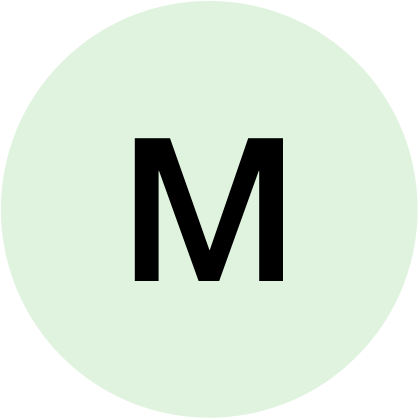}}}
\newcommand{\datasizemheadline}{\includegraphics[bb=0pt 20pt 100pt 110pt, height=1.0em]{img/glyphs/data_m.png}}
\newcommand{\datasizel}{\raisebox{-0.2\height}{\includegraphics[height=1em]{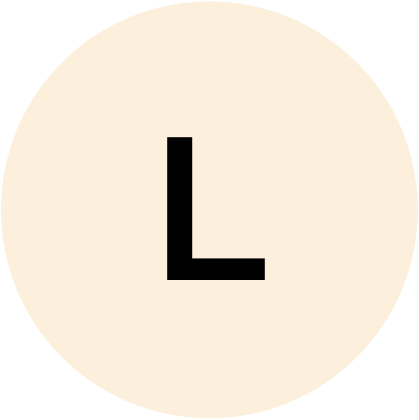}}}
\newcommand{\datasizelheadline}{\includegraphics[bb=0pt 20pt 100pt 110pt, height=1.0em]{img/glyphs/data_l.png}}
\newcommand{\datasizexl}{\raisebox{-0.2\height}{\includegraphics[height=1em]{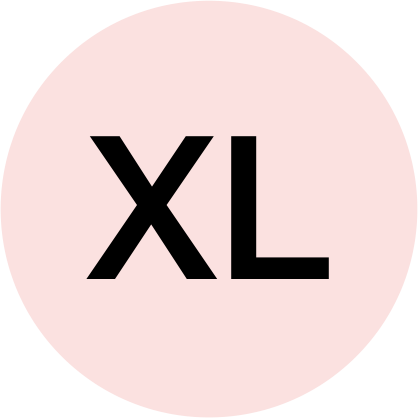}}}
\newcommand{\datasizexlheadline}{\includegraphics[bb=0pt 20pt 100pt 110pt, height=1.0em]{img/glyphs/data_xl.png}}

\newcommand{\goaldisplay}{\raisebox{-0.2\height}{\includegraphics[height=1em]{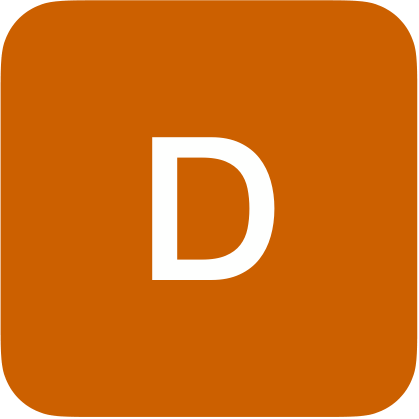}}}
\newcommand{\goaldisplayheadline}{\includegraphics[bb=0pt 20pt 100pt 110pt, height=1.0em]{img/glyphs/study_goals_d.png}}
\newcommand{\goalinteraction}{\raisebox{-0.2\height}{\includegraphics[height=1em]{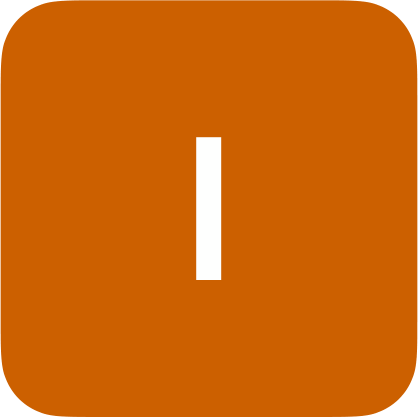}}}
\newcommand{\goalinteractionheadline}{\includegraphics[bb=0pt 20pt 100pt 110pt, height=1.0em]{img/glyphs/study_goals_i.png}}
\newcommand{\goalnavigation}{\raisebox{-0.2\height}{\includegraphics[height=1em]{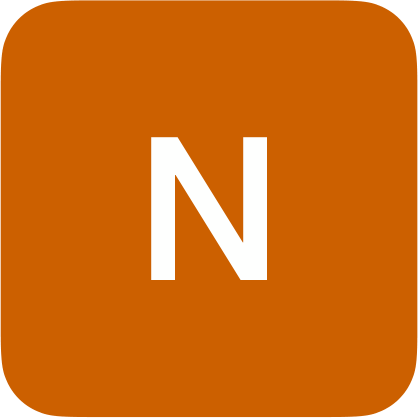}}}
\newcommand{\goalnavigationheadline}{\includegraphics[bb=0pt 20pt 100pt 110pt, height=1.0em]{img/glyphs/study_goals_n.png}}
\newcommand{\goalencoding}{\raisebox{-0.2\height}{\includegraphics[height=1em]{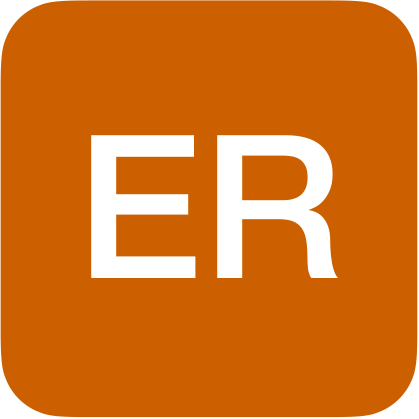}}}
\newcommand{\goalencodingheadline}{\includegraphics[bb=0pt 20pt 100pt 110pt, height=1.0em]{img/glyphs/study_goals_er.png}}
\newcommand{\goallayout}{\raisebox{-0.2\height}{\includegraphics[height=1em]{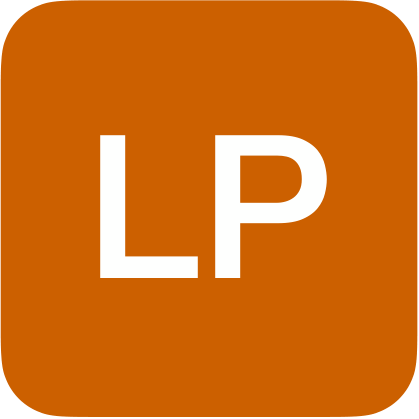}}}
\newcommand{\goallayoutheadline}{\includegraphics[bb=0pt 20pt 100pt 110pt, height=1.0em]{img/glyphs/study_goals_lp.png}}
\newcommand{\goalcollab}{\raisebox{-0.2\height}{\includegraphics[height=1em]{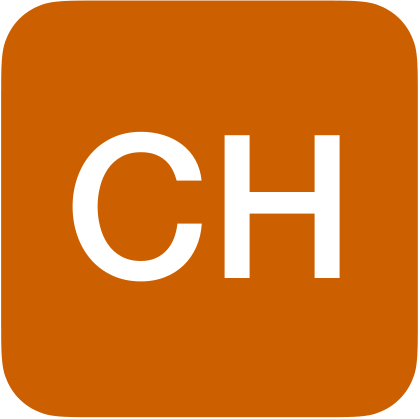}}}
\newcommand{\goalcollabheadline}{\includegraphics[bb=0pt 20pt 100pt 110pt, height=1.0em]{img/glyphs/study_goals_ch.png}}
\newcommand{\goalscale}{\raisebox{-0.2\height}{\includegraphics[height=1em]{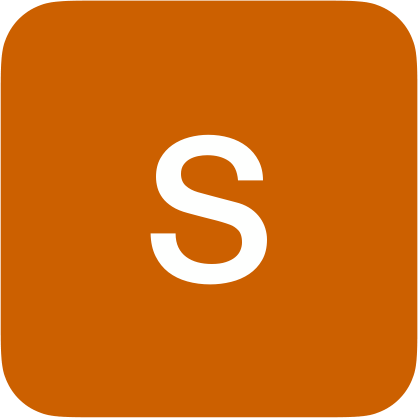}}}
\newcommand{\goalscaleheadline}{\includegraphics[bb=0pt 20pt 100pt 110pt, height=1.0em]{img/glyphs/study_goals_s.png}}

\title{Visual Network Analysis in Immersive Environments: A Survey}


\author{Lucas Joos}
\affiliation{%
  \institution{University of Konstanz}
  \city{Konstanz}
  \country{Germany}}
\email{lucas.joos@uni-konstanz.de}
\orcid{0000-0001-7049-5203}

\author{Maximilian T. Fischer}
\affiliation{%
  \institution{University of Konstanz}
  \city{Konstanz}
  \country{Germany}
}
\email{max.fischer@uni-konstanz.de}
\orcid{0000-0001-8076-1376}

\author{Julius Rauscher}
\affiliation{%
  \institution{University of Konstanz}
  \city{Konstanz}
  \country{Germany}
}
\email{julius.rauscher@uni-konstanz.de}
\orcid{0000-0003-1318-9642}

\author{Daniel A. Keim}
\affiliation{%
  \institution{University of Konstanz}
  \city{Konstanz}
  \country{Germany}
}
\email{keim@uni-konstanz.de}
\orcid{0000-0001-7966-9740}

\author{Tim Dwyer}
\affiliation{%
  \institution{Monash University}
  \city{Melbourne}
  \country{Australia}
}
\email{tim.dwyer@monash.edu}
\orcid{0000-0002-9076-9571}

\author{Falk Schreiber}
\affiliation{%
  \institution{University of Konstanz}
  \city{Konstanz}
  \country{Germany}
}
\affiliation{%
  \institution{Monash University}
  \city{Melbourne}
  \country{Australia}
}
\email{falk.schreiber@uni-konstanz.de}
\orcid{0000-0002-9307-3254}

\author{Karsten Klein}
\affiliation{%
  \institution{University of Konstanz}
  \city{Konstanz}
  \country{Germany}
}
\email{karsten.klein@uni-konstanz.de}
\orcid{0000-0002-8345-5806}

\renewcommand{\shortauthors}{Joos et al.}

\begin{abstract}
The increasing complexity and volume of network data demand effective analysis approaches, with visual exploration proving particularly beneficial.
Immersive technologies, such as augmented reality, virtual reality, and large display walls, have enabled the emerging field of immersive analytics, offering new opportunities to enhance user engagement, spatial awareness, and problem-solving.
A growing body of work has explored immersive environments for network visualisation, ranging from design studies to fully integrated applications across various domains.
Despite these advancements, the field remains fragmented, lacking a clear description of the design space and a structured overview of the aspects that have already been empirically evaluated.
To address this gap, we present a survey of visual network analysis in immersive environments, covering 138 publications retrieved through a structured pipeline.
We systematically analyse the key aspects that define the design space, investigate their coverage in prior applications (n=87), and review user evaluations (n=59) that provide empirical evidence for essential design-related questions.
By synthesising experimental findings and evaluating existing applications, we identify key achievements, highlight research gaps, and offer guidance for the design of future approaches.
Additionally, we provide an online resource to explore our results interactively, which will be updated as new developments emerge.
\end{abstract}

\ccsdesc[500]{General and reference~Surveys and overviews}
\ccsdesc[500]{Human-centered computing~Graph drawings}
\ccsdesc[500]{Human-centered computing~Virtual reality}
\ccsdesc[500]{Human-centered computing~Mixed / augmented reality}

\keywords{Visual Network Analysis, Network Visualisation, Immersive Analytics, Graph Drawing, Survey}

\maketitle

\section{Introduction}
\label{sec:introduction}

Analysing networks is essential in many application domains, such as social sciences~\cite{Correa2011,wasserman1994social}, biology~\cite{JS08,SchreiberG22}, and civil security~\cite{ferrara2014detecting,XuC2005}.
While graph structures can be described quantitatively using mathematical measures, visual investigation, e.g., in the form of node-link diagrams and matrices~\cite{nobre2019the}, is fundamental in facilitating the understanding of structure, incorporating domain knowledge, and often leading to new hypotheses~\cite{venturini2021what,gamper2020visual}.
While many visual network analysis approaches rely on classical 2D setups with mouse and keyboard interaction, a growing body of research is exploring the potential of \textbf{immersive analytics} approaches.

Immersive analytics (IA) investigates advantages and challenges of utilising immersive environments (IE) for data analysis and explores how these need to be designed to improve the efficiency and quality of the data analysis process~\cite{klein2022immersive,marriott2018immersive}.
As immersive display and interaction hardware (e.g., virtual reality (VR) and augmented reality (AR) head-mounted displays (HMDs) and large display walls) continue to improve and become more widely available, the number of applications and studies in the IA field has drastically increased over the last years~\cite{kraus2022immersive,friedl2024a}.
Due to this trend, IA research is now an essential part of almost every visualisation or human-computer interaction conference.
Therefore, it is not surprising that immersive solutions have also been applied to \textbf{network analysis}, with developments beginning as early as the 1990s~\cite{p_130}, proving beneficial when compared to classical 2D applications, for instance, regarding community detection~\cite{p_136}, path tracing~\cite{p_101}, or understanding of the graph structure~\cite{p_10}.

However, given the vast amount of publications presenting approaches with highly varying designs, conditions, and evaluation objectives, the design space, its coverage in previous approaches, and the empirical evaluation of alternatives remain unclear.
The \textbf{lack of a structured analysis} of the field and the essential aspects complicates the design of new applications and user evaluations targeting immersive network analysis.
Further, there is no overview of remaining gaps and open research directions.
With our work, we fill this gap by providing a comprehensive state-of-the-art survey on the topic of \textit{visual network analysis in immersive environments}, investigating three research questions:

\vspace*{1mm}

\textit{(1) How is the design space of visual network analysis in immersive networks characterised, (2) to what extent has it been realised, and (3) which aspects have been empirically evaluated in prior research?} 

\vspace*{1mm}

To assess these questions, we follow a \textbf{structured pipeline} to generate the corpus of relevant papers (\autoref{sec:methodology}), analyse the described approaches to determine the design space and its coverage, and identify to which extent design space alternatives have been experimentally evaluated (\autoref{sec:results}).
Based on this thorough overview of the \textbf{research landscape}, we derive overall \textbf{findings} and implications, and discuss future \textbf{research directions} (\autoref{sec:discussion}).
We provide an interactive \textbf{website} summarising our results, which will be continuously updated: \websiteurl.

\section{Related Work}
\label{sec:related-work}

The \textbf{visualisation and analysis of network structures} have long been subjects of investigation, resulting in numerous surveys covering various aspects of the topic.
Early works reviewed general graph visualisation and navigation techniques~\cite{herman2000graph}, while others focused on specialised challenges, such as visualising large-scale graphs~\cite{landesberger2011visual}, multivariate~\cite{kerren2014multivariate,nobre2019the}, dynamic~\cite{beck2014the}, or domain-specific networks, such as biological~\cite{SudermanH07,ehlers2025an} or geospatial ones~\cite{schoettler2021visualizing}.
Filipov et al.~\cite{filipov2023are} recently conducted a meta-review of surveys and taxonomies in network visualisation, synthesising trends and gaps across prior work.
However, most of these studies focus predominantly on 2D setups for visualisation and interaction, with only a few exceptions~\cite{burch2021the, yoghourdjian2018exploring}.

Research on \textbf{IA and 3D visualisation} is relatively new but rapidly expanding, and a growing number of surveys investigate aspects of IA, often touching upon network visualisation.
McIntire and Liggett~\cite{McIntire2014the} provide a critical review of 3D stereoscopic visualisation, highlighting its strengths and limitations, including its use with network data.
Fonnet and Prié~\cite{fonnet2021} offer a broad survey of IA, examining the role of technology, encoding, interaction, and collaboration.
While it includes some network-related studies, the broad IA perspective does not allow for a detailed comparison of network visualisation approaches.
Other works have taken more focused approaches.
Kraus et al.~\cite{kraus2022immersive} analysed abstract 3D visualisations in IA, categorising research by paper type, data, technology, environment, visual representation, and analysis tasks.
Liu et al.~\cite{liu2022interactive} reviewed interactive extended reality (XR) methods for information visualisation, including some network-related works, emphasising interaction and layout.
Fröhler et al.~\cite{froehler2022a} explored cross-reality analytics, examining technology, interaction, visualisation, collaboration, and evaluation, and included publications targeting network analysis.
Friedl-Knirsch et al.~\cite{friedl2024a} surveyed user evaluations in IA, focusing on evaluation design and methodology, including studies involving networks.
Belkacem et al.~\cite{belkacem2024interactive} reviewed visualisations for large display walls, addressing objectives, view layouts, data types, interaction, evaluation, and applications, with networks being one data type.
Lastly, Saffo et al.~\cite{saffo2024unraveling} examined existing IA approaches to outline the design space of the field.

While these surveys collectively provide a broad understanding of IA, none systematically focus on immersive network visualisation.
Existing surveys cover only a small subset of relevant papers and evaluate them from other perspectives.
We address this gap by providing a \textbf{detailed and systematic survey} on immersive network visualisation.

\section{Methodology}
\label{sec:methodology}

In this section, we present our methodology, starting by clarifying the scope, key terms, and the corpus retrieval pipeline.
Further, we derive the design space and additional categories for the subsequent analysis of the papers.

\subsection{Scope \& Paper Inclusion Criteria}
\label{sec:methodology-scope}

The survey's scope is dependent on key terms, which lack universal definitions and are interpreted differently across publications.
We, therefore, define these concepts in the following and establish paper inclusion criteria for clarification.

\subsubsection{Definitions}

Throughout this survey, we rely on the following definitions without asserting their broader validity.

\paragraph*{Visual Network Analysis} 
Visual network analysis is a subfield of \textbf{visual analytics}, which was defined as the \quot{analytical reasoning facilitated by interactive visual interfaces}~\cite{thomas2005illuminating}.
In the field of network analysis, these visual interfaces need to enable viewers to explore and interpret the structure (topology), attributes, and other characteristics of networks.
The analytical reasoning relates to questions regarding structural patterns and features, including the connectedness, density, clusters, paths, degrees, and others, as well as developing new hypotheses~\cite{lee2006task}.
As for the overall field of visual analytics, the visual network analysis incorporates \textbf{domain knowledge} of users and their visual perception and understanding capabilities~\cite{keim2008visual}.
Besides expressing the data visually, mostly by node-link or matrix representations with varying layouts and encodings, \textbf{interactivity} plays an essential role in the sense-making process~\cite{kerren2012toward}.
As for other visual interfaces, a broad range of interaction tasks can support the visual investigation of networks, such as selection, navigation, or filtering~\cite{brehmer2013taxononomy}.
Especially when networks become complex or visualised in 3D, where the camera perspective strongly influences the perceivability of features, navigation capabilities are indispensable to understanding the structure and perceiving details.
Thus, we require systems to support at least \textbf{navigation} to be considered as \emph{visual analysis} approaches, for instance, by zoom+pan (2D) or by rotation (3D).

\paragraph*{Immersive Environment}

There are diverging definitions and concepts referring to the term \textbf{immersive environment}~\cite{dwyer2018immersive,suh2018the,klein2022immersive,kraus2022immersive,kutak2023molecules}.
For our work, we focus on approaches incorporating \textit{technical immersion} by using immersive display and interaction technology, potentially leading to psychological \textbf{immersion}~\cite{dwyer2018immersive}, going beyond classical 2D desktop setups~\cite{klein2022immersive}.
Often, the \textit{reality-virtuality continuum} by Milgram and Kishino~\cite{milgram1994taxonomy} is used to characterise such environments, where a strong focus is set on mixed reality (MR) setups, covering everything between AR and VR, with the exception of pure, simulated VR without real-world elements.
With pure VR included, the continuum is typically referred to as extended reality (XR).
However, over time, technology and the interpretation of MR and immersive environments evolved, leading to new definitions that consider more dimensions than the degree of simulation, such as \textit{coherence}, \textit{world knowledge}, \textit{immersion}, or whether a simulation is focused on users or the world around them~\cite{skarbez2021revisiting,park2022metaverse,smart2007metaverse}.
Typically, \textbf{stereoscopic 3D} (S3D) environments are considered to address these dimensions, as they more realistically mimic the way we perceive our real world compared to 2D or projective 3D, reducing the impression of being in a simulated setting, especially when head tracking is used to simulate natural perspective changes.
Thus, S3D setups, such as HMD-based AR or VR, 3D monitors (either without or with spatial head tracking, also referred to as \textit{Fishtank VR}), or CAVE-like 3D environments, presenting S3D visualisations at room scale, are typically seen as immersive environments.
The degree of immersion can also be increased when users are actively \textbf{engaged} in the visualisation, for instance, by incorporating \textbf{natural interaction} techniques (e.g., users can touch and directly manipulate data) or by supporting physical locomotion (e.g., users can walk around the data representation), keeping them in a flow~\cite{elmqvist2011fluid,bueschel2018interaction}.
Besides stereoscopic vision, we also see two other factors of relevance creating the impression of \textit{spatial presence} leading to immersion in simulated environments~\cite{dwyer2018immersive}, namely the degree to which the user's \textbf{field of view} (FOV) is addressed by a visual representation and \textbf{spatial cues}, leading to the impression that the representation is present within actual space.
Hence, we also consider non-stereoscopic representations, such as large 2D walls or projections as \textit{immersive}, since they can lead to the impression of being situated within the data, and also spatially tracked mobile AR displays, allowing users to walk around a simulation, leading to the impression that it is actually part of their physical space.
Further, the feeling of immersion can be created by environments addressing further senses, like hearing, feeling, smelling, etc., especially in multi-sensory setups~\cite{marriott2018immersive}.
However, as our survey targets the \textit{visual} analysis of networks in such environments, we largely focus on the \textbf{visual sense} to create an immersive environment.

\subsubsection{Inclusion Criteria}

Based on these definitions and our research questions, we establish criteria to determine whether a paper is included in this survey, sharpening its scope and guiding the corpus retrieval.

\paragraph*{Paper Type} We include only publications that present either \textbf{(1) an implemented interactive system} designed for immersive network analysis or \textbf{(2) a user or expert study} evaluating immersive visual network analysis techniques (e.g., encodings, layouts, or interaction concepts) or the overall applicability with quantitative or qualitative results.
This ensures that included papers contribute academically either through a functional system or empirical insights.

\paragraph*{Visualisation Type}
Networks can be visualised in various ways, with node-link diagrams and matrices being the most common.
To maintain comparability and scope, we include only \textbf{visual node-link representations}.
These are widely used in immersive settings due to their natural 3D extension, unlike matrix visualisations.
While node-link representations are required, additionally using other representations (e.g., comparisons, hybrids) is permitted.

\paragraph*{Network Analysis}
Included papers must feature node-link representations with a level of \textbf{complexity} sufficient to enable meaningful visual network analysis (e.g., identifying shortest paths). 
Visualisations that connect only three or four objects are excluded, as are tools primarily designed to generate new data. 
Papers must also incorporate \textbf{interaction} to support exploration, with at least viewpoint adjustments meeting the minimum inclusion criteria.

\paragraph*{Visual Encoding}
Some methods employ complex representations for nodes or edges, often domain-specific and focused on attributes rather than network topology.
To ensure comparability, we include only node-link diagrams with \textbf{primitive representations} of nodes and edges (e.g., spheres, cubes, lines).
While variations in appearance (e.g., colour or size) are within our scope, highly complex node representations (e.g., UML classes) are excluded.

\paragraph*{Paper Presentation \& Quality}

To enable the retrieval of sufficient information for analysing and comparing approaches, included papers must describe the methods in \textbf{detail}.
Furthermore, we include only \textbf{scientific} papers written in \textbf{English}, \textbf{peer-reviewed}, and published in journals, conference proceedings, or books to ensure quality and comparability.
Only original work is considered, while re-publications or minor extensions of previously published work are excluded.

\subsection{Paper Corpus Retrieval}
\label{sec:methodology-corpus}

\begin{figure}
    \centering
    \begin{subfigure}{0.5\textwidth}
        \raisebox{0.2cm}{\includegraphics[height=3cm]{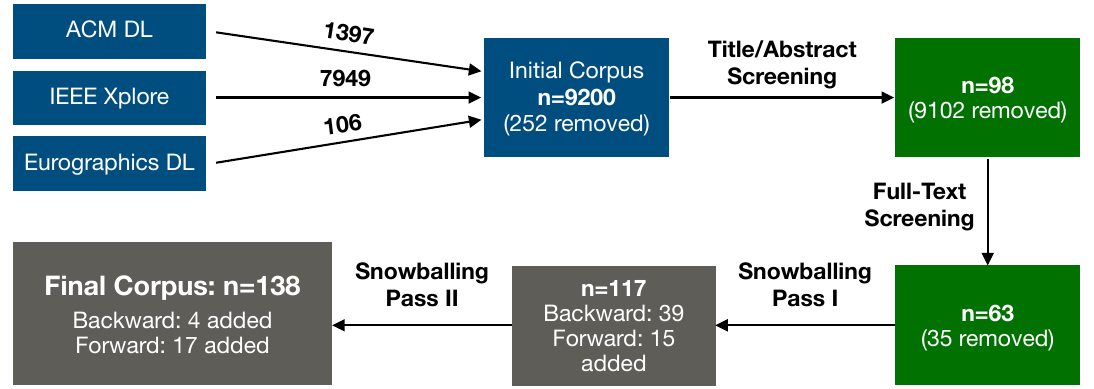}}
    \end{subfigure}%
    \hfill
    \begin{subfigure}{0.5\textwidth}
        \hfill
        \includegraphics[height=3.3cm]{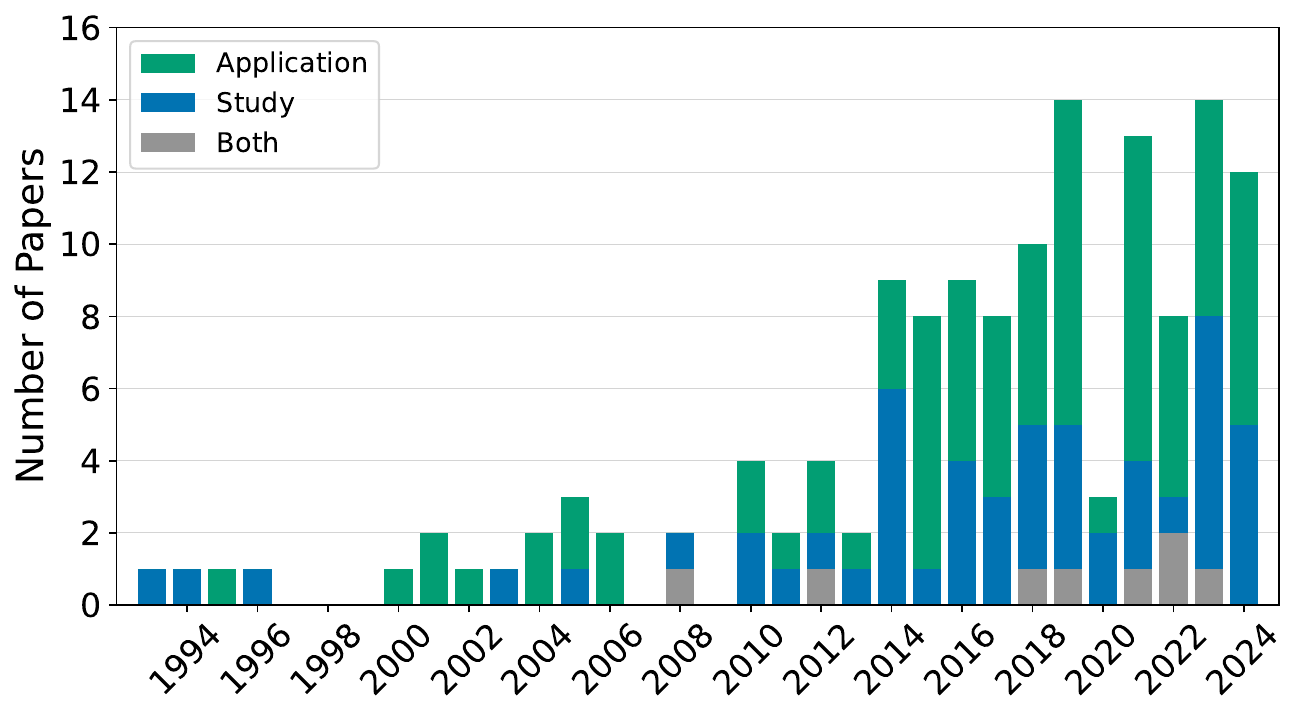}
    \end{subfigure}
    \caption{\textbf{Left:} Corpus retrieval was performed through keyword-based search in three online libraries (\textcolor{pipelineBlue}{$\blacksquare$}), followed by query filtering, duplicate removal, manual filtering (\textcolor{pipelineGreen}{$\blacksquare$}), and snowballing (\textcolor{pipelineGray}{$\blacksquare$}). \textbf{Right:} The final corpus encompasses 138 evaluated papers published between 1993 and 2024, of which 51 present a user study, 79 describe an application, and 8 contain both.}
    \label{fig:corpus-pipeline-statistics}
\end{figure}

To retrieve the set of publications included in this survey, we followed a structured pipeline that began with the established \textbf{PRISMA}~\cite{PRISMA2020} literature retrieval method.
This approach relies on keyword-driven searches in online publication databases, followed by iterative reduction steps such as duplicate removal and manual screening.
However, our initial screening revealed that publications relevant to our topic are widely distributed across numerous conferences and journals, making them difficult to retrieve through a fixed set of online libraries.
Therefore, we complemented the {PRISMA} scheme with the \textbf{snowballing method} proposed by Wohlin~\cite{wohlin2014guidelines}.
In this method, the references (\textit{backward pass}) and citations (\textit{forward pass}) of the papers retrieved through {PRISMA} were examined, and additional relevant publications were incorporated into the corpus.
This two-stage pipeline allowed us to collect a core set of papers via keyword-based search while ensuring that relevant publications from other domains and venues were also included (unless they were entirely disconnected from the research landscape).

For our paper corpus retrieval (see~\autoref{fig:corpus-pipeline-statistics} left), we chose three major Computer Science \textbf{publication libraries} for the initial search: \emph{IEEE Xplore}, \emph{ACM Digital Library}, and \emph{Eurographics Digital Library}.
These libraries encompass journals and conferences relevant to our topic (e.g., IEEE TVCG, Vis, PacificVis, ISMAR, VR; ACM TOG, CHI, SUI, VRST, VINCI; Eurographics EuroVis, CGF).
To query these databases, we defined keywords reflecting both the \datacolor{data} type (networks) and the presentation \techcolor{technology} for creating immersive environments, aligned with our scope and definitions (see~\autoref{sec:methodology-scope}).
Keywords within each category were combined using the \texttt{OR} operator, while the two Boolean formulas were linked with \texttt{AND}.
The resulting query formula ensured that papers in the result set included at least one relevant keyword in their \textit{title} or \textit{abstract} for both \datacolor{data} type and \techcolor{technology}.

\noindent
(\techkeyword{immersive}
\texttt{OR}
\techkeyword{immersion}
\texttt{OR}
\techkeyword{vr}
\texttt{OR}
\techkeyword{virtual reality}
\texttt{OR}
\techkeyword{ar}
\texttt{OR}
\techkeyword{augmented reality}
\texttt{OR}
\techkeyword{mixed reality}
\texttt{OR}
\techkeyword{extended reality}
\texttt{OR}
\techkeyword{cave}
\texttt{OR}
\techkeyword{wall display}
\texttt{OR}
\techkeyword{wall-display}
\texttt{OR}
\techkeyword{display-wall}
\texttt{OR}
\techkeyword{wall-size display}
\texttt{OR}
\techkeyword{wall-sized display})
\texttt{AND}
(\datakeyword{graph}
\texttt{OR}
\datakeyword{graphs}
\texttt{OR}
\datakeyword{network}
\texttt{OR}
\datakeyword{networks}
\texttt{OR}
\datakeyword{node-link})

Using the query, we retrieved 7949 publications from IEEE Xplore, 1397 from ACM Digital Library, and 106 from Eurographics Digital Library.
After merging, deduplicating, and ensuring papers matched the query, 9200 candidate papers remained.
Due to the ambiguity of the keywords, many were false positives, especially \datakeyword{network}, which often refers to communication networks (e.g., 5G) or neural networks, and \techkeyword{ar}, which can relate to \quot{auto regression}, \quot{activity region}, or \quot{area ratio}.
We manually screened the titles and abstracts of all candidates, removing 9102 papers that did not fit our survey's topic.
Of the remaining 98 publications, 35 were excluded after full-text review and further checks against the criteria defined in~\autoref{sec:methodology-scope}, such as publication type and language.
With the 63 retrieved papers in the first pipeline stage, we continued employing bi-directional snowballing, adding 54 papers (39 backward, 15 forward).
Repeating this process yielded another 21 papers (4 backward, 17 forward).
The final corpus comprises \textbf{138 papers} published between \textbf{1993 and 2024} (see~\autoref{fig:corpus-pipeline-statistics} right).
A dip in 2020 may reflect the COVID-19 pandemic, which likely hampered user studies.
With our pipeline, we included relevant papers published prior to November 15, 2024.

\subsection{Design Space \& Analysed Categories}
\label{sec:methodology-dimensions}

In the following, we describe how we developed the design space characterisation and additional categories (listed in \autoref{tab:dimensions-table} using icons%
\footnote{
    \emph{Icon License Declaration}: \href{https://www.svgrepo.com/page/licensing/\#Apache}{\textcolor{blue}{Apache License}} for \href{https://www.svgrepo.com/svg/339330/machine-learning-03}{\textit{Machine Learning 03 SVG Vector}} by \href{https://www.svgrepo.com/author/Carbon\%20Design/}{\textcolor{blue}{Carbon Design}}
and \href{https://www.svgrepo.com/svg/449088/group}{\textit{Group SVG Vector}} by \href{https://www.svgrepo.com/author/UXAspects/}{\textcolor{blue}{UXAspects}}; 
\href{https://www.svgrepo.com/page/licensing/\#CC\%20Attribution}{\textcolor{blue}{CC BY License}} for 
\href{https://www.svgrepo.com/svg/321565/teleport}{\textit{Teleport SVG Vector}} by \href{https://www.svgrepo.com/author/game-icons.net/}{\textcolor{blue}{game-icons.net}} 
and \href{https://www.svgrepo.com/svg/434396/idea}{\textit{Idea SVG Vector}} by \href{https://www.svgrepo.com/author/gonzodesign/}{\textcolor{blue}{gonzodesign}}; 
\href{https://www.svgrepo.com/page/licensing/\#GPL}{\textcolor{blue}{GNU GPL License}} for 
\href{https://www.svgrepo.com/svg/499204/hand}{\textit{Hand SVG Vector}} by \href{https://www.svgrepo.com/author/nagoshiashumari/}{\textcolor{blue}{nagoshiashumari}}; 
\href{https://www.svgrepo.com/page/licensing/\#MIT}{\textcolor{blue}{MIT License}} for \href{https://www.svgrepo.com/svg/444984/category-solid}{\textit{Category Solid SVG Vector}} by \href{https://www.svgrepo.com/author/Denali\%20Design/}{\textcolor{blue}{Denali Design}}, 
\href{https://www.svgrepo.com/svg/455118/virtual-reality-view}{\textit{Virtual Reality View SVG Vector}} by \href{https://www.svgrepo.com/author/Vectopus/}{\textcolor{blue}{Vectopus}}, 
\href{https://www.svgrepo.com/svg/395511/screen-desktop}{\textit{Screen Desktop SVG Vector}} by \href{https://www.svgrepo.com/author/thesabbir/}{\textcolor{blue}{thesabbir}}, 
\href{https://www.svgrepo.com/svg/473368/screen-users}{\textit{Screen Users SVG Vector}} by \href{https://www.svgrepo.com/author/jtblabs/}{\textcolor{blue}{jtblabs}}, 
\href{https://www.svgrepo.com/svg/455071/controller}{\textit{Controller SVG Vector}} by \href{https://www.svgrepo.com/author/Vectopus/}{\textcolor{blue}{Vectopus}}, 
\href{https://www.svgrepo.com/svg/486497/touch-filled}{\textit{Touch Filled SVG Vector}} by \href{https://www.svgrepo.com/author/Siemens/}{\textcolor{blue}{Siemens}}, 
\href{https://www.svgrepo.com/svg/440327/voice-fill}{\textit{Voice Fill SVG Vector}} by \href{https://www.svgrepo.com/author/emblemicons/}{\textcolor{blue}{emblemicons}}, 
\href{https://www.svgrepo.com/svg/509372/info}{\textit{Info SVG Vector}} by \href{https://www.svgrepo.com/author/afnizarnur/}{\textcolor{blue}{afnizarnur}}, 
\href{https://www.svgrepo.com/svg/372096/tablet}{\textit{Tablet SVG Vector}} by \href{https://www.svgrepo.com/author/vmware/}{\textcolor{blue}{vmware}}, 
\href{https://www.svgrepo.com/svg/509171/move}{\textit{Move SVG Vector}} by \href{https://www.svgrepo.com/author/Orchid/}{\textcolor{blue}{Orchid}}, 
\href{https://www.svgrepo.com/svg/488277/magnet}{\textit{Magnet SVG Vector}} by \href{https://www.svgrepo.com/author/Neuicons/}{\textcolor{blue}{Neuicons}} and 
\href{https://www.svgrepo.com/svg/509821/circle-tool}{\textit{Circle Tool SVG Vector}} by \href{https://www.svgrepo.com/author/zest/}{\textcolor{blue}{zest}}.
}%
), which are the basis for the paper analysis and comparison.
Some of these apply to all papers, others only to application or evaluation publications.

To analyse the design space, the paper corpus was iteratively scanned and coded by one author, checking for the essential design aspects described in the papers and whether distinct categories could be formed describing them.
Through multiple discussions with the other authors and a thorough consultation of related taxonomies and analyses, the initial characterisation was iteratively refined.
The authors individually cross-checked and discussed the result, leading to a consensus that was supported by all authors and matched the characteristics found in the papers without contradicting existing taxonomies.
Our design dimension categories are largely in accordance with the work by Saffo et al.~\cite{saffo2024unraveling}, investigating the design space in the entire IA field.
Besides the design space, the literature coding revealed essential information and differences that do not contribute to the design space, but are still interesting and thus added as additional analysis categories.

\begin{table}
    \centering
    \small
    \renewcommand{\arraystretch}{1.3} 
    \setlength{\tabcolsep}{5pt}
    \caption{Analysed categories and design space aspects with the values we found for applications, studies, and both (common). The icons defined for the expressions are used consistently in the text and summary tables.\vspace*{0mm}}
    \begin{tabular*}{\linewidth}{@{\extracolsep{\fill}}p{0.01\linewidth}p{0.17\linewidth}p{0.83\linewidth}}
    \toprule
     & \textbf{Categories} & \textbf{Values} \\
    \midrule
    \multirow{11.5}{*}{\rotatebox{90}{\textbf{Common}} } 
    & Domain & \makebox[1.5cm][l]{\domainabstract \, Abstract} \;
    \mbox{\domainbio \, Biomedical} \;
    \mbox{\domainknowledge \, Knowledge} \;
    \mbox{\domainnetwork \, Computer Networks} \;
    \mbox{\domainsocial \, Social} \;
    \mbox{\domainother \, Other} \par
    \makebox[1.50cm][l]{\domainsoftware \, Software} \;
    \mbox{\domainaai \, Artificial Intelligence}
\\
    & Display\; \mbox{Technology} & \makebox[3.5cm][l]{\displayhmd \, Head-Mounted Display} \; \makebox[2.65cm][l]{\displaythreed \, 3D Screens} \;
    \mbox{\displaycave \, CAVE-Like 3D Environment} \par
    \makebox[3.5cm][l]{\displaymobile \, Tracked Mobile Display} \;
    \makebox[2.65cm][l]{\displaylargetwod \, Large 2D Display} \;
    \mbox{\displaytwodsmall \, 2D Screens} \;
    \mbox{\displayphysicalisation \, Physicalisation} \\
    & Interaction Modality & \mbox{\interactioncontroller \, Controller} \;
    \mbox{\interactiongestures \, Gestures} \;
    \mbox{\interactiontouch \, Touch} \;
    \mbox{\interactionmousekeyboard \, Mouse/keyboard} \;
    \mbox{\interactiongaze \, Gaze} \;
    \mbox{\interactionspeech \, Speech} \;
    \mbox{\interactionmovement \, Movement} \\
    & Node Encoding & 
    \mbox{\nodescircle \, Circles} \;
    \mbox{\nodeplane \, Planes} \;
    \mbox{\nodessphere \, Spheres} \;
    \mbox{\nodescube \, Cubes} \;
    \mbox{\nodesother \, Other} \\
    & Edge Encoding & \mbox{\edgeslines \, Lines} \;
    \mbox{\edgestubes \, Tubes} \\
    & Layout\;(Dimension) & \mbox{\layoutdimtwo \, 2D} \;
    \mbox{\layoutdimthree \, 3D} \\
    & Layout\;(Perspective) & \mbox{\layoutperspexo \, Exocentric} \;
    \mbox{\layoutperspego \, Egocentric} \\
    & Layout (Method) & 
    \mbox{\layoutmethforce \, Force-Based} \;
    \mbox{\layoutmethsemantic \, Semantic} \;
    \mbox{\layoutmethcircular \, Circular} \;
    \mbox{\layoutmethprocess \, Data-Derived} \;
    \mbox{\layoutmethlayered \, Layered} \;
    \mbox{\layoutmethinput \, Input} \;
    \mbox{\layoutmethcustom \, Custom} \\
    & Navigation & 
    \makebox[2.85cm][l]{\mbox{\navigationhead \, Head Movement}} \;
    \mbox{\navigationbody \, Body Movement} \;
    \mbox{\navigationvirtual \, Virtual Movement} \;
    \mbox{\navigationmanipulation \, Graph Manipulation} \par
    \makebox[2.85cm][l]{\navigationdisplay \, Display Movement} \;
    \mbox{\navigationguided \, Guided}
     \\
    \midrule
    \multirow{4}{*}{\rotatebox{90}{\textbf{Application}}}
    & Collaboration & \mbox{\filledRec \, Sync+Co-Located} \;
    \mbox{\unfilledRec \, Async+Co-Located} \;
    \mbox{\filledTri \, Sync+Remote} \;
    \mbox{\unfilledTri \, Async+Remote} \\
    & Code Access & \mbox{$\bullet$ Yes} \;
    \mbox{$\circ$ No} \\
    & Task Support & \makebox[1.65cm][l]{\taskselect \, Select} \;
    \mbox{\taskchange \, Change} \;
    \mbox{\taskfilter \, Filter} \;
    \mbox{\taskimport \, Import} \;
    \mbox{\taskarrange \, Arrange} \;
    \mbox{\taskaggregate \, Aggregate} \;
    \mbox{\taskrecord \, Record} \par
    \makebox[1.65cm][l]{\taskannotate \, Annotate} \;
    \mbox{\taskderive \, Derive} \\
    \midrule
    \multirow{5.8}{*}{\rotatebox{90}{\textbf{Studies}}} 
    & Goal & \makebox[3.8cm][l]{\goaldisplay \, Display \& Dimensionality} \;
    \makebox[1.855cm][l]{\goalinteraction \, Interaction} \;
    \mbox{\goalnavigation \, Navigation} \;
    \mbox{\goalencoding \, Encoding \& Representation} \par
    \makebox[3.8cm][l]{\goallayout \, Layout \& Perspective} \;
    \makebox[1.855cm][l]{\goalscale \, Scalability} \;
    \mbox{\goalcollab \, Collaboration \& Hybrid} \\
    & Study Type & \mbox{$\boxtimes$ Controlled Quantitative Experiment} \;
    \mbox{$\boxdot$ Qualitative User Study} \\
    & Data Size & \mbox{\datasizes \, $< 60$} \;
    \mbox{\datasizem \, 61-120} \;
    \mbox{\datasizel \, 121-249} \;
    \mbox{\datasizexl \, $\geq 250$} \\
    & Study Tasks & 
    \makebox[1.95cm][l]{\tasktopology \, Topological} \;
    \makebox[2.34cm][l]{\taskattributes \, Attributes} \;
    \makebox[1.65cm][l]{\taskoverview \, Overview} \;
    \makebox[1.65cm][l]{\taskbrowsing \, Browsing} \;
    \mbox{\taskhighlevel \, High-Level/Domain} \par
    \makebox[1.95cm][l]{\taskmemory \, Memory} \;
    \makebox[2.34cm][l]{\taskdetectchange \, Detect Change} \;
    \makebox[1.65cm][l]{\taskselect \, Select} \;
    \makebox[1.65cm][l]{\tasknavigate \, Navigate} \;
    \mbox{\taskchange \, Change} \;
    \mbox{\taskaggregate \, Aggregate} \\
    \bottomrule
    \end{tabular*}
    \label{tab:dimensions-table}
    \end{table}

\subsubsection{Design Space}

\paragraph{Display Technology}

The display technology highly impacts the user \textbf{perception} and the degree of \textbf{immersion}.
In accordance with~\cite{friedl2024a}, we observe five dominant categories: \mbox{\displayhmd \, head-mounted displays} (HMDs), \mbox{\displaythreed \, 3D screens}, \mbox{\displaycave \, CAVE-Like 3D environments}, \mbox{\displaymobile \, tracked mobile display}, and \mbox{{\displaylargetwod \, large display walls}}.
Besides immersive technologies, we also observe evaluations incorporating classical \mbox{\displaytwodsmall \, 2D screens} and data \mbox{\displayphysicalisation \, physicalisations} for comparison.

\paragraph{Interaction Modality}

Interaction modality concerns what type of interaction device or channel is incorporated, allowing users to \textbf{manipulate} and \textbf{explore} the data, as required for effective analysis.
We follow the categorisation by Friedl-Knirsch et al.~\cite{friedl2024a}: tracked hand-held \mbox{\interactioncontroller \, controllers}, \mbox{\interactiongestures \, gestures}, \mbox{\interactiontouch \, touch}, \mbox{\interactiongaze \, gaze}, \mbox{\interactionspeech \, speech}, and \mbox{\interactionmovement \, movement}, where the representation is changed based on the tracked user position.
Additionally, some applications and studies also incorporate the classical \mbox{\interactionmousekeyboard \, mouse/keyboard} paradigm. 

\paragraph{Encoding}

One essential aspect is how nodes and edges are \textbf{visually expressed}, i.e., the \textit{encoding}.
For nodes, we distinguish multiple shape and shading categories, i.e., 2D \mbox{\nodescircle \, circles} or \mbox{\nodeplane \, planes} and 3D \mbox{\nodessphere \, spheres}, \mbox{\nodescube \, cubes}, or \mbox{\nodesother \, other} objects.
For edges, we observe 2D \mbox{\edgeslines \, lines} and 3D \mbox{\edgestubes \, tubes}.
Besides the basic representation of nodes and edges, their appearance is sometimes adapted to convey additional information, e.g., the colour, size, curvature, added text labels, and additional objects like arrow tips.

\paragraph{Layout}

The Layout addresses the \textbf{spatial arrangement} of nodes, playing a critical role in the clarity and interpretability of a visualisation.
In immersive environments, this aspect gains additional importance, as not only the \textit{layout method} is of relevance, but also the \textbf{dimensionality} (i.e., whether the layout is \mbox{\layoutdimtwo \, two-} or \mbox{\layoutdimthree \, three-dimensional}).
Further, we observe differences regarding the graph-user arrangement, i.e., whether users view the layout from an outside \textbf{perspective} (\mbox{\layoutperspexo \, exocentric}), or view the graph from inside (\mbox{\layoutperspego \, egocentric}).
Regarding the layout method itself, the approaches can be distinguished in seven categories: \mbox{\layoutmethforce \, force-based} algorithms (such as the one by Fruchterman and Reingold~\cite{fruchterman1991graph}), \mbox{\layoutmethcircular \, circular} arrangements, \mbox{\layoutmethlayered \, layered} structures, \mbox{\layoutmethsemantic \, semantic} placement (i.e., when the node positions match real-world observations), \mbox{\layoutmethprocess \, data-derived} positions (i.e., a calculation based on the underlying data itself, e.g., dimensionality reduction), non-semantical positions given by the  \mbox{\layoutmethinput \, input}, and other \mbox{\layoutmethcustom \, custom} placements.

\paragraph{Navigation}

Navigation defines how users \textbf{change their perspective} in relation to the data representation, which is essential for a comprehensive analysis, where overview-only representations are not sufficient.
We identify several distinct techniques: \mbox{\navigationhead \, head movement} and \mbox{\navigationbody \, body movement} (both typical for HMD- and CAVE-based environments), \mbox{\navigationvirtual \, virtual movement} (i.e., the virtual camera moves in the environment), \mbox{\navigationmanipulation \, graph manipulation} (i.e., the camera remains fixed, but the node and edge positions change), \mbox{\navigationdisplay \, display movement} (e.g., an AR tablet is moved), or \mbox{\navigationguided \, guided} navigation, where, for instance, a set of pre-defined viewpoints can be selected.

\paragraph{Collaboration}

One central goal of IA is to support collaborative work~\cite{marriott2018immersive}.
For all approaches, we explore whether they support \textbf{collaborative work}, and if so, categorise it according the Isenberg et al.'s~\cite{isenberg2011collaborative} matrix, defining \mbox{\filledRecText~{Synchronous+Co-Located}}, {\filledTriText~{Synchronous+Remote}}, {\unfilledRecText~{Asynchronous+Co-Located}}, and {\unfilledTriText~{Asynchronous+Remote}} collaboration.

\paragraph{Task Support}
\label{sec:task-support-paragraph}

Task support refers to how the system facilitates core \textbf{analytical interactions} required for an effective visual analysis.
While there are no taxonomies covering these specifically for networks (only for high-level tasks, as in \textit{Study Tasks}), we use the taxonomy by Brehmer and Munzner~\cite{brehmer2013taxononomy}, excluding \textit{navigate}, as it is analysed separately due to its crucial role in immersive settings.
The taxonomy leads to nine interaction tasks: \mbox{\taskselect \, select}, \mbox{\taskchange \, change}, \mbox{\taskfilter \, filter}, \mbox{\taskimport \, import}, \mbox{\taskarrange \, arrange}, \mbox{\taskaggregate \, aggregate}, \mbox{\taskrecord \, record}, \mbox{\taskannotate \, annotate}, and \mbox{\taskderive \, derive}.

\subsubsection{Additional Categories}

\paragraph{Domain}

The application domain provides context for the design choices and evaluation criteria while showing how prevalent immersive network analysis approaches are in the \textbf{different fields}.
We group the approaches into several domain categories, including \mbox{\domainabstract \, abstract} (i.e., not targeting any specific domain), \mbox{\domainbio \, biomedical}, \mbox{\domainknowledge \, knowledge}, \mbox{\domainnetwork \, computer networks}, \mbox{\domainsocial \, social}, \mbox{\domainsoftware \, software}, \mbox{\domainaai \, artificial intelligence}, and \mbox{\domainother \, other}.

\paragraph{Code Access}

Code \textbf{accessibility} is an important factor for reproducibility, extensibility, and community uptake.
For all applications, we check whether the code is available ($\bullet$) or not ($\circ$).

\paragraph{Study Objective}

Study objectives frame the evaluation design and the kind of \textbf{insights sought}.
We observe seven different study objectives investigating \mbox{\goaldisplay \, display \& dimensionality}, \mbox{\goalinteraction \, interaction}, \mbox{\goalnavigation \, navigation}, \mbox{\goallayout \, layout \& perspective}, \mbox{\goalencoding \, encoding} \& representation, \mbox{\goalscale \, scalability}, and \mbox{\goalcollab \, collaboration \& hybrid}.

\paragraph{Data Complexities}
\label{sec:categories-data-complexity}

Especially for user studies, the data used for the evaluation is of high relevance as it affects how \textbf{scalable} and applicable the results are.
While the number of edges is also important, we explore the number of nodes as an approximation for data complexity, as typically this number is given in the publications.
Based on the reported numbers, we define four data complexity sizes (further described in \autoref{sec:study-conditions}): \mbox{\datasizes \, {small}} ($\leq$ 60), \mbox{\datasizem \, {medium}} (61--120), \mbox{\datasizel \, {large}} \mbox{(121--249)}, and \mbox{\datasizexl \, {very large}} \mbox{($\geq$ 250)}.

\paragraph{Study Tasks}

For user studies, a central aspect is which \textbf{tasks} users were asked to solve during the experiments.
Analysing the task is crucial to compare study results and evaluate their impact.
Screening publications that present a user study, we identified three different classes of tasks, where each class can be subdivided according to different taxonomies.
The three classes encompass tasks that relate to classical \textbf{graph analysis}, \textbf{spatial understanding}, and basic \textbf{interaction}.
For graph analysis, we found the taxonomy by Lee et al.~\cite{lee2006task} (refined by Pretorius et al.~\cite{pretorius2014tasks}) to match well with the tasks we found, defining \mbox{\tasktopology \, topological}, \mbox{\taskattributes \, attribute-related}, \mbox{\taskoverview \, overview}, \mbox{\taskbrowsing \, browsing}, and \mbox{\taskhighlevel \, high-level/domain-dependent} tasks.
For spatial understanding targeting the mental map of users, Archambault and Purchase~\cite{archambault2013the} tasks exploring \mbox{\taskmemory \, memory} and \mbox{{\taskdetectchange \, change detection}} capabilities.
Lastly, for tasks targeting basic interactions with the graph, we use the taxonomy by Brehmer and Munzner~\cite{brehmer2013taxononomy} for categorisation (see \hyperref[sec:task-support-paragraph]{\textit{Task Support}}).

\section{Results}
\label{sec:results}

In the following, we present our findings for the design space dimensions and additional analysis categories (see \autoref{sec:methodology-dimensions}) for the \textbf{87 applications}~(\apps{}) in the corpus and incorporating the results of \textbf{59} papers describing \textbf{user studies}~(\studies{}).
While we describe the findings in an aggregated way to retrieve an overview of core insights, the results for each individual paper can be found in \autoref{tab:app-table} (applications), \autoref{tab:study-table} (user evaluations), and our online tool.

\subsection{Application Domains \& Code Availability}

Among the surveyed papers, most address domain-agnostic \mbox{\domainabstract \, \textbf{abstract}} network analysis (\appstudycount{26}{31}).
Otherwise, the \mbox{\domainbio \, \textbf{biomedical}} domain is the most prevalent (\appstudycount{25}{9}), covering metabolic pathways, genomic correlations with diseases, protein--protein interactions, and brain network analysis~\cite{p_16,p_28}.
\mbox{\domainsoftware \, \textbf{Software}} systems form another major area (\appstudycount{12}{6}), where networks represent functions, class hierarchies, dependencies, or co-changing code files~\cite{p_39,p_149}.
In \mbox{\domainsocial \, \textbf{social}} networks (\appstudycount{7}{8}), graphs capture relations between individuals, artists, inventors, or online profiles~\cite{p_22,p_70}.
Smaller groups address \mbox{\domainknowledge \, \textbf{knowledge}} graphs (\appstudycount{4}{2}), often for educational or semantic analysis~\cite{p_14,p_85}, \mbox{\domainaai \, \textbf{artificial intelligence}} (\texttt{A:}~4), visualising neural network structures~\cite{p_4,p_15}, and 
\mbox{\domainnetwork \, \textbf{computer networks}} (\appstudycount{3}{2}), focusing on traffic analysis and cybersecurity~\cite{p_87,p_127}.
Furthermore, there are \mbox{\domainother \, \textbf{other}} (n=6) use cases, targeting fields such as criminology, geospatial analysis, manufacturing, and enterprise processes.

Besides the application domain, we also checked the public availability ($\bullet$) of source code for applications or lack thereof ($\circ$), and when it was last updated.
This indicates whether immersive network analysis systems are currently ready to use or extend for interested researchers or end users, and if the systems remain compatible.
In total, we were able to access publicly available code (through valid links) for \textbf{17 papers} and found none for 70 papers.
We checked the last publication date for bundles or the last commit date for repositories.
Four applications have been published or updated in \textbf{2024} using GitHub~\cite{p_109,p_142,p_155} or Zenodo~\cite{p_59} (see \autoref{fig:paper-bhavnani-kerle-malcharek} right).
For five applications, the latest state is from \textbf{2023} and accessible via GitHub~\cite{p_16,p_39,p_93,p_119,p_123}.
Three further applications from \textbf{2021} are available on GitHub~\cite{p_91,p_118} or via a university website~\cite{p_133}.
The remaining available applications have received their latest update \textbf{at least five years} ago:~\cite{p_100} (2019), \cite{p_67,p_97} (2018, same application), \cite{p_65} (2017), and~\cite{p_27} (2015).

\begin{summarybox}
We observe that most applications and studies are not specifically tailored to the use cases of a specific domain.  
In instances where they are, the adaptations typically concern aspects such as the visual encoding, layout (e.g., based on brain regions~\cite{p_97}), study tasks (e.g., estimating cyber attacks~\cite{p_3}), or context objects (e.g., a biological cell~\cite{p_64}).  
As these factors often do not substantially limit the applicability to other contexts, one surprising observation is that publications in many cases predominantly cite work from within the same domain.  
While this pattern may be partly explained by a \textbf{lack of awareness} of research conducted in other fields, this work aims to help \textbf{bridge} findings across different domains.
Further, a large majority of applications are not \textbf{available} or not \textbf{maintained} anymore.
We thus recommend that future approaches publish code in online repositories, so it remains accessible and could also be maintained or forked by the community.
\end{summarybox}

\subsection{Study Conditions}
\label{sec:study-conditions}

Among the 59 study papers, 53 describe \textbf{controlled quantitative experiments}, where at least one variable is systematically changed, and the effect is measured in a quantitative way, while six studies report a \textbf{qualitative user study}, systematically collecting qualitative user feedback~\cite{friedl2024a}.
Some papers contain the results of multiple studies, for instance, when a pilot study has been conducted prior to the main evaluation.
The papers report \textbf{participant numbers} between 3 and 60 with a mean of 20.04.
The median is 18, and the standard deviation is 10.71.

\begin{figure*}
    \centering
    \includegraphics[width=\linewidth]{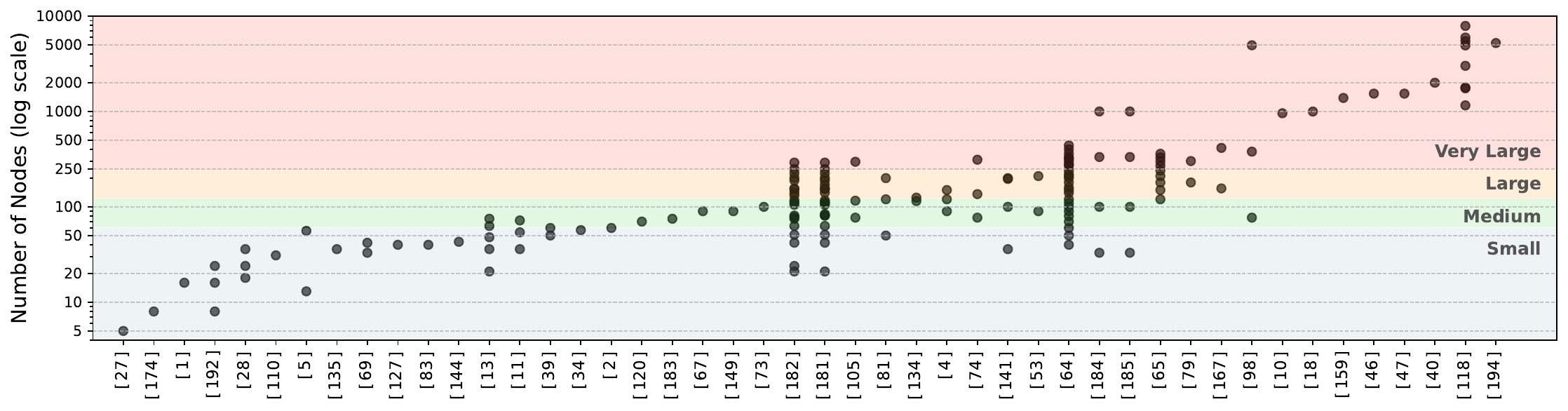}
    \caption{The graph sizes (number of nodes) for each user study (sorted by the median of the reported node numbers), ranging from 5 to 7885 (mean: 449.82, median: 120). A logarithmic scale is applied to ensure that differences can still be seen for small node numbers.}
    \label{fig:node-distribution}
\end{figure*}

As the \textbf{complexity} of networks plays a crucial role, highly affecting the applicability of approaches and generalisability, we also analyse the complexity of networks used in evaluations.
While also characteristics such as the number of edges, components, and clusters are important, we focus on the number of nodes, as this information was mostly present in the papers and gives an impression of the overall complexities.
For 46 of the 59 study papers, the number of nodes was given (see \autoref{fig:node-distribution}), ranging between 5 and 7885, with a mean of 449.82, and a median of 120.
24 of the papers report multiple numbers of nodes that were used, 22 used only a single graph size.
Based on a statistical analysis of the provided node numbers (considering the spread, quartiles, and interquartile ranges), we distinguish four classes of network sizes: \mbox{\datasizes \, {small}} ($\leq$ 60), \mbox{\datasizem \, {medium}} (61--120), \mbox{\datasizel \, {large}} \mbox{(121--249)}, and \mbox{\datasizexl \, {very large}} \mbox{($\geq$ 250)}.
In total, we found 24 studies using \textit{small}, 21 \textit{medium}, 12 \textit{large}, and 19 \textit{very large} networks.
18 studies incorporate data from more than one of these classes, with only three studies covering all.

While many papers report multiple graph sizes used in their studies (see~\autoref{fig:node-distribution}), only one investigation focuses on the \textbf{scalability} regarding immersive network analysis, i.e., how large graphs can get to still be analysable meaningfully with an immersive visualisation approach.
Ware and Mitchell~\cite{p_146} extended their original study~\cite{p_107} and investigated the effect of an increasing number of nodes and edges on the ability to trace paths in a S3D setting.
Their user study finds that the error rate linearly increases with the difference in the number of edges and nodes.

\begin{summarybox}
In the analysed studies (mostly \textbf{quantitative} evaluations), we find slightly more than 50\% of the studies testing multiple sizes, while the others did not consider data \textbf{complexity} as a factor.
When multiple sizes were tested, they were often in a similar range  (see~\autoref{fig:node-distribution}), while only three studies covered all four complexity categories.
Further, with a median of 120 and a maximum of less than 8000 nodes, the tested networks are all rather \textbf{small} compared to real-world datasets.
Studies explicitly investigating the effects of data complexity are rare and need to be investigated further in future research.
\end{summarybox}

\subsection{Display Technology}
\label{sec:res-display-tech}

\paragraph*{\displayhmd \ Head-Mounted Displays}
HMDs are by far the most prevalent technology across applications (n=49) and studies (n=32).
We differentiate between \textbf{VR} devices, which provide a fully virtual environment, and \textbf{AR/MR} systems, which blend virtual elements with the real world.  
AR/MR is dominated by the Microsoft HoloLens device (v1 and v2), occasionally combined with VR headsets~\cite{p_144}.
Its main advantage is the ability to situate visualisations in real-world contexts (e.g., networks overlaid on physical tables~\cite{p_156}), though surprisingly few applications exploit this potential.  
Most HMD applications use VR (n=46), frequently targeting specific devices such as the HTC Vive, Oculus Rift, and Meta Quest.
While this offers deep immersion and strong spatial cues, many systems are tied to outdated hardware (29 prior to 2020, 9 after), with only a handful adopting device-independent frameworks like SteamVR (e.g.,~\cite{p_59}).
Several studies comparing VR applications with \textbf{immersive or non-immersive conditions} generally report \textbf{positive} outcomes for VR, though results vary depending on the task.
Kotlarek et al.~\cite{p_10} find VR enhances structural interpretation compared to 2D desktops, yet spatial memory tasks favour 2D setups.
Beltrán and Geng~\cite{p_29} show VR significantly improves speed and decision-making accuracy in collaborative network tasks over 2D projections. Huang et al.~\cite{p_84} compare different displays and interactions, observing advantages in embodied conditions, especially for interaction-intensive tasks, though results are inconclusive.
McGuffin et al.~\cite{p_101} highlight fewer errors and higher user preference for VR over 2D in path tracing, with no significant difference from physical 3D models augmented by AR.
Erra et al.~\cite{p_132} find VR and gesture-based interfaces boost engagement and immersion but involve steeper learning curves.
Schaller and Schreiber~\cite{p_143} report higher effectiveness and satisfaction with their 2D interface, but suggest VR's potential warrants further investigation.
Sin et al.~\cite{p_121} demonstrate improved comprehension and engagement when analysing knowledge graphs in immersive VR versus desktop conditions.
Comparisons between 2D setups and stereoscopic HMD-based AR reveal \textbf{varied results} across studies.
Ask et al.~\cite{p_3} find stereoscopic AR improves situational awareness, confidence, and reduces communication demands in collaborative cybersecurity tasks, though 2D setups outperform AR for topology recognition.
Ablett et al.~\cite{p_52}, evaluating a hybrid AR-high-resolution display versus laptop setups for navigation tasks, show non-immersive conditions being faster and less error-prone.
Belcher et al.~\cite{p_128} find tangible rotation interactions significantly aid graph link analysis in AR, while stereoscopy alone offers no advantage over desktop setups.
Mehra et al.~\cite{p_124} report significant benefits of AR for software comprehension tasks, including improved efficiency, recall, and reduced cognitive load compared to traditional 2D visualisations.
Lastly, Yu et al.~\cite{p_83} examine AR display properties, concluding that lower viewing angles and fewer nodes facilitate faster performance, with users preferring rotating diagrams rather than physical navigation.

\paragraph*{\displaycave \ CAVE-Like Environments}
The second most common immersive display technology in applications (n=14) is CAVE-like environments, but less frequent in studies (n=5).
These setups typically use \textbf{S3D projection} on multiple \textbf{walls} with user tracking to cover a large field of view.  
Applications often employ room-scale CAVEs or miniCAVEs with shutter or polarised glasses.
Studies show that increased immersion (multi-wall vs. single-wall) and active head tracking \textbf{improve} topological and navigation tasks.
Comparisons with VR suggest that CAVE systems offer similar accuracy, while HMDs can achieve faster task completion at lower cost.
Some studies compare CAVE-like environments with HMD-based VR and non-immersive setups, generally finding \textbf{positive} effects of immersion.
Cordeil et al.~\cite{p_69} report similar accuracies for collaborative network tasks between VR and CAVE, but faster completion times with VR.
Bacim et al.~\cite{p_116} show higher display fidelity (four-wall stereoscopic CAVE with head tracking) significantly improves efficiency in topological tasks.
Henry and Polys~\cite{p_117} demonstrate increased immersion and egocentric navigation within CAVE systems, benefiting overview and local topological tasks, respectively.

\paragraph*{\displaylargetwod \ Large 2D Displays}
While S3D CAVE-like systems are much more common, we also found applications incorporating displays largely \textbf{covering} the user's \textbf{field of view} (n=9).
These consist of flat display walls~\cite{p_41}, spherical displays with the user in its centre~\cite{p_94}, or multiple displays arranged in a semi-circle around the user, creating a large homogeneous display~\cite{p_38}.
We further found three approaches using a room-scale multi-wall display (similar to a CAVE) without stereoscopic vision~\cite{p_28}.
Two-dimensional displays without stereoscopic vision have been incorporated into nine user studies.
These were either large display walls (n=5), covering a large amount of the user's vision at one side, or room-scale (n=4), mostly with projector-based displays covering multiple walls.
With few exceptions, most studies do not compare large 2D displays to other display modalities or only in hybrid settings~\cite{p_52}.
Betella et al.~\cite{p_126} find room-scale immersive 2D displays enhance users' structural comprehension of networks compared to 2D visualisations. 

\paragraph*{\displaymobile \ Tracked Mobile Displays}

\begin{figure}
    \centering
    \begin{subfigure}{0.4\textwidth}
      \includegraphics[height=3.7cm]{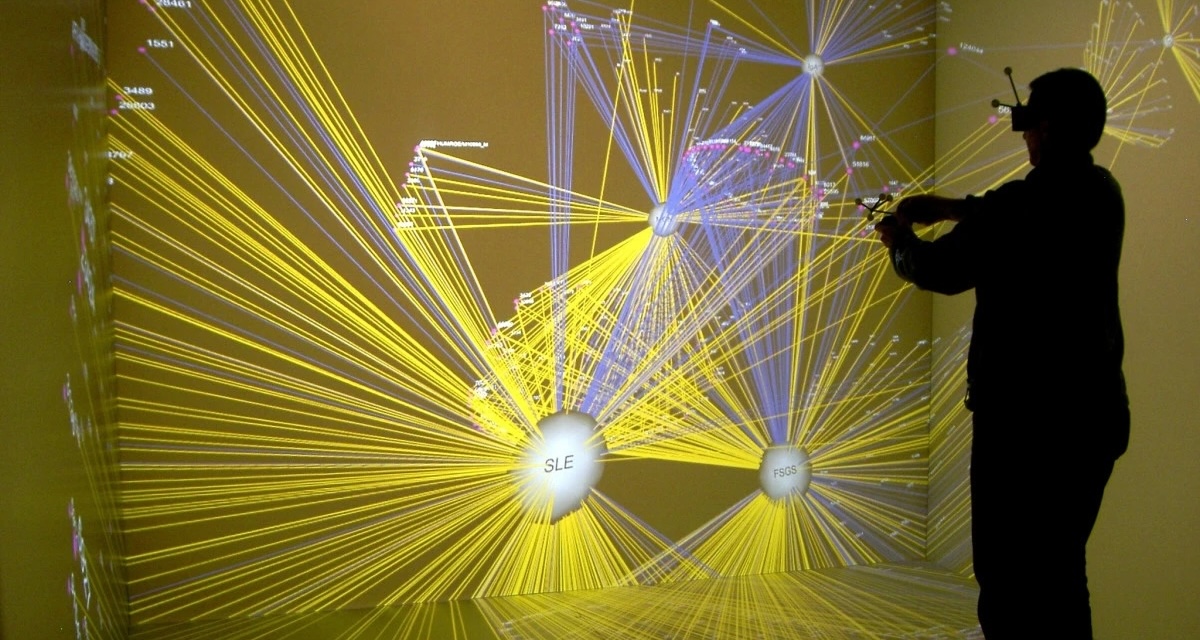}
    \end{subfigure}%
    \hfill
    \begin{subfigure}{0.525\textwidth}
        \includegraphics[height=3.7cm]{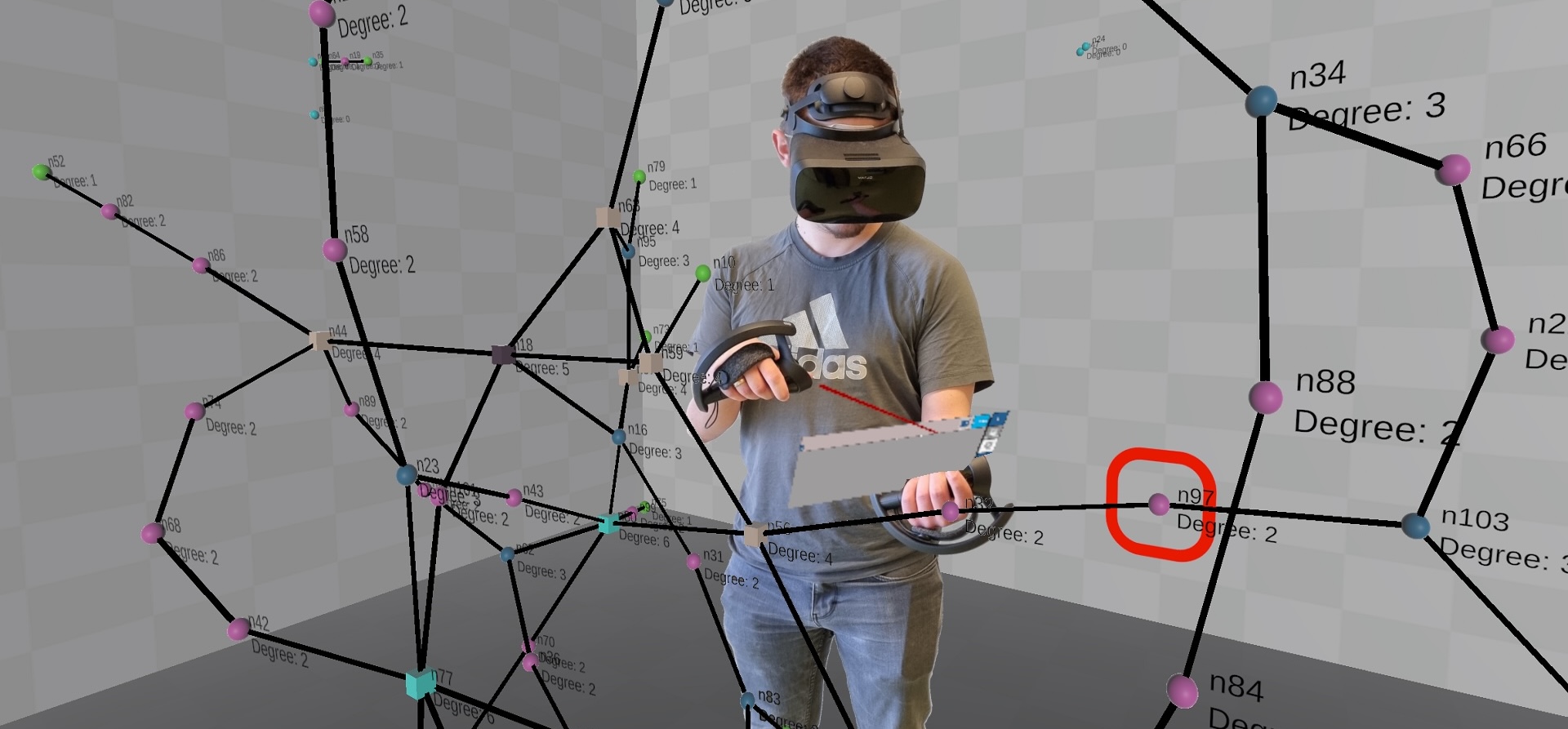}
    \end{subfigure}
    \caption{\textbf{Left:} A user explores a biomedical network with nodes corresponding to diseases or genes and edges to up or down regulations. A room-scale CAVE with head tracking and S3D vision immerses the user in the data. Retrieved from~\cite{p_40} (\href{https://creativecommons.org/licenses/by/2.0/}{CC BY 2.0 License}). \textbf{Right:} GAV-VR is an open-source framework for immersive network analysis supporting most state-of-the-art VR headsets. The software is designed to be extensible (e.g., for user studies) and supports handheld interaction controllers. Retrieved from~\cite{p_59} (\href{https://creativecommons.org/licenses/by/4.0/}{CC BY 4.0 License}).}
    \label{fig:paper-bhavnani-kerle-malcharek}
\end{figure}

Tracked mobile displays (\appstudycount{4}{5}) include smartphones and tablets providing \textbf{see-through AR} and \textbf{spatial tracking}. 
The network visualisation is projected within the displayed environment, allowing the user to walk around or move the device to regions of interest (even into the graph)~\cite{p_22}.
As for most previously discussed HMD-based AR applications, real-world positions (e.g., relating to nodes) were not incorporated.
Studies investigating mobile AR find mostly \textbf{positive} effects.
Stanko and Kriglstein~\cite{p_11} present a mobile AR educational game for learning Dijkstra's shortest path algorithm, comparing AR with 3D and 2D models against non-AR visualisations, finding the AR versions to be more engaging, though less effective for understanding.
Li and Wang~\cite{p_22} design and evaluate an interactive mobile AR application for topological analysis tasks and compare it to a 2D variant. The AR condition led to higher answer times and lower accuracy, but it was considered more exciting, motivating, and innovative.
Abdelfattah et al.~\cite{p_36} compare 2D and AR visualisations for software systems between novice and experienced developers, finding that AR only helps novices (not experts) to improve dependency identification in large systems.

\paragraph*{\displaythreed \ 3D Screens}
The least common display technology among the surveyed applications is 3D screens.
We only found three approaches making use of a S3D display, where the display is not tracked (compared to HMD setups) or the visual field is not largely covered (compared to CAVE-like setups).
The stereoscopic effect is achieved through colouring (Anaglyph)~\cite{p_70}, shutter glasses~\cite{p_87}, or an optical see-through AR display based on a projector and a transparent glass-like medium~\cite{p_135}.
The papers and corresponding technologies, dating back to 1995, 2012, and 2015, reinforce the impression that fixed 3D monitors have gradually been \textbf{replaced} by \textbf{HMD or large-scale devices}.
However, there are multiple user studies (n=14) evaluating the use of 3D screens, either without head tracking or with perspective changes based on the tracked user (Fishtank VR).
In most cases, monitors were used, incorporating either polarisation glasses (n=3) or shutter glasses (n=6).
Other studies use projectors with synced shutter glasses or polarisation.
Two studies used mirrors to present each eye with the image of a distinct monitor.

The majority of studies compare non-immersive 2D settings with 3D screens, observing mostly \textbf{positive effects} for \textbf{S3D} screens.
Roberts et al.~\cite{p_17} evaluate the impact of reconstruction filters for visualisation and stereoscopic displays on 3D path-searching tasks. They show that stereoscopic displays improve spatial accuracy without affecting response times compared to monoscopic settings, while filter choice significantly affects task performance.
Roberts and Ivrissimtzis~\cite{p_106} further evaluate reconstruction filters for network visualisations and path-searching tasks, finding that higher opacity and balanced filters optimise spatial understanding. S3D displays outperform the monoscopic condition in accuracy.
Hassaine et al.~\cite{p_82} analyse the impact of stereoscopic depth and viewpoint density in multiview 3D screens, showing that even small stereo depth (2 cm) significantly improves path-searching accuracy and reduces task latency.
Ware and Franck~\cite{p_48} investigate the impact of stereo and motion cues (via head coupling) on the perception of 3D network visualisations. 
Their findings indicate that head-coupled stereo significantly enhances graph comprehension (error rate), enabling a threefold increase in the size of graphs understood, with motion cues contributing more than stereo alone.
In another study~\cite{p_148}, the authors evaluate the effect of stereoscopy and perspective changes, e.g., graph rotation or head coupling, finding that these significantly improve the efficiency and accuracy of path tracing.
They present a further experiment~\cite{p_130}, evaluating path tracing tasks using 2D and 3D displays (with and without head coupling and stereoscopic vision), suggesting that head-coupled stereoscopic displays allow users to perceive three times more complex graphs than 2D representations, with head coupling being particularly responsible.
Ware and Mitchell~\cite{p_107} assess the effects of stereoscopy, rendering style, and motion cues on path tracing for 3D graphs, finding that a setup with stereoscopy and motion outperformed the other conditions, achieving 90~\% accuracy for 1000 nodes, whereas 33-node graphs led to the same accuracy in 2D. Notably, the motion cue was more helpful than stereoscopic vision alone.
However, some studies report \textbf{mixed or negative findings}.
Fabrikant et al.~\cite{p_1} compare monoscopic 3D, S3D, and 2D displays, finding no significant differences in analysis task solving and concluding that the added value of S3D is insufficient to counter the added cognitive demand.
Greffard et al.~\cite{p_26} examine the impact of stereoscopy for community detection, finding that S3D enhances accuracy in identifying complex graph structures compared to 2D and monoscopic 3D, though 2D remains faster for simpler graphs. They observe different interaction strategies based on the display condition.
Greffard et al.~\cite{p_136} compare 2D, 3D perspective, and S3D displays (projector-based) for community detection, showing that stereoscopy outperforms 3D perspective in most cases. Compared to 2D, S3D is only more efficient for data of higher complexity (i.e., more nodes, edges, and clusters), while 2D is faster in simpler cases.

\paragraph*{Combinations}
Several systems (n=7) combine devices to \textbf{merge strengths}: e.g., \textit{focus+context} setups with HoloLens plus large 2D walls~\cite{p_52}, mobile displays augmenting static screens~\cite{p_62} (see \autoref{fig:paper-kister-neurocave} left), or zSpace integrated into CAVE environments for linked overview and detail~\cite{p_64,p_76}.
Such combinations show promise for balancing \textbf{immersion} with \textbf{information density}, though evaluations are rare.
Also, CAVE-like settings were combined with 3D monitors~\cite{p_65} and handheld screens used mainly for navigation~\cite{p_77}.
While studies combining different display modalities to hybrid systems show great \textbf{potential} in other areas~\cite{feiner1991hybrid}, studies with mixed setups are rare for network data.

\begin{table*}
    \tiny
    \setlength{\tabcolsep}{1.1pt}
    \caption{The summarised analysis results for the \texttt{applications}. \autoref{tab:dimensions-table} provides a legend for the different symbols. If available, the $\bullet$ symbols link to the code repositories. An interactive variant of the table is provided at \websiteurl.}
\begin{tabular*}{\textwidth}{lllllllllllll|lllllllllllll}
    \toprule
\rotatebox{90}{\textbf{Reference}} & \rotatebox{90}{\textbf{Domain}} & \rotatebox{90}{\textbf{Display}} & \rotatebox{90}{\textbf{Interaction}} & \rotatebox{90}{\textbf{Node Encoding}} & \rotatebox{90}{\textbf{Edge Encoding}} & \rotatebox{90}{\textbf{Layout Dimensions}} & \rotatebox{90}{\textbf{Layout Perspective}}  & \rotatebox{90}{\textbf{Layout Method}}  & \rotatebox{90}{\textbf{Navigation}} & \rotatebox{90}{\textbf{Collaboration}} & \rotatebox{90}{\textbf{Code}} & \rotatebox{90}{\textbf{Task Support}} & \rotatebox{90}{\textbf{Reference}} & \rotatebox{90}{\textbf{Domain}} & \rotatebox{90}{\textbf{Display}} & \rotatebox{90}{\textbf{Interaction}} & \rotatebox{90}{\textbf{Node Encoding}} & \rotatebox{90}{\textbf{Edge Encoding}} & \rotatebox{90}{\textbf{Layout Dimensions}} & \rotatebox{90}{\textbf{Layout Perspective}}  & \rotatebox{90}{\textbf{Layout Method}}  & \rotatebox{90}{\textbf{Navigation}} & \rotatebox{90}{\textbf{Collaboration}} & \rotatebox{90}{\textbf{Code}} & \rotatebox{90}{\textbf{Task Support}}\\
\bottomrule
    \end{tabular*}
    \label{tab:app-table}
    \setlength{\tabcolsep}{6pt}
    \end{table*}

\begin{summarybox}
Studies on immersive display modalities largely emphasise the \textbf{advantages} of \textbf{immersive} technologies and \textbf{3D} visualisations.
Especially multiple older studies using S3D screens exhibit clear advantages in spatial accuracy when combined with head-tracking.
But also studies focusing on newer technology, particularly VR HMDs and CAVE-like environments, find advantages for tasks involving spatial understanding and user engagement.
Findings broadly support benefits associated with stereoscopy, motion cues, head-coupling, and spatial immersion for \textbf{comprehending} complex graph structures and enhancing user \textbf{motivation}.
However, these advantages are context-dependent, and comparative results vary depending on the specific \textbf{tasks} and display \textbf{conditions}, underscoring the necessity for additional research using modern hardware and \textbf{standardised} evaluation protocols.
AR and MR technologies offer considerable potential for \textbf{context-aware} visualisations, although their full benefits remain largely untapped.
CAVE-like systems provide performance comparable to VR HMDs but often entail higher costs, making VR headsets generally more attractive in terms of \textbf{affordability} and ease of deployment.
Large-scale 2D displays have shown value particularly for structural \textbf{comprehension} tasks, though comparative studies remain limited.
Tracked mobile displays and hybrid visualisation setups, integrating multiple display modalities, promise substantial potential by \textbf{balancing} immersion, spatial comprehension, and information density, yet remain relatively unexplored.
\end{summarybox}

\subsection{Interaction Modalities}

\paragraph*{\interactioncontroller \ Controllers}
Controllers are the most prevalent interaction modality, \textbf{extensively utilised} across both applications (n=51) and empirical studies (n=23).
Typically, handheld devices like tracked controllers, joysticks, or styluses provide interaction capabilities, such as selecting nodes, rotating visualisations, or activating menus~\cite{p_8,p_154}.
This modality is predominant in \textbf{HMD-based} setups~\cite{p_78} but also widely applied in CAVE-like systems~\cite{p_53,p_77} and S3D screens~\cite{p_64,p_76}.
Some studies incorporate alternative controller types like foot pedals for hands-free control~\cite{p_49}, or tracked physical objects~\cite{p_128}, but without focusing on their effect.

\paragraph*{\interactiongestures \ Gestures}
Gesture-based interaction is employed in 19 applications and evaluated in 13 user studies, using both older sensor-based methods (gloves or magnetic sensors~\cite{p_28,p_71}) and modern camera-tracking solutions like Leap Motion or Kinect~\cite{p_7,p_37}.
Infrared cameras in a 2D dome~\cite{p_94} and the camera-based hand tracking in early AR hardware (such as HoloLens 1) enable basic interaction, mostly based on predefined gestures~\cite{p_156}.
Advanced headsets (like HoloLens~2 or Meta Quest 2) leverage precise finger tracking, facilitating \textbf{natural and direct manipulations} such as grabbing or rotating graph elements~\cite{p_15,p_127}.
Multiple user studies investigate the effect of using gestures for interaction, often with \textbf{mixed results}, or aim to find natural gestures based on user feedback.
Huang et al.~\cite{p_5} present a gesture-based system for graph visualisation in virtual reality and evaluate its applicability compared to mouse input for graph analysis tasks. They find that gesture controls outperform mouse inputs for complex graph exploration despite longer learning times.
Erra et al.~\cite{p_7} evaluate a VR system using natural gestures for 3D graph exploration, comparing it to a desktop mouse/keyboard setup for information retrieval and aggregation tasks, finding that the 3D condition is more challenging and time-intensive but enhances engagement and playfulness.
Another study by Erra et al.~\cite{p_132} reaches similar conclusions.
Frey et al.~\cite{p_80} conducted a user study to identify intuitive hand gestures for overview, zoom, and detail-on-demand interactions with participants, suggesting tapping or grabbing a node, pinching, or swimming to zoom.
The study by Huang et al.~\cite{p_84} sees potential for embodied interaction modalities, especially for interaction-intensive tasks.

\begin{figure}
    \centering
    \begin{subfigure}{0.55\textwidth}
      \includegraphics[height=3.32cm]{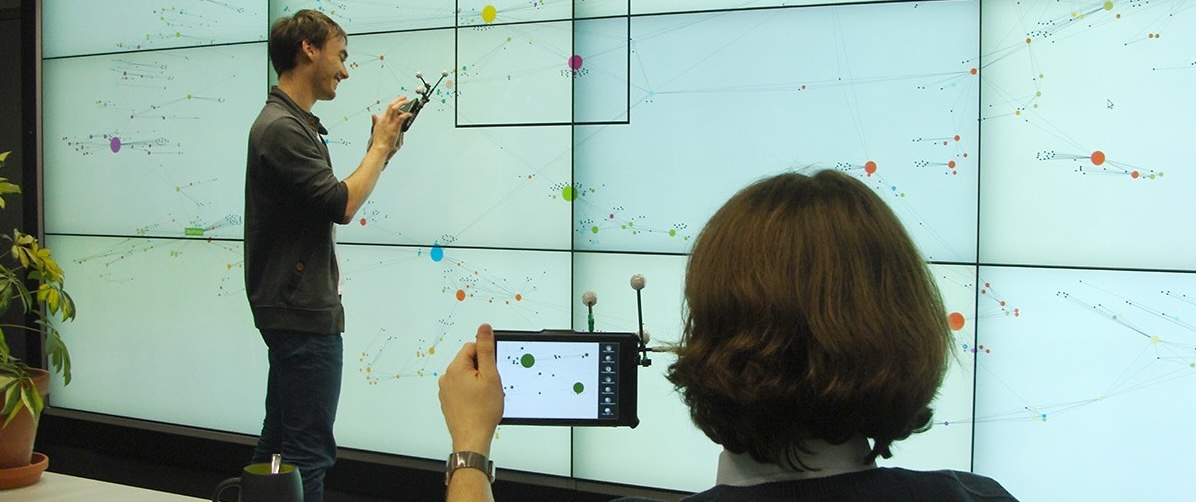}
    \end{subfigure}%
    \hfill
    \begin{subfigure}{0.45\textwidth}
        \includegraphics[height=3.32cm]{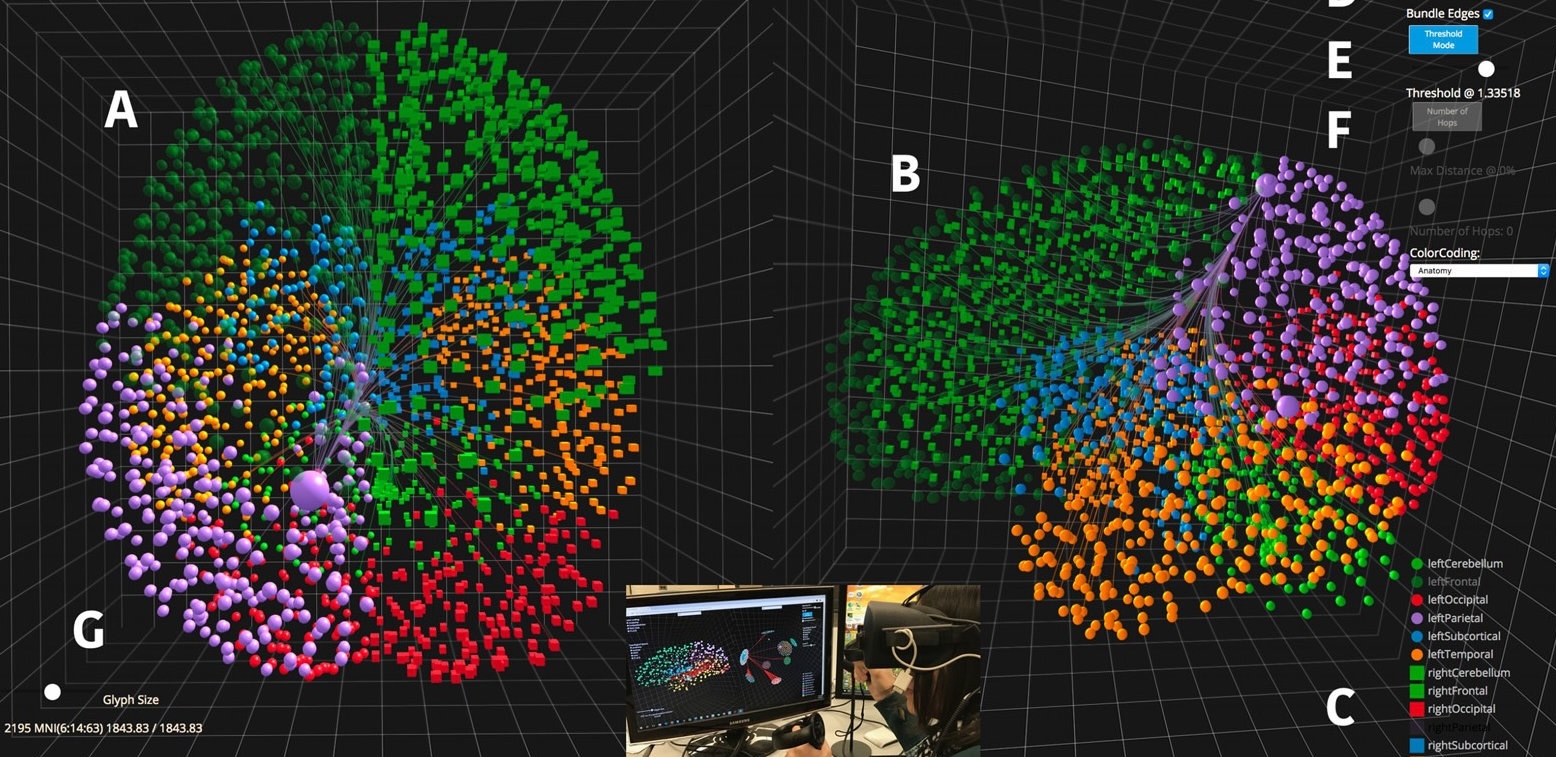}
    \end{subfigure}
    \caption{\textbf{Left:} A hybrid interface combining a 2D wall for overview and tracked handheld displays allows multiple users to physically move with the mobile device, changing the size and position of the focus region displayed on the handheld display. Retrieved from~\cite{p_62} (with permission from John Wiley \& Sons). \textbf{Right:} NeuroCave is a VR system for interactive brain data analysis. It uses colour and shape encodings to represent nodes from different brain regions, with edge bundling to enhance path perception. Users can select from multiple layouts, including semantic mapping (left) or cluster-based (right). Retrieved from~\cite{p_97} (\href{https://creativecommons.org/licenses/by/4.0/}{CC BY 4.0 License}).}
    \label{fig:paper-kister-neurocave}
\end{figure}

\paragraph*{\interactiontouch \ Touch}
Touch interaction (\appstudycount{11}{6}) is commonly integrated via handheld touchscreen devices (like tablets), either for \textbf{mobile AR} applications~\cite{p_22, p_93} or in combination with large 2D or 3D \textbf{screens}~\cite{p_18,p_24} (see \autoref{fig:paper-kister-neurocave} left).
Large touch displays, occasionally complemented by handheld devices, provide similar interactions~\cite{p_41,p_62} (see \autoref{fig:paper-kister-neurocave} left). 
Some studies typically employ direct touchscreen interaction for graph manipulations, especially in mobile settings, but without comparing the modality itself.

\paragraph*{\interactionmousekeyboard \ Mouse/Keyboard}
As multiple studies use 2D or 3D monitors for presenting graph structures, the classical mouse/keyboard interaction idiom is widely used in applications and studies (\appstudycount{10}{22}).
This is particularly true for applications using a S3D \textbf{screen} to display representations~\cite{p_65,p_70} and combined desktop-immersive interfaces with synchronised views~\cite{p_32,p_64}.
Three applications allow users to adjust the 3D perspective via keyboard and mouse while using a 3D stereo device~\cite{p_7,p_27,p_76}.
In one application, a mouse trackball and SpaceMouse are used for graph rotation and positioning~\cite{p_134}.
While studies highlight the \textbf{practicality} of mouse/keyboard setups for precise interactions, immersive modalities, such as gestures, typically \textbf{outperform} the classical setup~\cite{p_7,p_132}.

\paragraph*{\interactiongaze \ Gaze}
Gaze-based interaction (\appstudycount{8}{4}) leverages \textbf{head} or \textbf{eye tracking} for selection, detail retrieval, or camera manipulation tasks, usually by implementing virtual cursors controlled by head orientation or eye movements~\cite{p_35,p_153}.
In some approaches, gaze is used to dynamically adjust detail levels~\cite{p_127} or employ a \textit{detail lens} at the user's focus area.
The rare studies incorporating gaze employ it to control a virtual cursor as an auxiliary modality to \textbf{complement} controllers or gestures, but without evaluating this modality~\cite{p_53,p_103}.

\paragraph*{\interactionspeech \ Speech}
Speech interaction is less common but gaining interest (\appstudycount{7}{2}).
While simple \textbf{voice commands} trigger predefined actions (such as creating new nodes)~\cite{p_135} or annotations~\cite{p_29}, advanced approaches integrate \textbf{speech-to-text} and NLP, allowing natural language querying of databases, modifying visualisations dynamically~\cite{p_14,p_98,p_144}.
Studies primarily implement speech input in combination with other modalities for annotation creation.

\paragraph*{\interactionmovement \ Movement}
While head or body movement is commonly used to naturally change perspective, it is \textbf{rarely employed }in the applications for further interaction with the visual representation (\appstudycount{5}{3}).
If incorporated, it is used to control the graph rotation~\cite{p_19}, steer free-fly navigation~\cite{p_78}, modify the level of detail~\cite{p_103}, or change the detail view shown on a separate mobile device~\cite{p_62} (see \autoref{fig:paper-kister-neurocave} left).
In a similar way, movement is only part of a few user evaluations, where room-scale walking, body leaning, or head shaking~\cite{p_49} are part of the interaction.
However, regarding empirical evidence concerning the effects, this modality remains \textbf{underexplored}.

\begin{summarybox}
Most reviewed approaches did not explicitly focus on the selection or evaluation of interaction modalities, often opting for those requiring the \textbf{least implementation effort}, such as mouse/keyboard for screen-based setups or controllers for immersive environments.
Empirical evidence on the effects of different modalities remains sparse, though immersive interactions, notably gestures, tend to demonstrate \textbf{promising results} regarding \textbf{naturalness} and \textbf{engagement}.
Future research should further investigate immersive modalities and apply the results from previous research~\cite{bueschel2018interaction}, aiming for more intuitive and natural interactions, and systematically analyse their impact.
\end{summarybox}

\subsection{Encoding}

For the encoding of networks in immersive settings, we identified multiple factors influencing the \textbf{appearance}.
Besides visual aspects, such as the appearance of nodes, edges, or contextual objects, some approaches also investigate additional \textbf{senses} or the representation type itself.

\paragraph*{Nodes}
Nodes are most commonly visualised using \textbf{shaded} 3D elements in immersive applications, especially \mbox{\nodessphere \, \textbf{spheres}} (n=54) and \mbox{\nodescube \, \textbf{cubes}} (n=16), although 2D primitives such as \mbox{\nodescircle \, \textbf{circles/points}} (n=16/2) and \mbox{\nodeplane \, \textbf{rectangles}} (n=5) are also used.
Visual differentiation between node classes or attributes is typically achieved via shape, colour, size, text labels, and occasionally embedded images.
These encoding techniques are similarly reflected in study environments, with studies most often using spheres (n=29), cubes (n=9), and circles (n=17).
Experimental results indicate that visual attributes can influence \textbf{perception and performance}.
For instance, Hensen et al.~\cite{p_93} find that image-based node representations in mobile AR significantly improve memorisation and usability.
Colour and text are commonly used for attribute mapping in both applications (n=49, n=43) and studies (n=19, n=18), though size and images are rarely evaluated explicitly.

\begin{figure}
    \centering
    \begin{subfigure}{0.59\textwidth}
      \includegraphics[height=3.07cm]{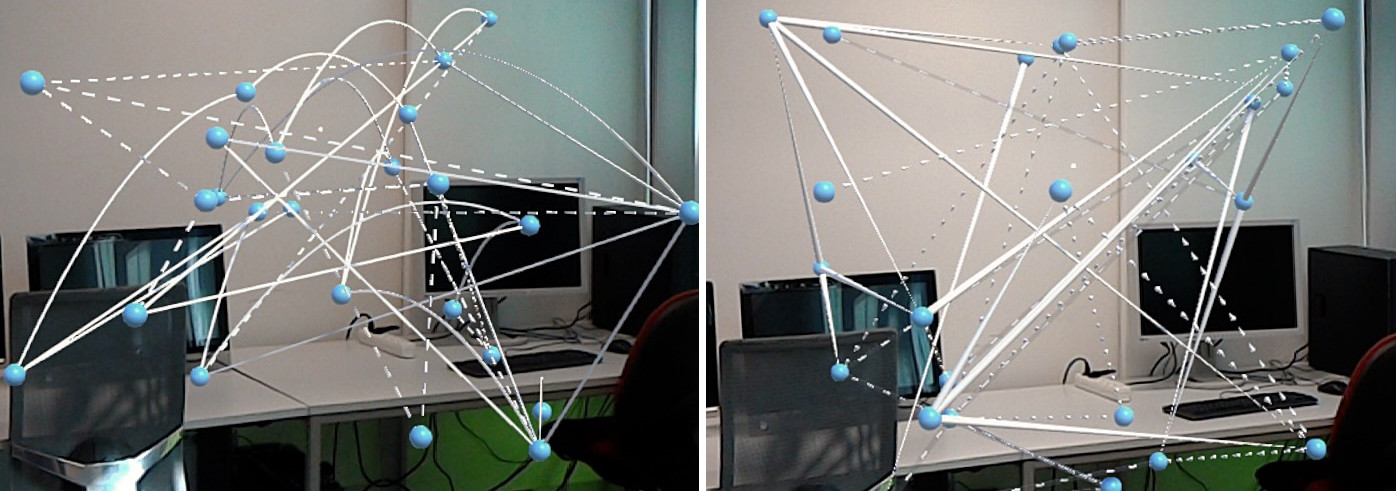}
    \end{subfigure}%
    \hfill
    \begin{subfigure}{0.41\textwidth}
        \includegraphics[height=3.07cm]{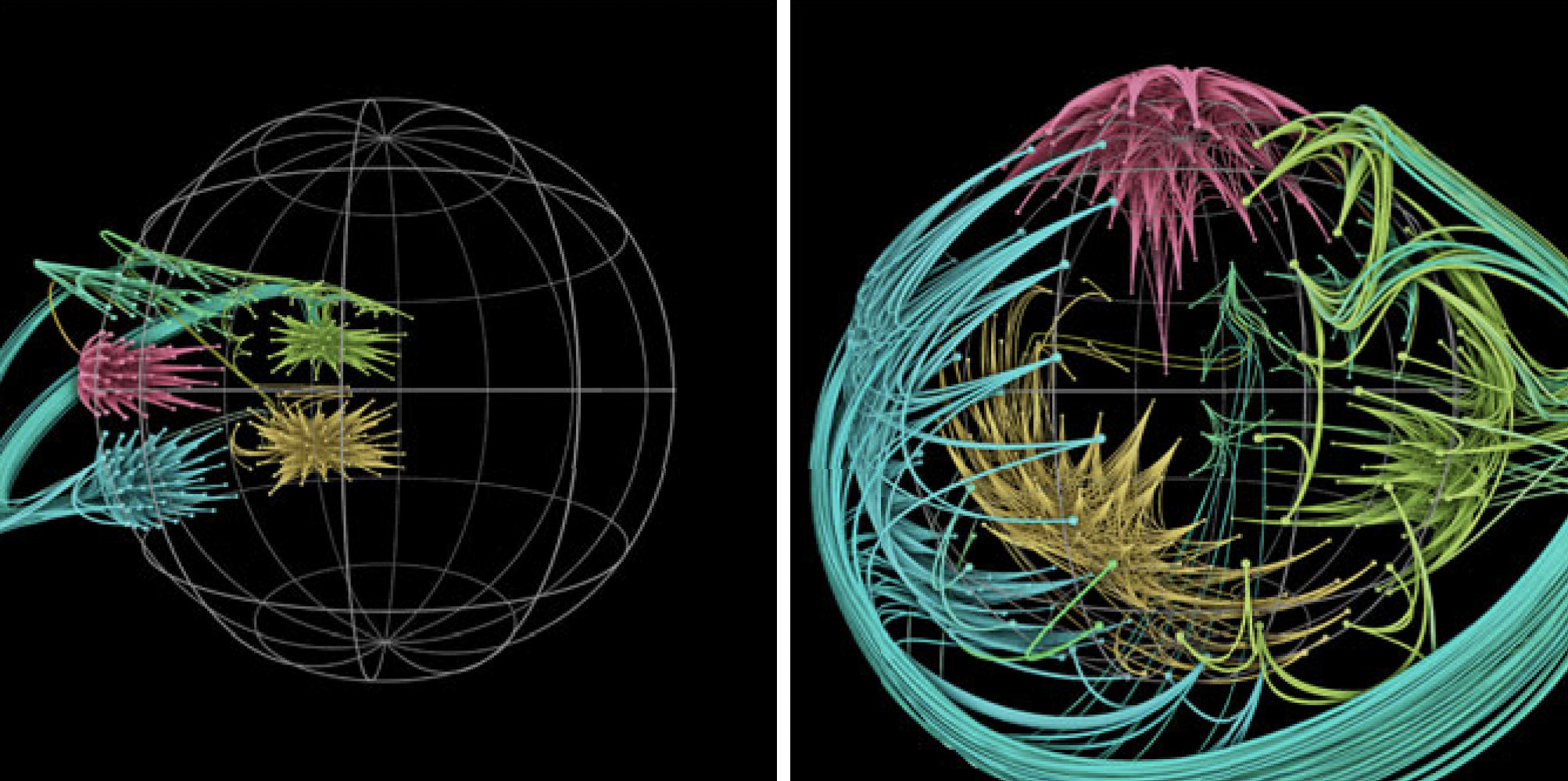}
    \end{subfigure}
    \caption{\textbf{Left:} Büschel et al. investigate the applicability of different edge encodings for undirected (first) and directed (second) graphs in an AR setup. Retrieved from~\cite{p_23} (with permission from IEEE). \textbf{Right:} Kwon et al. investigate spherical layouts for graph structures in VR, perceived with an egocentric perspective. The amount of the sphere surface covered can be varied, covering different portions of the user's FOV. Retrieved from~\cite{p_9} (with permission from IEEE).}
    \label{fig:paper-bueschel-kwon}
\end{figure}

\paragraph*{Edges}
Edges in immersive network applications are predominantly shown as \textbf{non-shaded} 2D \mbox{\edgeslines \, \textbf{lines}} or rectangles (n=71), though a smaller subset uses shaded 3D \mbox{\edgestubes \, \textbf{tubes}} (n=14).
Applications often map additional information to edge colour, thickness, direction (via arrows or gradients), and, less commonly, text labels.
These patterns are echoed in user studies, which overwhelmingly use lines (n=38), with only limited use of tubes or cuboids (n=17).
Some empirical studies evaluate the effect of different edge encodings.
Drogemuller et al.~\cite{p_13} investigate the effect of Lombardi-inspired curved edges compared to straight ones, finding straight edges preferred by users and leading to better task-solving performance.
Büschel et al.~\cite{p_23} explore edge styles for 3D graphs in an AR setup. For path-tracing tasks with undirected graphs, curved edges perform worse than straight and dashed ones (see \autoref{fig:paper-bueschel-kwon} left). For directed graphs, animated and tapered edges work better than glyph-based edges.
Büschel et al.~\cite{p_81} evaluate eight edge encodings for AR node-link diagrams, asking users to map ordinal or nominal attributes. Colour performs best for nominal attributes, while animated patterns and stippling are preferred for ordinal data.
Joos et al.~\cite{p_137} study encodings for 3D node-link diagrams to compare two weighted networks, finding glyph-based edges most effective. A subsequent study comparing node-link and matrix representations shows node-link diagrams outperform matrices for given tasks.
Kwon et al.~\cite{p_9} investigate different shadings for edges (see \autoref{fig:paper-bueschel-kwon} right).

\paragraph*{Visual Context}
While most systems visualise graphs in otherwise empty 3D space (n=67), some incorporate \textbf{contextual backdrops}.
These range from data-specific environments like brain models~\cite{p_134}, cells~\cite{p_76}, and CAD models~\cite{p_21}, to abstract references such as virtual spheres~\cite{p_12}, planes~\cite{p_138}, and metaphoric islands~\cite{p_150}. 
One AR system situates the graph on a real physical table~\cite{p_156}.
The role of context objects has not been evaluated so far.

\paragraph*{Additional Senses}
Although immersive network visualisation is primarily visual, some systems leverage \textbf{additional senses}.
Three applications incorporate \textbf{sonification}~\cite{p_15, p_28, p_108}, and one adds olfactory feedback~\cite{p_75}.
In the \textbf{olfactory} system, a scent corresponding to a node’s class is emitted when it is selected, supporting memorability. 
Sound has been used to map properties like node count, edge strength, and proximity.
Study-based evaluations confirm that \textbf{multisensory} encoding can be \textbf{beneficial}.
Papachristodoulou et al.~\cite{p_25, p_110} demonstrate that congruent soundscapes improve spatial accuracy and facilitate estimation of network characteristics in brain connectomes.
Another study~\cite{p_101} employs data physicalisation to offer haptic feedback, allowing users to explore network structure through touch, which led to similar results as a VR condition.
While there is not much research yet on incorporating additional senses for immersive graph analysis, findings suggest that audio, scent, and touch can \textbf{supplement visual analysis} in these settings, particularly when encodings are meaningfully aligned with the data.

\paragraph*{Representation}

Beyond visualisation of specific representations, some studies compare \textbf{representations} or evaluate their \textbf{retrieval}.
Roberts et al.~\cite{p_17} and Roberts and Ivrissimtzis~\cite{p_106} investigate reconstruction filters used to derive node-link diagrams, finding links between the filter and users' spatial understanding.
Pan et al.~\cite{p_56} introduce a 3D adjacency matrix for visualising triads in networks, combining it with a 2D node-link diagram.
Their study shows that adding the matrix improves task accuracy and efficiency by reducing visual clutter compared to a 2D NL diagram alone.
Similarly, Joos et al.~\cite{p_137} compare node-link and matrix representations in VR, finding node-link diagrams superior for edge weight comparison.
AlTarawneh et al.~\cite{p_30} use stereoscopic depth to visualise 3D layered graphs and analyse the effect of graph size and transparency on user performance.
Their results suggest that graph size did not significantly influence the ability to detect variations and relationships, while transparency negatively affected accuracy but improved efficiency.
Alper et al.~\cite{p_112} present stereoscopic highlighting, a technique for 2D graph visualisation on stereo displays, which projects regions of interest closer to the viewer.
Results show that combining stereoscopic and static visual highlighting improves adjacency and connectivity task performance.

\begin{summarybox}
Immersive network visualisations commonly use \textbf{familiar encodings}: nodes as shaded 3D \textbf{spheres} or \textbf{cubes}, and edges as simple 2D \textbf{lines}.
The perceptual effects of shaded 3D versus flat 2D elements remain underexplored.
Visual distinctions are typically conveyed through colour, shape, size, and labels, which is reflected in both applications and user studies.
While straight edges often outperform more complex alternatives (e.g., curved or stippled edges) in clarity and task performance, several \textbf{novel encodings} for nodes, such as glyphs, images, and multisensory elements (e.g., sound, scent, touch), have shown promise.
These findings suggest a need to \textbf{revisit established conventions} and explore hybrid approaches, particularly for node and attribute representation in immersive settings.
\end{summarybox}

\subsection{Layout}

Regarding the graph layout (i.e., the node positions), we analyse whether two or three dimensions are used (\textit{dimensionality}), from which \textit{perspective} the user perceives the layout, and which \textit{method} is used to assign coordinates.

\paragraph*{Dimensionality}

Across applications and studies, \textbf{dimensionality} plays a key role in how networks are represented and perceived.
A majority of the applications and studies utilise \mbox{\layoutdimthree \, \textbf{three-dimensional}} layouts (\appstudycount{76}{54}), treating all dimensions either equally, for instance through force-based positioning, or distinctively, such as encoding an attribute in the third axis or allowing users to vertically raise nodes for further exploration~\cite{p_53}.
Multiple applications and studies are restricted to \mbox{\layoutdimtwo \, \textbf{two-dimensional}} layouts (\appstudycount{11}{28}).
23 studies examine \mbox{\layoutperspexo \,\layoutperspego} \textbf{both} dimensionalities, frequently using them as experimental conditions for comparison.
Most studies comparing the layout dimensionality also change the display modalities between the conditions, typically \textbf{favouring 3D conditions} for task performance and spatial understanding (see \autoref{sec:res-display-tech}).
Additionally, Feyer et al.~\cite{p_0} compare 2D, 2.5D, and 3D representations for multi-layer networks in VR, finding that no single configuration universally outperforms others, but task-specific benefits emerge

\paragraph*{Perspective}

The perspective from which users view a network (either \mbox{\layoutperspexo \, \textbf{exocentric}} or \mbox{\layoutperspego \, \textbf{egocentric}}) has been shown to significantly affect \textbf{navigation} and \textbf{search performance}~\cite{p_43}.
For both applications and studies, exocentric perspectives dominate (\appstudycount{43}{34}), where users observe the network externally (e.g., \autoref{fig:paper-joos-belcher} left).
While three applications provide purely egocentric views, placing users inside the network as if they were part of it, none of the studies do.
Further, a notable portion of applications (n=42) and studies (n=25) support \textbf{both} perspectives, often by enabling scale changes and interactive navigation, such as zooming into the network~(e.g., \autoref{fig:paper-bhavnani-kerle-malcharek}), or as comparative conditions. 
Sorger et al.~\cite{p_43} evaluate egocentric network exploration techniques, revealing that egocentric perspectives improve local search efficiency and navigation while maintaining spatial orientation and reducing cybersickness.
The study by Henry and Polys~\cite{p_117} finds potential for egocentric perspectives.
Sun et al.~\cite{p_54} compare room-scale and table-scale environments for logical reasoning tasks, finding three different types of spatial arrangements by users, where some prefer overview perspectives, and others favour stepping inside the network.
Kwon et al.~\cite{p_9} compare three conditions in a VR setup, namely a 2D layout and two spherical 3D arrangements perceived by an egocentric perspective, with one 3D condition incorporating depth routing and advanced shading (see~\autoref{fig:paper-bueschel-kwon} right).
Their results show that 3D layouts facilitate faster interaction and improved correctness for analysis tasks, where especially the condition with additional depth rendering succeeded.

\paragraph*{Method}

The methods used to assign \textbf{spatial coordinates }to network nodes vary widely across applications and studies.
Most applications and user studies (\appstudycount{40}{23}) employ \mbox{\layoutmethforce \, \textbf{force-based}} layouts, with the Fruchterman \& Reingold algorithm being the most common~\cite{fruchterman1991graph}.
Alternatives include the Unity physics engine and algorithms from OGDF~\cite{chimani2013open}.
\mbox{\layoutmethsemantic \, \textbf{Semantic}} layouts, used in 16 applications and seven studies, position nodes based on real-world references such as brain regions~\cite{p_97} (see \autoref{fig:paper-kister-neurocave} right) or geospatial locations~\cite{p_19}.
Other approaches (\appstudycount{8}{1}) rely on \mbox{\layoutmethprocess \, \textbf{data-derived}} node placement, achieved by transforming data using techniques like dimensionality reduction or clustering (\mbox{\layoutmethprocess}).
Additional configurations include layouts provided as \mbox{\layoutmethinput \, \textbf{input}} (\appstudycount{8}{2}), show nodes in \mbox{\layoutmethlayered \, \textbf{layered}} structures (\appstudycount{6}{2}), align them in \mbox{\layoutmethcircular \, \textbf{circular}} or spherical patterns (\appstudycount{8}{4}), or use other \mbox{\layoutmethcustom \, \textbf{custom}} or hybrid strategies (\appstudycount{6}{14}), such as random placement, multiple methods combined, or an initial layout determined by an algorithm is manually adapted to fulfil certain criteria~\cite{p_148}.
Razi et al.~\cite{p_72} evaluate six VR layouts for hierarchical pedigree networks and observe higher accuracy and satisfaction with spherical and vase-like layouts, while flat floor-based arrangements underperform.
Otherwise, studies on layout methods (without changing display or dimensionality) are rare.

\begin{summarybox}
Most applications and studies adopt \textbf{three-dimensional} layouts, typically leveraging \textbf{force-based} positioning algorithms, often combined with \textbf{exocentric} perspectives.
A notable number of works compare dimensionalities or perspectives (but mostly in combination with other display modalities), often favouring 3D and egocentric conditions for improved interaction and comprehension.
Alternative layout methods include \textbf{semantic} positioning based on real-world references, \textbf{data-derived} transformations (e.g., dimensionality reduction), and various \textbf{custom or hybrid} strategies.
Evaluations of these methods are limited, with only a few studies isolating layout generation from display modality or dimensionality effects.
\textbf{Gaps} remain in systematic evaluations of layout \textit{methods} across different tasks and user goals. Especially underexplored are \textbf{egocentric perspectives} in user studies, and how layout choices affect performance independently of other factors.
\end{summarybox}

\begin{table*}
    \tiny
    \setlength{\tabcolsep}{0.85pt}
    \caption{A summary of the analysed user evaluations with their conditions and goals. \autoref{tab:dimensions-table} provides a legend for the different symbols. An interactive variant of the table is provided at \websiteurl.}
\begin{tabular*}{\textwidth}{llllllllllllll|llllllllllllll}
    \toprule
\rotatebox{90}{\textbf{Reference}} & \rotatebox{90}{\textbf{Study Focus}} & \rotatebox{90}{\textbf{Study Type}} & \rotatebox{90}{\textbf{Domain}} & \rotatebox{90}{\textbf{Display}} & \rotatebox{90}{\textbf{Interaction}} & \rotatebox{90}{\textbf{Data Size}} & \rotatebox{90}{\textbf{Node Encoding}} & \rotatebox{90}{\textbf{Edge Encoding}} & \rotatebox{90}{\textbf{Layout Dimensions}} & \rotatebox{90}{\textbf{Layout Perspective}}  & \rotatebox{90}{\textbf{Layout Method}}  & \rotatebox{90}{\textbf{Navigation}} & \rotatebox{90}{\textbf{Tasks}} & \rotatebox{90}{\textbf{Reference}} & \rotatebox{90}{\textbf{Study Focus}} & \rotatebox{90}{\textbf{Study Type}} & \rotatebox{90}{\textbf{Domain}} & \rotatebox{90}{\textbf{Display}} & \rotatebox{90}{\textbf{Interaction}} & \rotatebox{90}{\textbf{Data Size}} & \rotatebox{90}{\textbf{Node Encoding}} & \rotatebox{90}{\textbf{Edge Encoding}} & \rotatebox{90}{\textbf{Layout Dimensions}} & \rotatebox{90}{\textbf{Layout Perspective}}  & \rotatebox{90}{\textbf{Layout Method}}  & \rotatebox{90}{\textbf{Navigation}} & \rotatebox{90}{\textbf{Tasks}}\\
\bottomrule
    \end{tabular*}
    \label{tab:study-table}
    \setlength{\tabcolsep}{6pt}
    \end{table*}

\subsection{Task}

We analysed which tasks the different applications support (interaction tasks) and were used in user studies for the evaluation (interaction, graph analysis, and spatial understanding tasks).

\paragraph*{Interaction Tasks}

Almost all applications (n=77) support some kind of \mbox{\taskselect \, \textbf{selection}}, mainly of nodes, edges, context objects, or clusters.
This can be achieved \textbf{directly} by selecting the corresponding visual representations or \textbf{indirectly}, for instance, by selecting or searching for certain attributes (e.g., in menus).
In most systems, selected objects are highlighted and can be used to retrieve details or to modify the graph (e.g., by deleting the selected node), but it is also used as input to query the graph using LLMs~\cite{p_98}.
Six user evaluations required participants to select a graph element, mostly a single node (e.g.,~\cite{p_47}).
But only in two studies was the design of different \textbf{selection techniques} evaluated.
Prouzeau et al.~\cite{p_45} evaluate multi-user interaction techniques for exploring graph topology on touch wall displays, focusing on selection methods enabling collaborative graph analysis. They find that propagation-based selection supports tighter coordination and improves performance on complex graphs, whereas basic selection promotes parallel work but reduces accuracy.
Joos et al.~\cite{p_47} evaluate six techniques for node selection in VR, addressing challenges like clutter and occlusion. Their study shows that traditional ray-based selection suffices for simple networks, while complex networks benefit from methods like fisheye techniques or filter-plane approaches.

In contrast to static presentations, interactive visual analysis tools often contain functions to give users control over some aspects of the representations.
Among the surveyed applications, 35 support some sort of user-controlled \mbox{\taskchange \, \textbf{change}}.
This mainly relates to the \textbf{appearance} of nodes and edges (e.g., which visual primitive to be used), the visual encoding (e.g., colour scales or which visual variables to use for mapping attributes), the \textbf{interaction} mode, \textbf{layout}, or the \textbf{size} of the entire visualisation.
In one study, users were asked to manually untangle a graph layout~\cite{p_60} using a physics-based method to manually change a graph layout, where clusters behave like rigid bodies connected by joints.
When trying to untangle a given graph on a large 2D display, this system required less effort than a standard baseline.
Halpin et al.~\cite{p_53} incorporate a 3D CAVE environment with a 2D graph layout and let users extrude nodes into the third dimension to see additional details and perceive the structure in another way.
With the extrusion technique, users solving graph analysis tasks were faster for localised tasks, while there was no benefit for overview tasks.

\mbox{\taskfilter \, \textbf{Filtering}} is essential to find information of interest, especially when the visualisation is cluttered due to a high amount of data.
This type of analysis task is supported by 30 systems, which often relate to hiding edges or nodes with certain properties.
While most applications use a menu or direct interaction with the elements, one system incorporates a 3D filtering cube, allowing the application of multiple attribute filters interactively~\cite{p_154}.

As visualisations of large datasets can become highly cluttered when all information is shown, it is sensible to show details only on demand.
Thus, \mbox{\taskimport \, \textbf{importing}} is an essential visual analysis task supported by 30 systems.
Besides loading new datasets or showing additional information for selected entities, the creation of new data is also part of it, e.g., creating additional nodes or edges.

In total, 24 systems provide the option to \mbox{\taskarrange \, \textbf{arrange}} parts of the visual representation.
This either relates to an entire graph, e.g., when multiple of them are shown, and users can change their relative position to each other or to the position change of individual graph elements.
Especially, manually repositioning nodes is a commonly supported task, mostly only resulting in a visual adaptation of edges to match new node positions, but sometimes also changing the data, e.g., the weights of the edges~\cite{p_15}.

It is often beneficial to show graphs at different \textbf{levels of detail} and let users change them not only for hierarchical networks but also for those containing clusters or motifs.
This is summarised with the \mbox{\taskaggregate \, \textbf{aggregate}} task, which is supported by 12 systems.
A group of nodes (e.g., a cluster or nodes in the same hierarchy level) can be replaced by a single one, also aggregating the edges, or a node representing multiple others can be expanded.
This task was part of two studies, e.g., asking users to manually cluster nodes~\cite{p_132}.

To effectively work with visual analysis systems and to keep track of the results, it can be beneficial to \mbox{\taskrecord \, \textbf{record}} the state and history, as supported by eleven applications.
This encompasses storing or exporting views, generating a history log of actions, resetting the view, and allowing users to switch back to another state in the history (undo).

Especially for \textbf{multi-user} collaborations or \textbf{repeated usage} at different times, \mbox{\taskannotate \, \textbf{annotations}} are beneficial and are supported by six of the surveyed applications, mostly storing and displaying comments at nodes, edges, or clusters.

While most systems are restricted to viewing existing data, four applications allow for \mbox{\taskderive \, \textbf{deriving}} and visualising new information.
This is done by allowing users to execute \textbf{algorithms} on the current visualisation and view the result, or by \textbf{manipulating} the underlying data (depending on the selected elements).

Three studies evaluate the \mbox{\tasknavigate \, \textbf{navigation}} capabilities, for instance, asking users to change the orientation of a graph, minimising or maximising the number of visible nodes~\cite{p_52}.

\paragraph*{Graph Analysis Tasks}
This task category mostly focuses on the retrieval of structural or other information from a given graph~\cite{lee2006task}, which is the most common type of incorporated task in the study papers (n=53).
The \textbf{most prevalent} analysis tasks are the \mbox{\tasktopology \, \textbf{topological}} ones (n=48), where users need to analyse the structure, for instance, by finding the shortest path between two given nodes, identifying neighbours, node degrees, or clusters.
Besides structural tasks, multiple studies (n=15) also include tasks targeting the \mbox{\taskattributes \, \textbf{attributes}} of nodes or edges.
Users are asked, for instance, to find a graph element with a certain value or label or to retrieve the value of the given element, such as the weight of an edge.
Further tasks require good \mbox{\taskoverview \, \textbf{overview}} capabilities (n=6) to estimate the number of nodes in general or with a given attribute, for instance, to \mbox{\taskbrowsing \, \textbf{browse}} the graph (n=3), following a path through the network, or to answer more \mbox{\taskhighlevel \, \textbf{high-level or domain-specific}} questions (n=3), e.g., regarding the severeness or type of a cyber attack~\cite{p_3}.

\paragraph*{Spatial Understanding Tasks}
In addition to the classical analysis tasks, some studies focus on the \textbf{spatial understanding} and \textbf{mental map} of users~\cite{archambault2013the}.
Six require the users to \mbox{\taskmemory \, \textbf{memorise}} features of a graph, for instance, to rediscover a certain node, as highlighted previously~\cite{p_9,p_10}.
Furthermore, five user evaluations ask the users to \mbox{\taskdetectchange \, \textbf{detect changes}}, which could relate to newly added nodes~\cite{p_10} or the edge weight differences between two networks~\cite{p_137}.

\begin{summarybox}
The analysis reveals coverage of \textbf{fundamental interaction} tasks across the applications, particularly \emph{selection}, \emph{changing}, and \emph{filtering}.
However, \textbf{advanced tasks} such as \emph{aggregation}, \emph{recording}, \emph{annotating}, and \emph{deriving} are supported far less frequently.
While most applications support only a subset of analysis interaction tasks (see \autoref{tab:app-table}), some also provide a broad spectrum~\cite{p_59,p_33}.
To allow for effective and powerful visual analysis, future applications should aim for \textbf{broad task support}.
While some studies have already investigated how tasks can be supported by immersive applications, for instance, for selection, changes, or navigation, there are still many open questions regarding the optimal support for other tasks or conditions.
Further, in user studies, \emph{graph analysis tasks}, especially topological analysis, dominate user studies (48 out of 53 studies), whereas tasks involving attributes, overview, browsing, or domain-specific reasoning are less explored, leaving \textbf{potential gaps} in evaluating more \textbf{complex analytical workflows}.
\emph{Spatial understanding tasks} are only sparsely studied, with a handful of evaluations targeting memorisation (6 studies) or change detection (5 studies), suggesting that cognitive aspects are still \textbf{underrepresented} in user evaluations.
\end{summarybox}

\subsection{Navigation}

\begin{figure}
    \centering
    \begin{subfigure}{0.56\textwidth}
      \includegraphics[height=3.47cm]{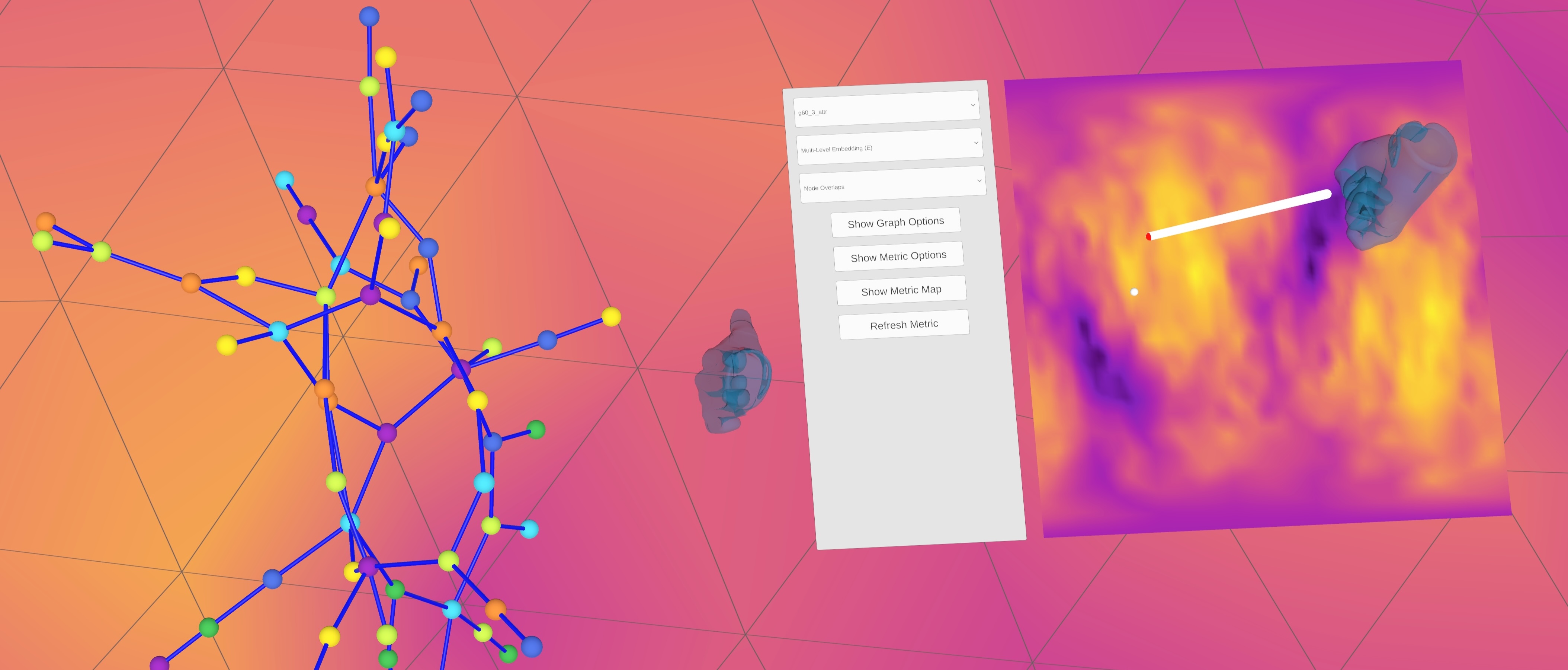}
    \end{subfigure}%
    \hfill
    \begin{subfigure}{0.44\textwidth}
        \includegraphics[height=3.47cm]{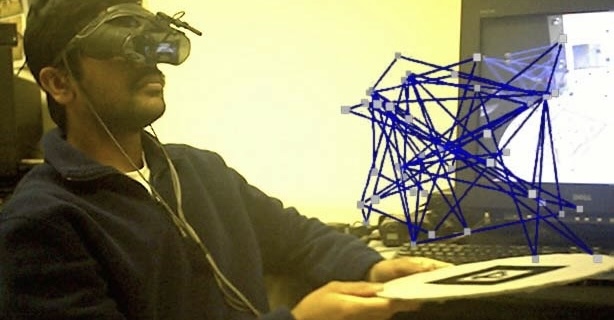}
    \end{subfigure}
    \caption{\textbf{Left:} Users stand inside a sphere perceived by VR, where a colour coding shows aesthetics-based qualities of different viewpoints. The sphere surface or a 2D map can be used to select optimal viewpoints (yellow), providing guided navigation. Retrieved from~\cite{p_12} (\href{https://creativecommons.org/licenses/by/4.0/}{CC BY 4.0 License}). \textbf{Right:} Belcher et al. incorporate a physical plate with a marker to naturally move and rotate a 3D graph perceived with an AR HMD. Retrieved from~\cite{p_128} (with permission from IEEE).}
    \label{fig:paper-joos-belcher}
\end{figure}

Navigation is a crucial feature for visual analysis systems, enabling data viewing at various scales (\textbf{overview+detail}) and adjusting \textbf{perspectives} in 3D settings.
Among the surveyed papers, most report multiple incorporated navigation techniques (\appstudycount{76}{38}), but also some only make use of one method (\appstudycount{10}{17}).

\paragraph*{\navigationhead \ Head \& \navigationbody \ Body Movement}

The most supported navigation technique is \textbf{Head Movement} (\appstudycount{77}{42}), followed by \textbf{Body Movement} (\appstudycount{55}{26}).
VR and AR systems \textbf{natively} incorporate \textbf{head tracking}, enabling users to change perspectives naturally. 
To allow perspective changes beyond physical head rotation, the visualisation itself can be rotated based on head direction~\cite{p_19}. 
In CAVE-like setups and 3D monitors with head tracking, such as the zSpace, the same principle as HMDs applies for \textbf{natural perspective adaptation}. 
For large-scale 2D or 3D screens surrounding users, like 2D domes~\cite{p_24,p_94}, users can adjust perspectives and visualisation views by rotating their heads, even without explicit tracking.
Similar to perspective changes in HMDs or CAVE-like systems, relative user position can be tracked and translated into \textbf{virtual camera movement}, enabling walking or interaction. 
In room-scale setups, users can navigate and adjust viewpoints by \textbf{physically walking}, even without explicit tracking. 
In some 2D setups, user positions influence a handheld view~\cite{p_62} or control a correspondingly moved virtual camera~\cite{p_44}.
Zielasko et al.~\cite{p_49} evaluate five hands-free VR navigation techniques for seated immersive data analysis, including leaning metaphors and accelerometer-based pedals.
They highlight body leaning and accelerometer pedals as the best performers for 3D graph navigation tasks.
Betella et al.~\cite{p_44} use a 3D visualisation environment with large 2D room-scale screens and compare modes of camera adjustment based on body tracking.
The results favour an absolute mapping of floor coordinates to the camera over relative displacements, with the latter performing worse, especially at room scale.
James et al.~\cite{p_102} examine head movement for AR-based hands-free navigation on a large shared display for path-following tasks, moving a map with different facilitating techniques.
They find persistent AR connections to improve accuracy in high-precision tasks, while transient links are more effective for low-precision tasks.

\paragraph*{\navigationvirtual \ Virtual Movement}

Instead of requiring physical movement, 33 applications and eleven studies support \textbf{virtual movement}, modifying the camera's position or angle. 
We analysed the papers for methods enabling free camera control, excluding approaches that rely on predefined viewpoints (see \textit{Guided Change of Perspective}). 
The two primary methods identified are \textbf{teleportation}, where users can select a target viewpoint (e.g., by pointing), and \textbf{fly navigation}, allowing the camera to move continuously along axes.
Drogemuller et al.~\cite{p_46} compare VR navigation using teleportation, flying (one- and two-handed), and a miniature-based approach.
Their study finds a two-handed flying mechanic, where users indicate a movement vector with their hands, to be the fastest and most preferred for topological tasks.

\paragraph*{\navigationmanipulation \ Graph Manipulation}

Especially monitor-based, but also other immersive settings allow users to navigate through direct \textbf{manipulation} of the graph visualisation, including \textbf{translation}, \textbf{rotation}, and \textbf{scaling} (\appstudycount{56}{23}). 
These interactions are typically facilitated by predefined gestures~\cite{p_14}, the \textit{grab} interaction, where users press controller buttons to apply the controller's relative position and rotation to the graph, either upon direct contact with the visualisation~\cite{p_31} or indirectly from any position~\cite{p_154}. 
Additionally, rotation can be performed around one or more axes, such as via joystick control~\cite{p_12} (see \autoref{fig:paper-joos-belcher} left). 
In 2D systems, this approach aligns with the \textit{zoom+pan} paradigm, commonly used in mobile devices~\cite{p_33,p_38} and large touch displays~\cite{p_41}.

\paragraph*{\navigationguided \ Guided Change of Perspective}

Besides navigation by virtual movement, where a user can freely choose a perspective without support from the system, there are also approaches providing \textbf{guidance} to choose a new point of view (\appstudycount{10}{6}).
The most common guided navigation technique lets users select a node, for instance, using a menu~\cite{p_8}, or by more direct selection~\cite{p_43} (including the eye gaze~\cite{p_58}), changing the viewpoint, such that the centre is clearly visible or centred.
The same technique can also be applied to select individual \textbf{predefined perspectives}~\cite{p_100} or a path along multiple of these~\cite{p_25}.
Further, a 3D \textbf{overview} sphere has been used to define the point of view of a large, 2D detail view~\cite{p_52}.
Other approaches rotate the network \textbf{continuously} to allow for interaction-free perspective changes~\cite{p_107}.
For the zSpace, the stylus can be used to select a region of interest, focusing it in the 3D visualisation itself~\cite{p_64} or in a \textbf{linked view} in a CAVE-like setting~\cite{p_76}.
Lastly, an approach has been introduced~\cite{p_12} that visualises the \textbf{aesthetics-based} quality of different perspectives and lets the user select one (see \autoref{fig:paper-joos-belcher} left).

\paragraph*{\navigationdisplay \ Display Movement}

For mobile, spatially tracked displays, navigation can also be achieved by \textbf{moving the display} (\appstudycount{5}{5}).
This is mostly the case for AR settings, where a spatially \textbf{tracked mobile device} with camera-based AR is used to visualise a graph in 3D~\cite{p_22,p_34}.
Users can move the device to change the viewpoint and the distance to the graph, and often move into the representation itself.
In~\cite{p_62}, moving a mobile display linked with a large 2D wall (see \autoref{fig:paper-kister-neurocave} left) changes the focus of the detail view, adapting the alternative representation visualised on the device.

\begin{summarybox}
Most applications implement \textbf{multiple navigation techniques}, with head and body movement being most common, particularly in VR and AR settings, where they support \textbf{natural perspective changes}.
Virtual movement methods, such as teleportation and flying, are also widely used, with \textbf{two-handed flying} shown to perform best for spatial tasks.
While direct graph manipulation is prevalent across many setups, \textbf{guided} perspective changes and navigation via display movement remain less common and underexplored, though initial studies suggest \textbf{potential benefits}.
Despite broad adoption of various navigation methods, comprehensive empirical comparisons across techniques remain limited, as the few existing studies mostly investigate alternatives for one navigation category.
In particular, \textbf{guided and display-based navigation} approaches are \textbf{underexplored}, and there is a need for more systematic evaluations to assess their effectiveness across diverse tasks and system setups.
\end{summarybox}

\subsection{Collaboration}

Collaboration is a central aspect of data analysis and a key goal of \textbf{immersive analytics}~\cite{billinghurst2018collaborative}.
We found 57 applications provide no support for collaborative work, 30 allow for one (n=26) or multiple (n=4) collaboration types.
For the others, the most common supported type is \mbox{\filledRecText~\textbf{Synchronous+Co-Located}} (n=24), where people work together at the same site and time.
This is \textbf{implicitly} the case for applications using large displays that can be viewed by multiple persons at the same time.
Some applications extend this implicit support by more \textbf{explicit} methods, like providing mobile devices~\cite{p_33,p_62} (see \autoref{fig:paper-kister-neurocave} left) or HMDs~\cite{p_35} to each user with an \textbf{individual view}.
For HMD-based applications, \textbf{synchronisation} (e.g., using a server) can be incorporated to let multiple users analyse and modify the same visualisation at the same time \cite{p_29}, a concept that has also been applied to Mobile AR~\cite{p_34}.
Bonyuet et al.~\cite{p_2} allow for collaborative work where one user works in a CAVE and another person works with a PC, viewing and controlling the same interlinked visualisation.
{\filledTriText~\textbf{Synchronous+Remote}} collaboration capabilities are provided by nine systems, mostly HMD-based tools that already support \textit{Synchronous+Co-Located} collaboration by server-based synchronisation.
Four approaches support the {\unfilledRecText~\textbf{Asynchronous+Co-Located}} collaboration type, mainly letting users create shared \textbf{annotations} or images~\cite{p_38} or making the individual views of users accessible to others and dynamic \textbf{interlinking}~\cite{p_21}.
In a similar way, the created views and annotations can also be shared across sites, allowing for {\unfilledTriText~\textbf{Asynchronous+Remote}} collaboration \cite{p_33}, as for instance, shown by ContextuWall~\cite{p_38}.

Some studies specifically investigate collaborative settings, often incorporating \textbf{hybrid} interfaces with promising results.
Sun et al.~\cite{p_35} present a collaborative data analysis system combining wall-size displays and optical see-through head-mounted displays for privacy and multi-level data sharing.
Their study suggests this hybrid setup supports role-specific data access and collaboration, with users primarily working in private areas.
Nishimoto and Johnson~\cite{p_57} enhance a CAVE environment with an AR HMD, extending the view, and test the effect compared to the CAVE-only condition.
While no significant differences in efficiency or accuracy were observed, the integration promotes better spatial awareness and physical navigation.
Schwajda et al.~\cite{p_125} investigate hybrid transformations of graph data from 2D displays into 3D AR spaces, focusing on transition methods and visual link techniques.
Their study shows that user-controlled transitions improve task efficiency, while constant transformations reduce error rates.
Two studies examine collaborative selection on a multi-touch surface~\cite{p_45} and a hybrid combination of a 2D wall and an AR HMD, offering an overview and private visualisation space~\cite{p_102}.

\begin{summarybox}
While immersive analytics also aims to improve \textbf{collaboration support}, a large majority of analysed applications do not support it yet or only \textbf{implicitly}.
However, the applications incorporating \textbf{explicit} collaboration support show \textbf{promising} results, even though there are only a few evaluations investigating collaboration in immersive network analysis approaches.
Thus, applying collaboration support to applications and investigating how users can be supported best, and how the analysis is affected by collaboration support, should be part of future research.
\end{summarybox}

\section{Discussion}
\label{sec:discussion}

The field of visual network analysis in immersive environments has demonstrated significant progress over the last decades regarding applications and evaluations.
Immersive approaches have been adopted across a wide range of \textbf{domains}, from biomedical research and software engineering to social networks, knowledge graphs, AI model inspection, and cybersecurity, covering diverse datasets, use cases, and task types.
However, these efforts often remain isolated within their respective domains, with limited cross-referencing or integration of findings from other fields, despite many solutions being potentially transferable.
This fragmentation reduces opportunities for building on shared insights and hampers the development of broadly applicable, reusable tools.
Our review of approaches not restricted to any specific domain aims to overcome this and show connections between the fields.

Numerous empirical evaluations compared \textbf{immersive and non-immersive environments} for network analysis, covering a broad range of hardware setups, data, tasks, and additional conditions.
In the majority of those evaluations, immersive settings have been shown to be beneficial, e.g., regarding the error rate and efficiency in task solving (mostly topological tasks), with particular advantages observed for tasks requiring spatial understanding, navigation, or complex structural interpretation.
This was often boosted by stereoscopy, head tracking, and motion cues.
However, in some circumstances, non-immersive setups were shown to be advantageous, for instance, for simpler graph structures, topology recognition, or time-critical navigation tasks.
There, 2D or monoscopic displays were faster and sometimes more accurate, particularly in cases where immersive setups introduced higher cognitive load or steeper learning curves.
As apparent, these results are conflicting and show the essential difficulty in the field: the design space of visual network analysis in immersive environments is much larger than its 2D pendant, blowing up the degrees of freedom when designing an application or evaluation, and lowering the comparability of studies significantly.

With our survey, we approached the research questions, \textit{how the design space of visual network analysis in IE is characterised, to what extent it has been realised, and which aspects have been empirically evaluated in prior research}.
In contrast to non-immersive 2D settings, where mostly the visualisation itself (i.e., the graph layout and visual mapping of attributes) and interaction providing visual analytics capabilities form the design space, the literature coding showed a much more \textbf{complex design space} for immersive environments:
It spans multiple interconnected dimensions, with \textbf{display technology}, ranging from HMDs and CAVE-like systems to tracked mobile devices, large 2D walls, and 3D screens, \textbf{interaction modality}, including controllers, gestures, touch, gaze, speech, movement, and traditional mouse/keyboard, \textbf{encoding} of nodes and edges through 2D/3D shapes, colour, size, and other (visual) variables, \textbf{layout}, defined by spatial arrangement, dimensionality, and user perspective, \textbf{navigation} techniques, such as head/body movement, virtual movement, graph manipulation, and guided viewpoints, \textbf{collaboration} support, from synchronous co-located to asynchronous remote, and \textbf{task} support, covering analytical actions like selecting, filtering, changing, or annotating.
Apart from the design space, we explored the results of user \textbf{evaluations}, which were shown to highly vary in the study objectives, but also the conditions (e.g., what kind of data or tasks researchers included).
As the design space not only encompasses a significant set of dimensions, the options for the individual categories are often extensive.

Hence, we observe that large parts of the design space are rather untouched, with applications and studies often following straightforward implementations.
In terms of coverage, certain dimensions are dominated by a small set of well-established choices: VR HMDs are by far the most common \textbf{display technology}, with CAVE-like systems, large 2D walls, tracked mobile devices, and 3D screens appearing far less frequently.
\textbf{Interaction} is most often implemented through handheld controllers, followed by gestures and touch, while gaze, speech, and body movement are rarely adopted or evaluated in depth.
For \textbf{encoding}, shaded 3D spheres for nodes and simple 2D lines for edges remain the default, with limited uptake of richer glyphs, dynamic styles, or multisensory encodings such as audio, haptics, or scent.
\textbf{Layouts} are overwhelmingly three-dimensional, force-based, and exocentric, with egocentric perspectives, semantic or data-derived positioning, and hybrid strategies underexplored.
\textbf{Navigation} is dominated by natural head/body movement and virtual camera controls, while guided perspective changes, display-based navigation, or hybrid techniques are uncommon and rarely compared systematically.
\textbf{Collaboration} support is largely absent or limited to synchronous co-located scenarios, with asynchronous or remote modes rarely realised.
\textbf{Task} support is similarly skewed towards basic operations such as selection, filtering, and simple changes, and study tasks are dominated by topological analysis, leaving higher-level reasoning, graph-specific interaction, and spatial understanding underrepresented.
Overall, this concentration on a narrow subset of options highlights how much of the potential design space remains unrealised, and points to substantial opportunities for innovation, diversification, and systematic comparative evaluation.
Future work could also benefit from more \textbf{hardware-independent} implementations to ensure long-term applicability and facilitate reproducibility, and from exploring \textbf{situated analytics}~\cite{thomas2018situated} by embedding network data within its real-world context.

When considering which aspects of the design space have been empirically evaluated, our analysis reveals a strong imbalance.
Most studies vary and compare \textbf{display technologies}, particularly immersive versus non-immersive setups, with numerous comparisons between VR HMDs, CAVE-like environments, large 2D displays, and 3D screens.
These evaluations often target their impact on \textbf{task performance} and \textbf{spatial understanding}, frequently highlighting benefits of stereoscopy, head tracking, and motion cues.
Other dimensions, however, receive far less systematic attention.
\textbf{Interaction modalities} are rarely compared directly, with exceptions for gestures versus mouse keyboard or controller-based input, and most applications simply adopt the modality most convenient for the hardware, often without integrating promising results from prior interaction research.
\textbf{Encoding choices}, especially node and edge styles, are occasionally studied, with some work on edge curvature, animation, or glyphs, but multisensory encodings~\cite{marriott2018immersive} remain almost entirely unexplored in controlled experiments.
\textbf{Layout} evaluations mainly focus on dimensionality (2D vs. 3D) or perspective (exocentric vs. egocentric), often confounded with changes in display technology, while layout methods themselves are seldom isolated as an experimental factor.
\textbf{Navigation techniques} are sometimes compared within a category, for example, teleportation versus flying, yet cross-category comparisons and studies of guided or display-based navigation remain rare.
\textbf{Collaboration} support is evaluated only in a handful of cases, often through bespoke hybrid systems, and its influence on analysis processes is not yet well understood.
Similarly, evaluations of \textbf{task support} beyond topological analysis, such as attribute-based reasoning, high-level domain-specific tasks, or spatial understanding, are rare.
Moreover, many studies still focus narrowly on accuracy or completion time, neglecting other relevant measures such as engagement, learning, spatial memory, or collaboration quality, which could better reflect \textbf{real-world usage}.
This uneven coverage suggests that while some design space aspects, especially \textbf{display technology}, have received considerable empirical attention, many others remain underexplored, limiting our understanding of how different design choices affect analytical performance, user experience, and collaboration in immersive network analysis.
A further challenge is that the lack of maintained and openly available systems hampers replicability and comparability, underlining the need for \textbf{publicly accessible, up-to-date, and multi-platform} tools to advance the field.

\subsection{Future Challenges \& Directions}

Based on the analysis of the literature and additional consideration of the current research landscape, we estimate that progress in immersive network analysis will require further effort in three central aspects.
Addressing these research priorities across domains will not only deepen our understanding of immersive network analysis but also pave the way for robust, transferable, and widely usable systems that can adapt to the evolving landscape.

\paragraph*{Design Space Exploration}

While questions regarding the design space in immersive environments for network analysis, such as encodings, interaction, navigation, etc., have been analysed already, the previous section shows that there are still many questions that have not been investigated or that should be revisited due to different conditions.
Unique opportunities provided by immersive systems, such as support for effective collaborative work, multi-sensory experiences, and hybrid interfaces, seamlessly combining the advantages of different modalities, are still underexplored.
To this end, \textbf{evaluation practices} must become more systematic, including controlled comparisons across a wider range of tasks, richer performance and user experience measures, and replication of promising findings across domains and datasets.
Further, metrics to measure the suitability of approaches should not only rely on task-solving metrics but also on user-centred measures, such as engagement, cognitive and physical demand, simulation sickness, and ensuring that users enjoy working with the system for a prolonged time.
Lastly, the development and long-term maintenance of \textbf{open, extensible, and hardware-agnostic platforms} (like SteamVR or WebXR) is essential to enable reuse, facilitate study replication, and support community-driven innovation.

\paragraph*{Data \& Scalability}

The analysis of network complexities used in the evaluations showed that, in a large proportion of studies, data \textbf{complexity} is not considered as an explicit factor.
With a median of only 120 nodes, the networks used in many evaluations are comparatively small.
Real-world datasets, however, can be orders of magnitude larger.
For example, in the Stanford Network Analysis Project (SNAP)~\cite{snapnets}, social networks have a mean size of around 310,000 nodes (with a maximum exceeding four million), network sizes in intelligence analysis can usually reach millions of nodes~\cite{Fischer.CommunicationAnalysis.2022}, while biomedical networks average over 233 million nodes, with the largest containing around 3.6 billion~\cite{biosnapnets}.
Given that clutter and occlusion effects grow rapidly with increasing graph size, directly impacting the perceivability of elements and the ability to interact with them, future studies should incorporate \textbf{larger} datasets and ideally employ \textbf{standardised} size categories to improve comparability across evaluations.
Moreover, there is a clear need for studies that explicitly investigate the effects of graph size and the limits of \textbf{perceivability}, as only addressed by one study focusing on this issue.
Scalability also remains an underrepresented challenge in the reviewed applications.
While a few approaches employ overview representations alongside detail views for handling larger graphs, \textbf{performance} optimisation strategies, such as GPU-based instancing, or visual \textbf{simplification} methods, including motif aggregation or other focus+context techniques, are still rarely adopted.
Immersive settings may offer unique advantages for working with large-scale data, but \textbf{targeted research} is required to fully exploit these capabilities.

\paragraph*{Intelligent System}

Current applications and studies often regard visual network analysis as a \textbf{highly manual} process, where users observe a visual representation with basic interaction capabilities and solve low-level tasks, such as verifying path existence.
However, with modern algorithms and user interfaces, such applications should provide more \textbf{guidance} and reduce the manual effort for users as far as possible.
As shown by an application, incorporating LLMs, for instance, can lead to more intelligent user interfaces, where users communicate what they are interested in and get presented with the right data.
Their expertise allows the users to evaluate the data and retrieve new information, forming a perfect symbiosis between human and computer systems.
Further, \textbf{intelligent systems} can guide users to potentially relevant anomalies or suggest adequate analysis methods, which is not part of immersive network analysis systems so far.
With the great results in other domains, but also in immersive analytics~\cite{wang2025immersive}, we see high potential in leveraging AI-based techniques for immersive network analysis, which could also contribute to higher accessibility.

\subsection{Limitations}

Despite careful efforts, this work has limitations.  
Our chosen definitions inherently constrain the scope, keywords, included online libraries (which we addressed, to some extent, through a snowballing approach), and the inclusion criteria.  
These \textbf{constraints} may have excluded relevant work, such as those written in other languages or published only as preprints.  
Furthermore, the manual filtering and classification processes relied on \textbf{human judgement}, which is inherently error-prone.  
In particular, both the literature coding and the derivation of the design space dimensions were conducted by the authors, based on our interpretation of the reviewed material.  
While we followed a structured approach and aimed for consistent application of our categories, this process is inevitably subjective, and different researchers might have drawn slightly different boundaries or groupings.  
To address this, we have developed an interactive website where users can report mistakes, suggest additions, and provide alternative interpretations.  
Regarding the discussion of key findings, we were limited to evaluating \textbf{published results}.
Unfortunately, negative or null findings are often under-reported despite their importance in guiding future research.

\section{Conclusion}
\label{sec:conclusion}

In this survey, we presented a comprehensive state-of-the-art report on visual network analysis in immersive environments. 
We systematically collected, filtered, and categorised relevant publications, resulting in 138 papers, 87 of which describe an application and 59 present a user study informing the applicability or design of such (8 both).
Our analysis and comparison of the different approaches and studies provide information on the coverage of this fragmented field, highlight gaps, and lead to guidelines for future research advancing this field.
Our companion website \websiteurl{} offers an interactive option to explore our results and will be continuously extended.

\begin{acks}
The authors gratefully acknowledge financial support by the Federal Ministry for Economic Affairs and Climate Action (BMWK, grant No. 03EI1048D) and the Deutsche Forschungsgemeinschaft (DFG) – Project-ID 251654672 – TRR 161.
\end{acks}

\bibliographystyle{ACM-Reference-Format}
\bibliography{bibliography}

\end{document}